\documentclass[12pt]{cernprep}
\usepackage{graphicx}
\usepackage{amssymb}
\usepackage{epsfig,color,rotating}
\begin{document}
\newcommand{\dedx}{\mbox{${\rm d}E/{\rm d}x$}}
\newcommand{\EcB}{$E \! \times \! B$}
\newcommand{\omt}{$\omega \tau$}
\newcommand{\omtsq}{$(\omega \tau )^2$}
\newcommand{\rphi}{\mbox{$r \! \cdot \! \phi$}}
\newcommand{\srphi}{\mbox{$\sigma_{r \! \cdot \! \phi}$}}
\newcommand{\dg}{\mbox{`durchgriff'}}
\newcommand{\mg}{\mbox{`margaritka'}}
\newcommand{\pT}{\mbox{$p_{\rm T}$}}
\newcommand{\GeVc}{\mbox{GeV/{\it c}}}
\newcommand{\MeVc}{\mbox{MeV/{\it c}}}
\def\kr{$^{83{\rm m}}$Kr\ }
\begin{titlepage}
\docnum{CERN--PH--EP--2010--026}
\date{13 July 2010}
%\begin{flushright}
%*** DRAFT 02 ***
%\end{flushright} 
%
\vspace{1cm}
\title{CROSS-SECTIONS OF LARGE-ANGLE HADRON PRODUCTION \\
IN PROTON-- AND PION--NUCLEUS INTERACTIONS VI: \\
CARBON NUCLEI AND BEAM MOMENTA 
FROM \mbox{\boldmath $\pm3$}~GeV/\mbox{\boldmath $c$} 
TO \mbox{\boldmath $\pm15$}~GeV/\mbox{\boldmath $c$}}

\begin{abstract}
We report on double-differential inclusive cross-sections of 
the production of secondary protons, charged pions, and deuterons,
in the interactions with a 5\% $\lambda_{\rm int}
$ 
thick stationary carbon target, of proton and pion beams with
momentum from $\pm3$~GeV/{\it c} to $\pm15$~GeV/{\it c}. Results are 
given for secondary particles with production 
angles $20^\circ < \theta < 125^\circ$. Cross-sections on carbon nuclei are
compared with cross-sections on beryllium, copper, tantalum and lead nuclei.
\end{abstract}

\vfill  \normalsize
\begin{center}
The HARP--CDP group  \\  

\vspace*{2mm} 

A.~Bolshakova$^1$, 
I.~Boyko$^1$, 
G.~Chelkov$^{1a}$, 
D.~Dedovitch$^1$, 
A.~Elagin$^{1b}$, 
D.~Emelyanov$^1$,
M.~Gostkin$^1$,
A.~Guskov$^1$, 
Z.~Kroumchtein$^1$, 
Yu.~Nefedov$^1$, 
K.~Nikolaev$^1$, 
A.~Zhemchugov$^1$, 
F.~Dydak$^2$, 
J.~Wotschack$^{2*}$, 
A.~De~Min$^{3c}$,
V.~Ammosov$^{4\dagger}$, 
V.~Gapienko$^4$, 
V.~Koreshev$^4$, 
A.~Semak$^4$, 
Yu.~Sviridov$^4$, 
E.~Usenko$^{4d}$, 
V.~Zaets$^4$ 
\\
\vspace*{8mm}
This publication is dedicated to the memory of our colleague V.~Ammosov 
 
\vspace*{8mm} 

$^1$~{\bf Joint Institute for Nuclear Research, Dubna, Russia} \\
$^2$~{\bf CERN, Geneva, Switzerland} \\ 
$^3$~{\bf Politecnico di Milano and INFN, 
Sezione di Milano-Bicocca, Milan, Italy} \\
$^4$~{\bf Institute of High Energy Physics, Protvino, Russia} \\

\vspace*{5mm}

\submitted{(To be submitted to Eur. Phys. J. C)}
\end{center}

\vspace*{5mm}
\rule{0.9\textwidth}{0.2mm}

\begin{footnotesize}

$^a$~Also at the Moscow Institute of Physics and Technology, Moscow, Russia 

$^b$~Now at Texas A\&M University, College Station, USA 

$^c$~On leave of absence at 
Ecole Polytechnique F\'{e}d\'{e}rale, Lausanne, Switzerland 

$^d$~Now at Institute for Nuclear Research RAS, Moscow, Russia

$^{\dagger}$~Deceased on 11 January 2010

$^*$~Corresponding author; e-mail: joerg.wotschack@cern.ch
\end{footnotesize}

\end{titlepage}

%\newpage
%\mbox{ }
%\tableofcontents
%\vspace{0.8cm}

\newpage 

\section{Introduction}

The HARP experiment arose from the realization that the inclusive differential cross-sections of hadron production in the interactions of few GeV/{\it c} protons with nuclei were known only within a factor of two to three, while more precise cross-sections are in demand for several reasons.
%
%The `neutrino factory' (see Ref.~\cite{neutrinofactory} and further references cited therein) is a serious contender for a future accelerator facility that addresses fundamental questions on neutrino oscillations. One of the neutrino factory's many technological challenges is the production of charged pions with sufficient intensity to achieve the required particle fluxes in the decay chain pions $\rightarrow$ muons $\rightarrow$ neutrinos. It is imperative that pion production cross-sections are under control.
%
These are the optimization of the design parameters of the proton driver of a neutrino factory (see Ref.~\cite{neutrinofactory} and further references cited therein), but also to the understanding of the underlying physics and the modelling of Monte Carlo generators of hadron--nucleus collisions, flux predictions for conventional neutrino beams, and more precise calculations of the atmospheric neutrino flux. 

The HARP experiment was designed to carry out a programme of systematic and precise (i.e., at the few per cent level) measurements of hadron production by protons and pions with momenta from 1.5 to 15~GeV/{\it c}, on a variety of target nuclei. It took data at the CERN Proton Synchrotron in 2001 and 2002.

The HARP detector combined a forward spectrometer with a large-angle spectrometer. The latter comprised a cylindrical Time Projection Chamber (TPC) around the target and an array of 
Resistive Plate Chambers (RPCs) that surrounded the TPC. The purpose of the TPC was track 
reconstruction and particle identification by \dedx . The purpose of the RPCs was to complement the particle identification by time of flight.

This is the sixth of a series of cross-section papers with results from the HARP experiment. In the first paper\cite{Beryllium1} we described the detector characteristics and our analysis algorithms, on the example of $+8.9$~GeV/{\it c} and $-8.0$~GeV/{\it c} beams impinging on a 
5\% $\lambda_{\rm int}$ Be target. The second paper~\cite{Beryllium2} presented results for all beam momenta from this Be target. The third~\cite{Tantalum}, fourth~\cite{Copper}, and fifth~\cite{Lead} papers presented results from the interactions with 5\% $\lambda_{\rm int}$ tantalum, copper, and lead targets.  In this paper, we report on the large-angle production (polar angle $\theta$ in the range $20^\circ < \theta < 125^\circ$) of secondary protons and charged pions, and of deuterons, in the interactions with a 5\% $\lambda_{\rm int}$ carbon target of protons and pions with beam momenta of $\pm3.0$, $\pm5.0$, $\pm8.0$, $\pm12.0$, and $\pm15.0$~GeV/{\it c}. 

Our work involves only the HARP large-angle spectrometer.

\section{The beams and the HARP spectrometer}

The protons and pions were delivered by the T9 beam line in the East Hall of CERN's Proton Synchrotron. This beam line supports beam momenta between 1.5 and 15~GeV/{\it c}, with a momentum bite $\Delta p/p \sim 1$\%.

The beam instrumentation, the definition of the beam particle trajectory, the cuts to select `good' beam particles, and the muon and electron contaminations of the particle beams, 
are the same as described, e.g., in Ref.~\cite{Beryllium1}.

The target was a disc made of high-purity (99.99\%) carbon, with a radius of 15.13~mm and a thickness of 18.95~mm (5\% $\lambda_{\rm int}$). The target weight had been measured to be 25.656~g, leading to an average target density of 1.88~g/cm$^3$, 17\% lower than the data book value of 2.27~g/cm$^3$. We used the measured value of 1.88~g/cm$^3$ for the cross-section normalization.

The finite thickness of the target leads to a small attenuation of the number of incident beam particles. The attenuation factor is $f_{\rm att} = 0.975$.

Our calibration work on the HARP TPC and RPCs is described in detail in Refs.~\cite{TPCpub} and \cite{RPCpub}, and in references cited therein.

The momentum resolution $\sigma (1/p_{\rm T})$ of the HARP--TPC is typically 0.2~(GeV/{\it c})$^{-1}$ and worsens towards small relative particle velocity $\beta$ and small polar angle $\theta$.
The absolute momentum scale is determined to be correct to better than 2\%, both for positively and negatively charged particles.
 
The polar angle $\theta$ is measured in the TPC with a 
resolution of $\sim$9~mrad, for a representative 
angle of $\theta = 60^\circ$. To this a multiple scattering
error has to be added which is on the average 
$\sim$7~mrad for a proton with 
$p_{\rm T} = 500$~MeV/{\it c} in the TPC gas and $\theta = 60^\circ$, 
and $\sim$4~mrad for a pion with the same characteristics.
The polar-angle scale is correct to better than 2~mrad.     

The TPC measures \dedx\ with a resolution of 16\% for a track length of 300~mm.

The intrinsic efficiency of the RPCs that surround the TPC is better than 98\%.

The intrinsic time resolution of the RPCs is 127~ps and the system time-of-flight resolution (that includes the jitter of the arrival time of the beam particle at the target) is 175~ps. 

To separate measured particles into species, we assign on the basis of \dedx\ and $\beta$ to each particle a probability of being a proton, a pion (muon), or an electron, respectively. The probabilities add up to unity, so that the number of particles is conserved. These probabilities are used for weighting when entering tracks into plots or tables.

A general discussion of the systematic errors can be found, e.g., in Ref.~\cite{Beryllium1}.  
For the data from the $-3$, $-5$ and $+5$~GeV/{\it c} beams, the systematic error
arising from the parametrization of proton and pion abundances in the respective Monte Carlo simulations was doubled.
All systematic errors are propagated into the momentum spectra of secondaries and then added in quadrature. They add up to a systematic uncertainty of our inclusive cross-sections at the few-per-cent level, mainly from errors in the normalization, in the momentum measurement, in particle identification, and in the corrections applied to the data.

\section{Monte Carlo simulation}

We used the Geant4 tool kit~\cite{Geant4} for the simulation of the HARP large-angle spectrometer.

Geant4's QGSP\_BIC physics list provided us with reasonably realistic spectra of secondaries from incoming beam protons with momentum below 12~GeV/{\it c}. For the secondaries from beam protons at 12 and 15~GeV/{\it c} momentum, and from beam pions at all momenta, we found the standard physics lists of Geant4 unsuitable~\cite{GEANTpub}. 

To overcome this problem, we built our own HARP\_CDP physics list. It starts from Geant4's standard QBBC physics list, but the Quark--Gluon String Model is replaced by the FRITIOF string fragmentation model for kinetic energy $E>6$~GeV; for $E<6$~GeV, the Bertini Cascade is used for pions, and the Binary Cascade for protons; elastic and quasi-elastic scattering is disabled. Examples of the good performance of the HARP\_CDP physics list are given in Ref.~\cite{GEANTpub}.

\section{Cross-section results}

In Tables~\ref{pro.proc3}--\ref{pim.pimc15}, collated in the Appendix of this paper, we give
the double-differential inclusive cross-sections ${\rm d}^2 \sigma / {\rm d} p {\rm d} \Omega$
for various combinations of incoming beam particle and secondary particle, including statistical and systematic errors. In each bin, the average momentum at the vertex and the average polar angle are also given.

The data of Tables~\ref{pro.proc3}--\ref{pim.pimc15} are available in ASCII format in Ref.~\cite{ASCIItables}.

Some bins in the tables are empty. Cross-sections are only given if the total error is not larger than the cross-section itself. Since our track reconstruction algorithm is optimized for tracks with $p_{\rm T}$ above $\sim$70~MeV/{\it c} in the TPC volume, we do not give cross-sections from tracks with $p_{\rm T}$ below this value. Because of the absorption of slow protons in the material between the vertex and the TPC gas, and with a view to keeping the correction for absorption losses below 30\%, cross-sections from protons are limited to $p > 450$~MeV/{\it c} at the interaction vertex. Proton cross-sections are also not given if a 10\% error on the proton energy loss in materials between the interaction vertex and the TPC volume leads to a momentum change larger than 2\%. Pion cross-sections are not given if pions are separated from protons by less than twice the time-of-flight resolution.

The large errors and/or absence of results from the $+15$~GeV/{\it c} pion beam are caused by scarce statistics because the beam composition was dominated by protons.

\begin{figure*}[h]
\begin{center}
\begin{tabular}{cc}
\includegraphics[height=0.30\textheight]{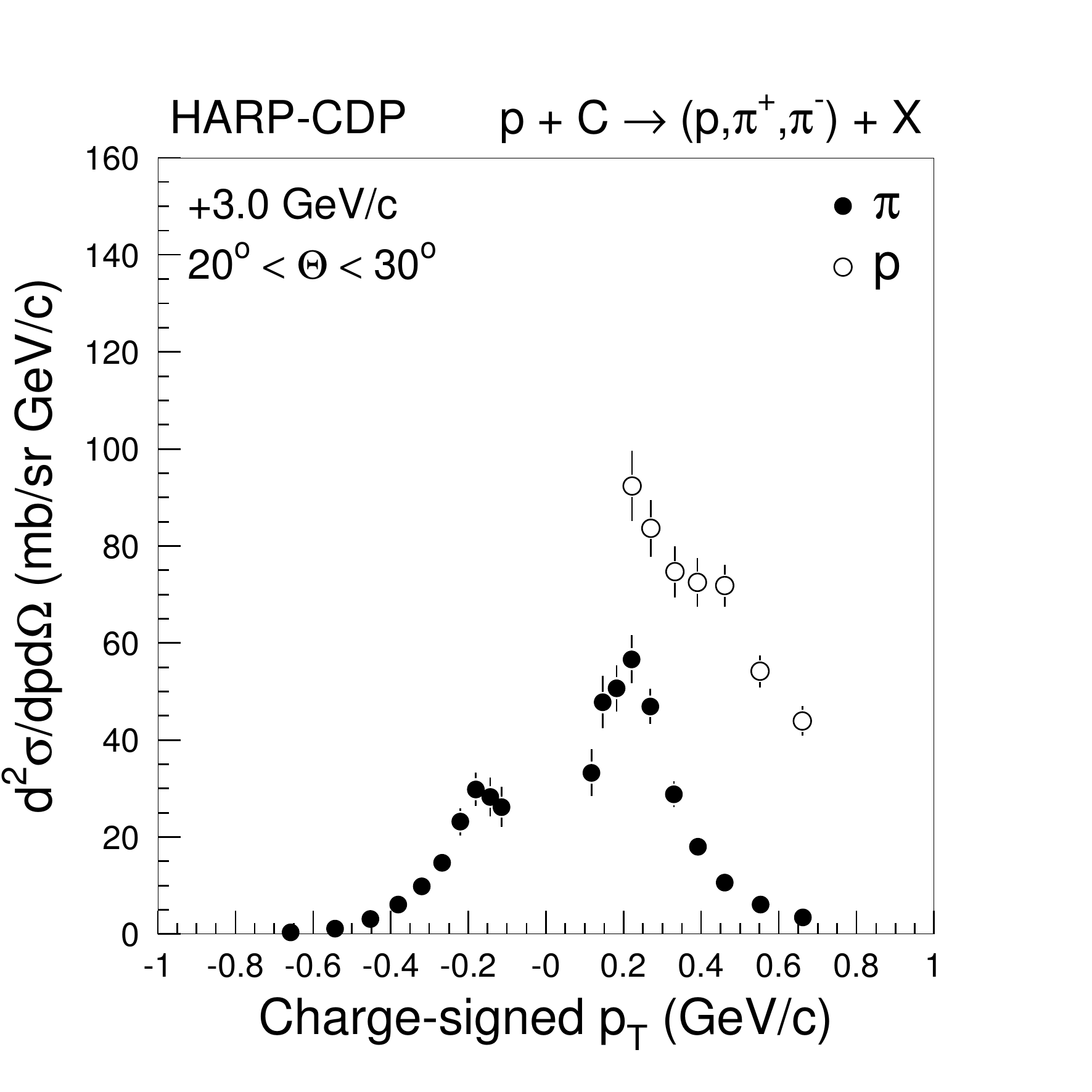} &
\includegraphics[height=0.30\textheight]{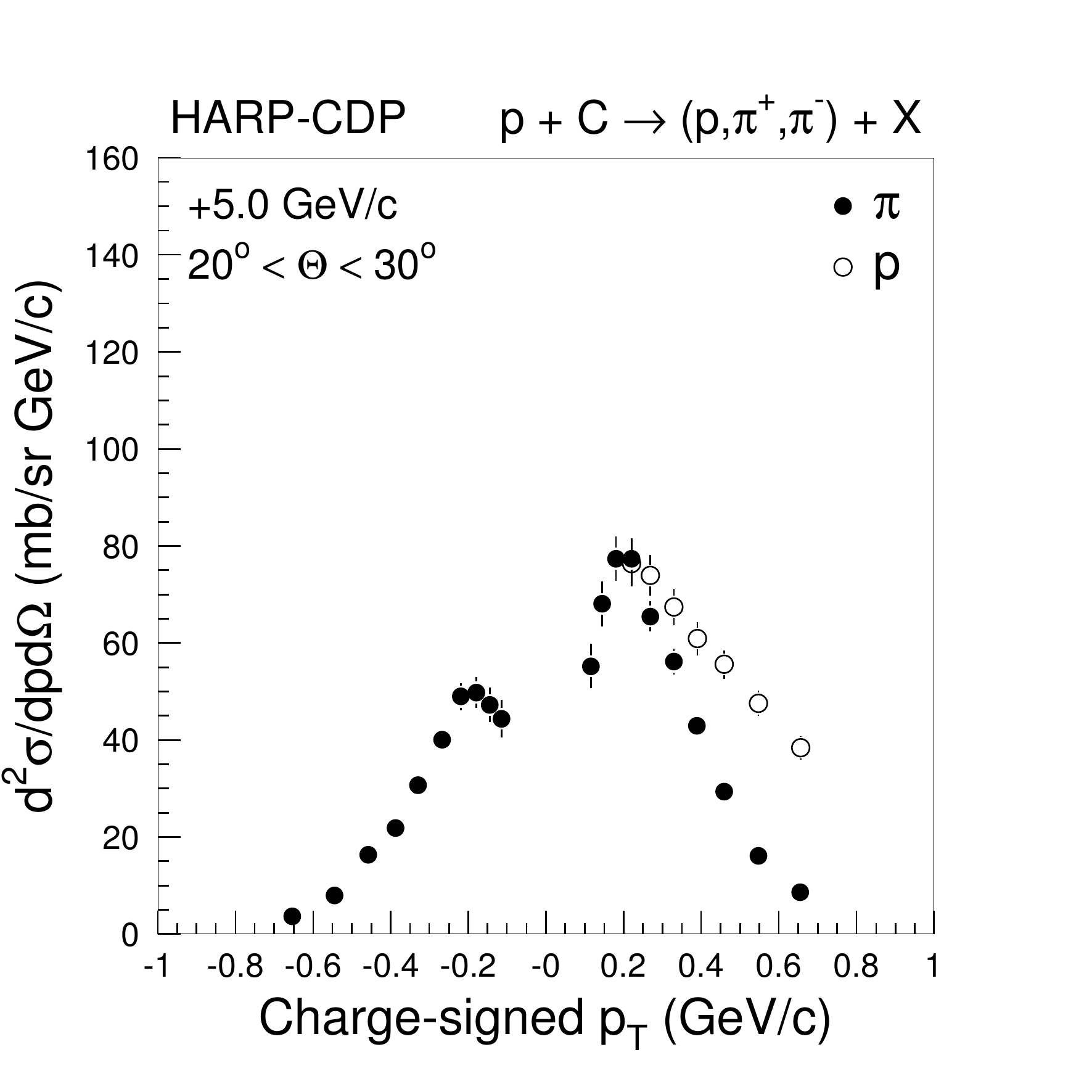} \\
\includegraphics[height=0.30\textheight]{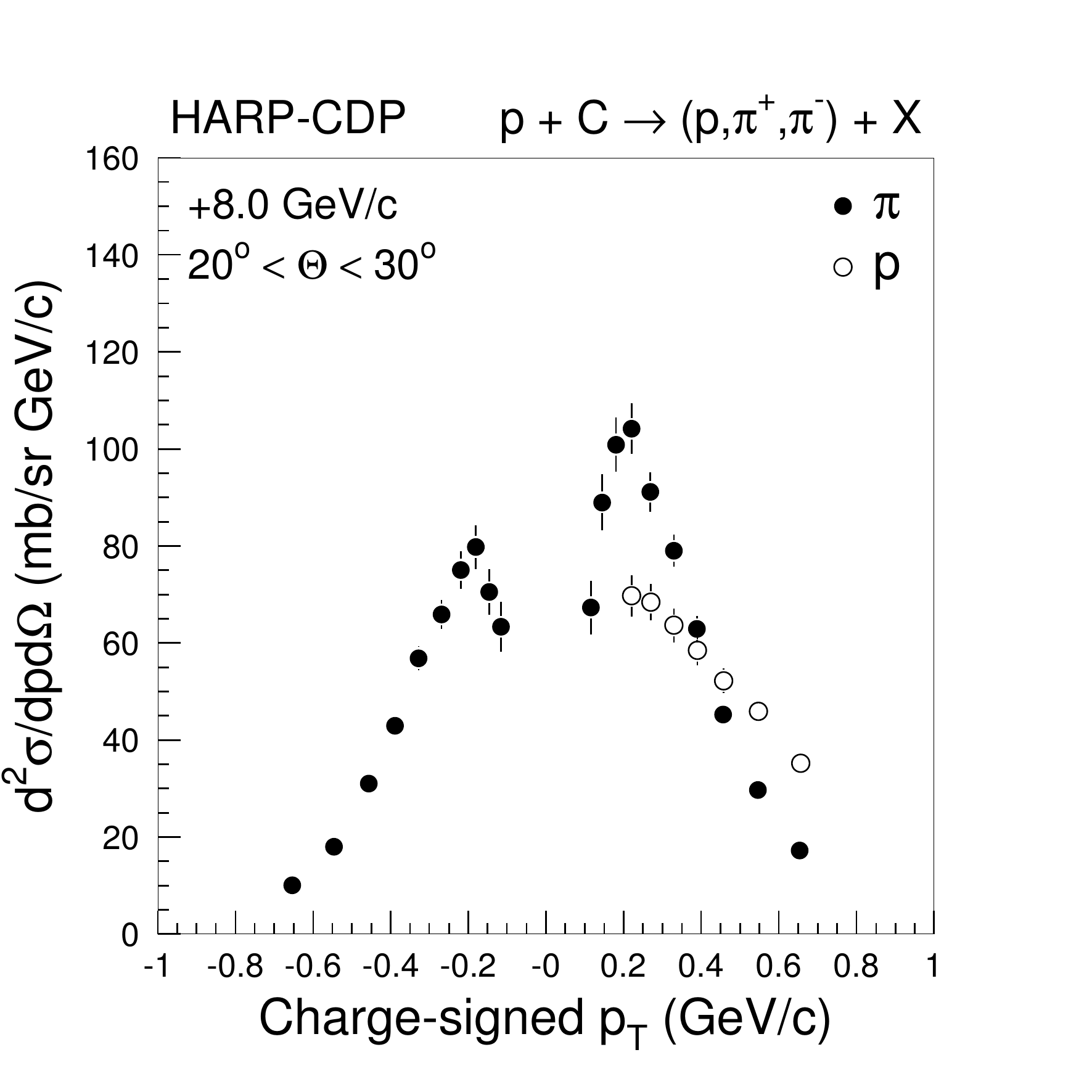} &
\includegraphics[height=0.30\textheight]{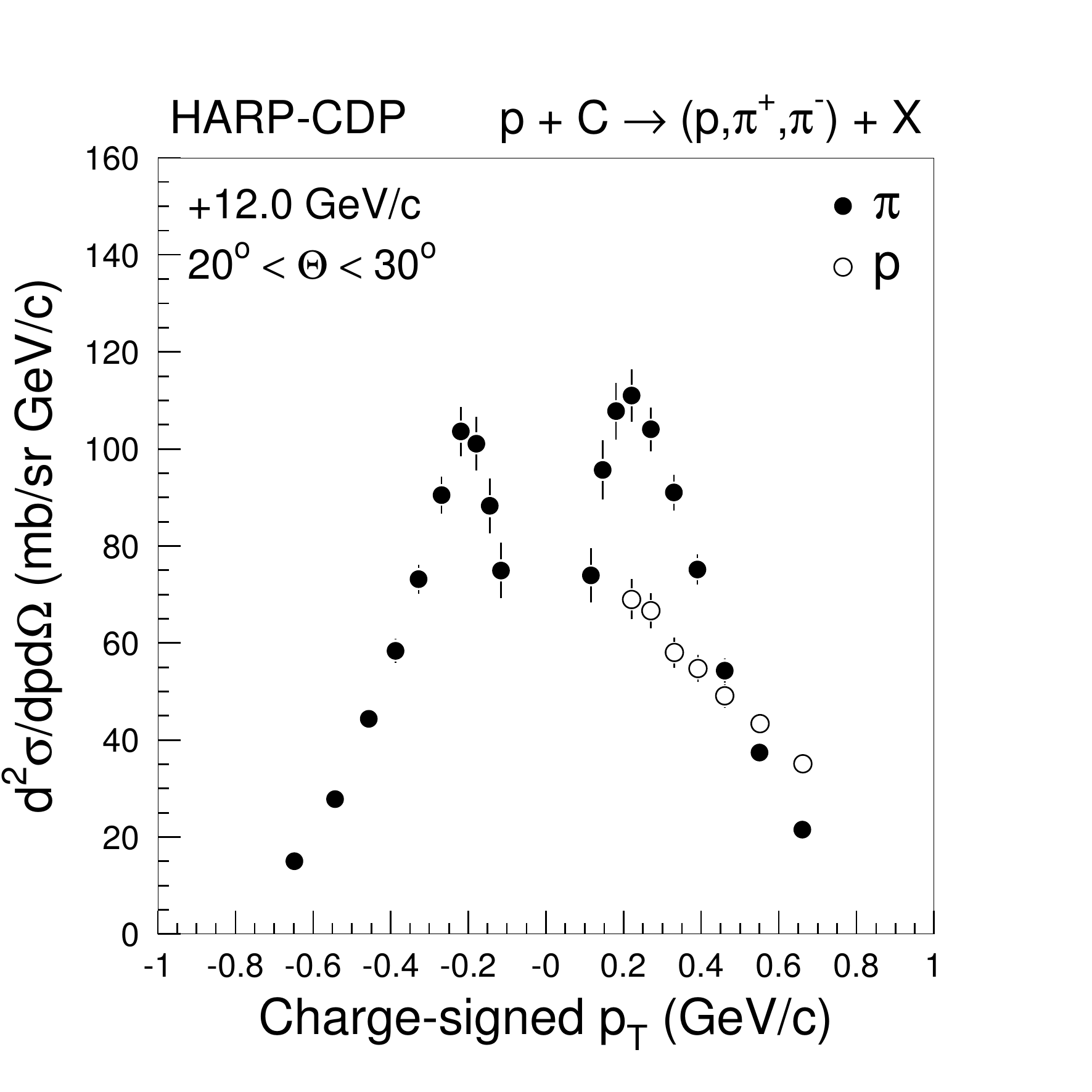} \\
\includegraphics[height=0.30\textheight]{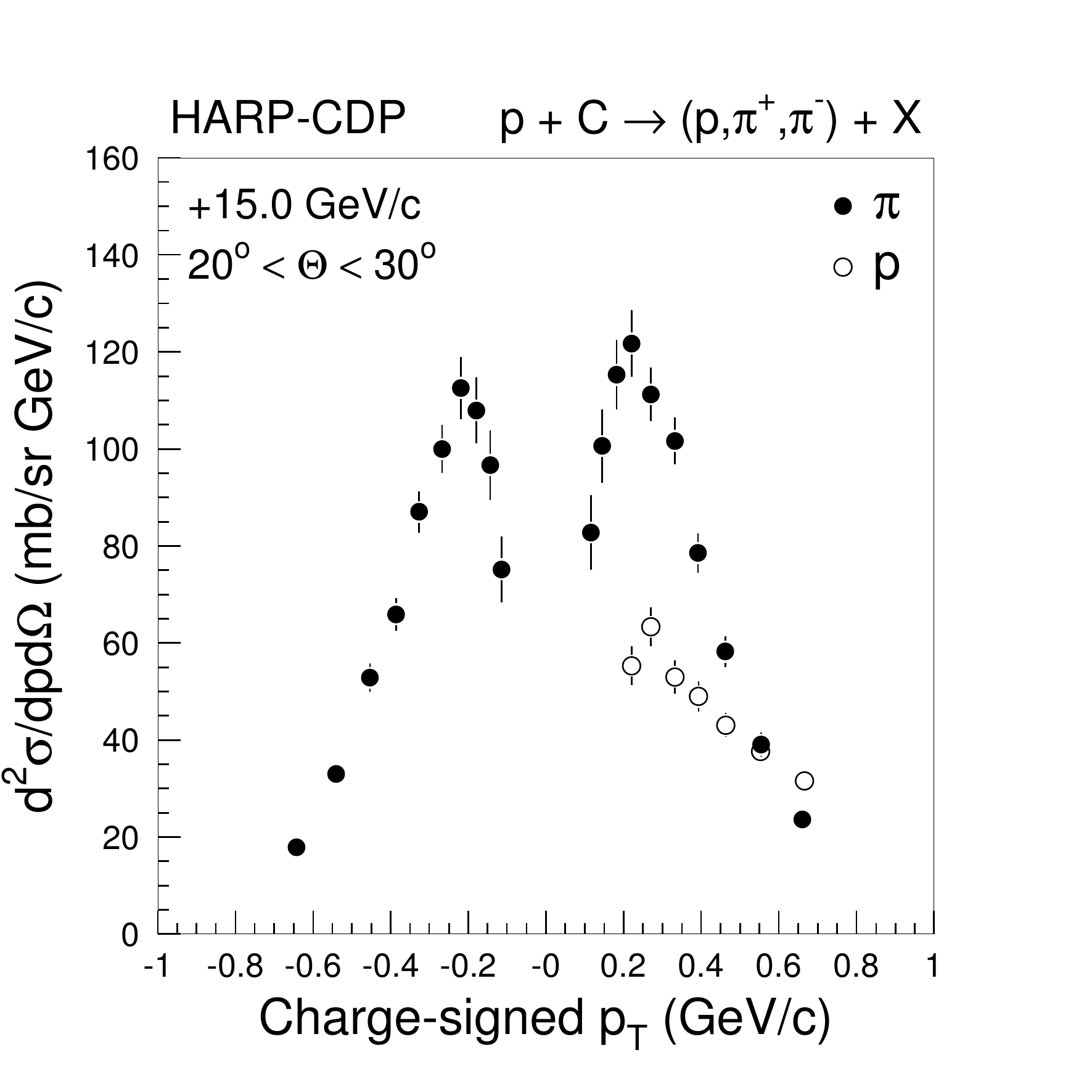} &  \\
\end{tabular}
\caption{Inclusive cross-sections of the production of secondary protons, $\pi^+$'s, and $\pi^-$'s, by protons on carbon nuclei, in the polar-angle range $20^\circ < \theta < 30^\circ$, for different proton beam momenta, as a function of the charge-signed $p_{\rm T}$ of the secondaries; the shown errors are total errors.} 
\label{xsvsmompro}
\end{center}
\end{figure*}

\begin{figure*}[h]
\begin{center}
\begin{tabular}{cc}
\includegraphics[height=0.30\textheight]{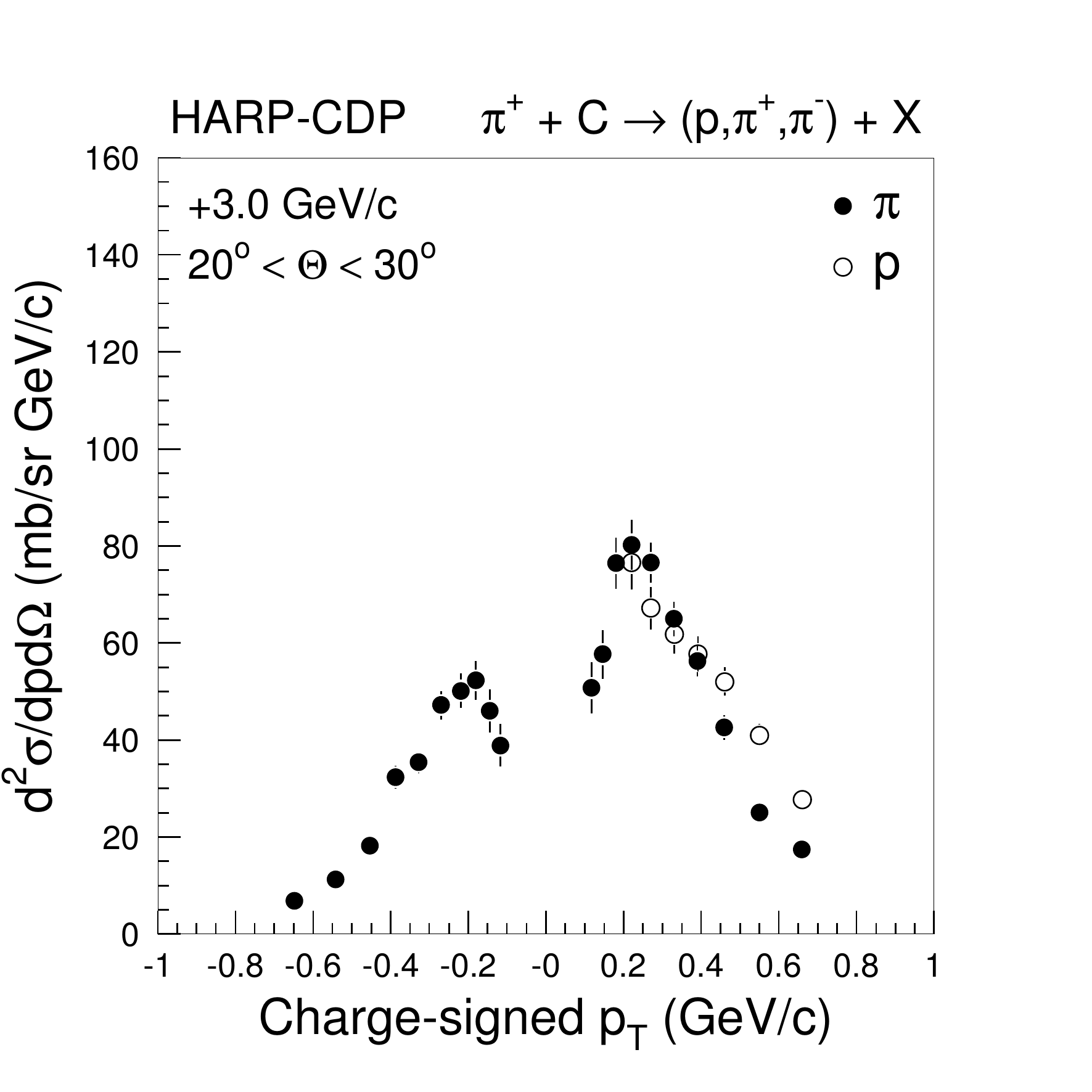} &
\includegraphics[height=0.30\textheight]{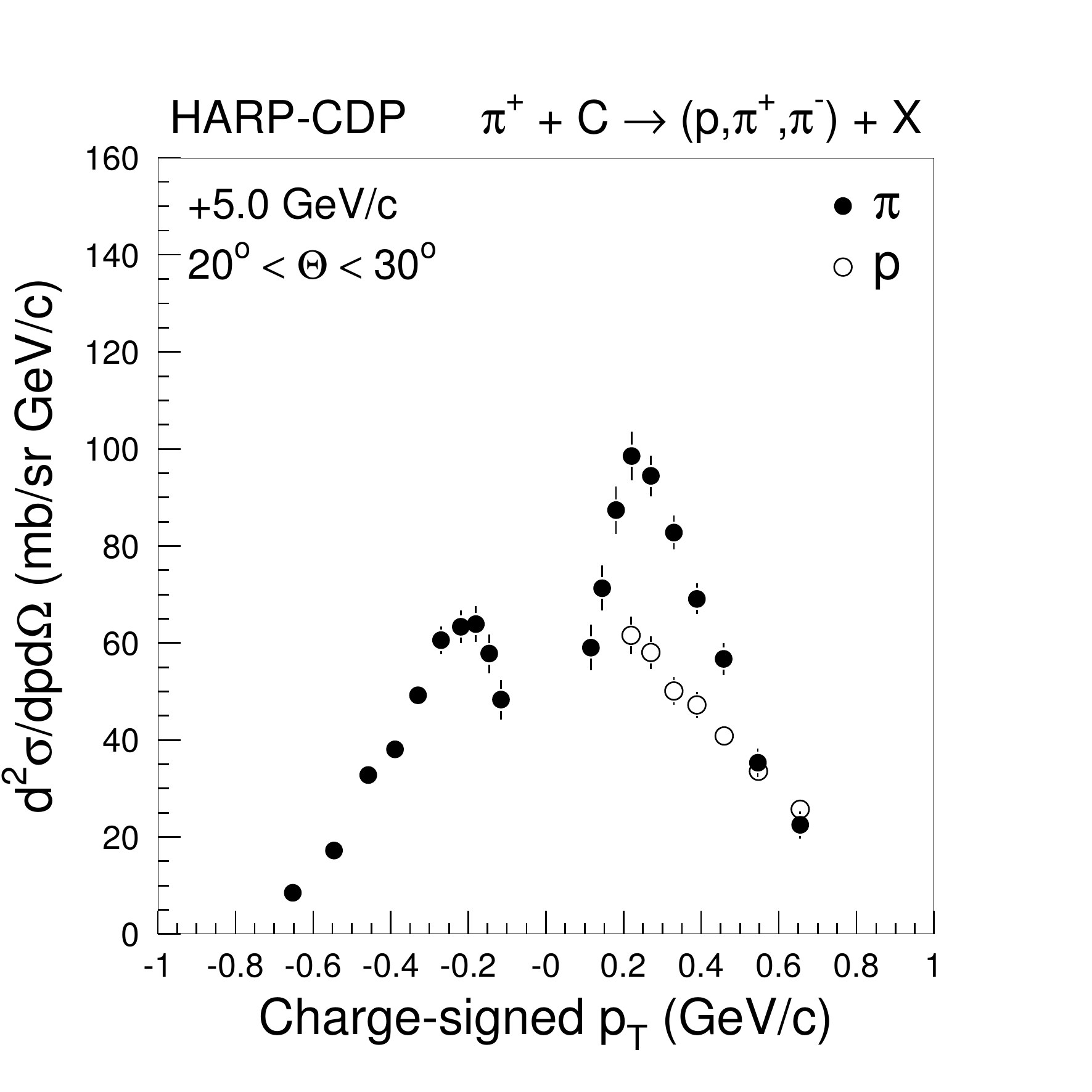} \\
\includegraphics[height=0.30\textheight]{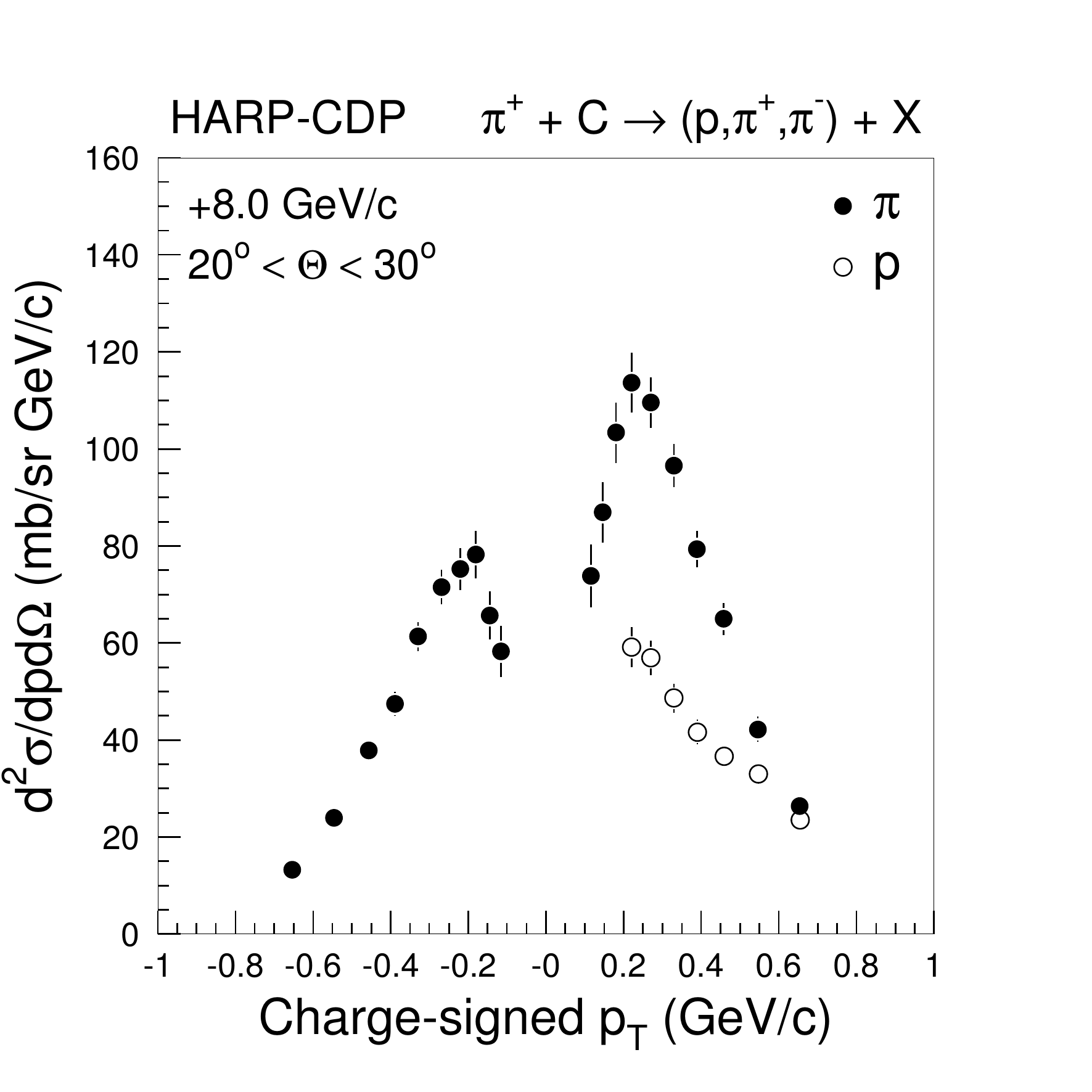} &
\includegraphics[height=0.30\textheight]{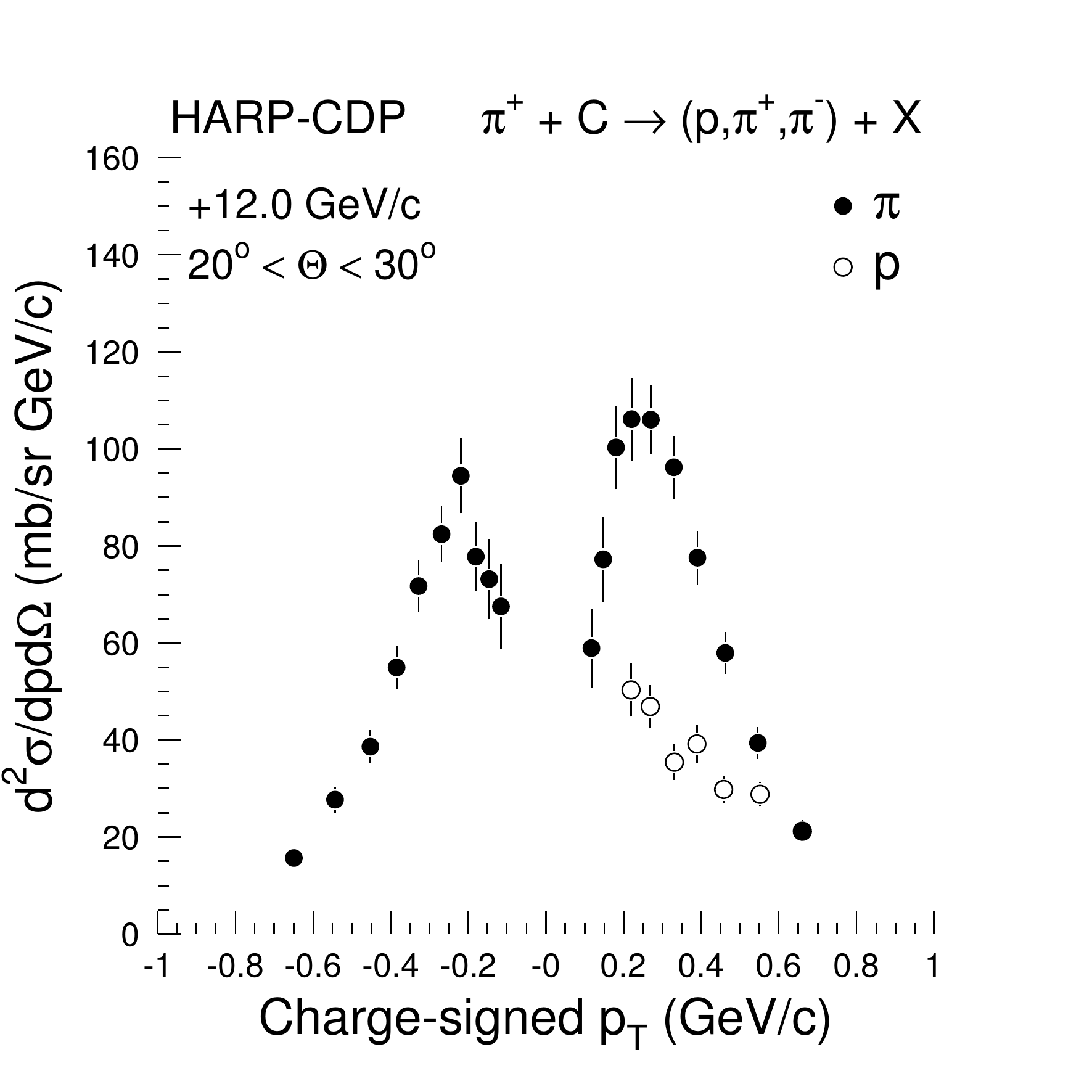} \\
\includegraphics[height=0.30\textheight]{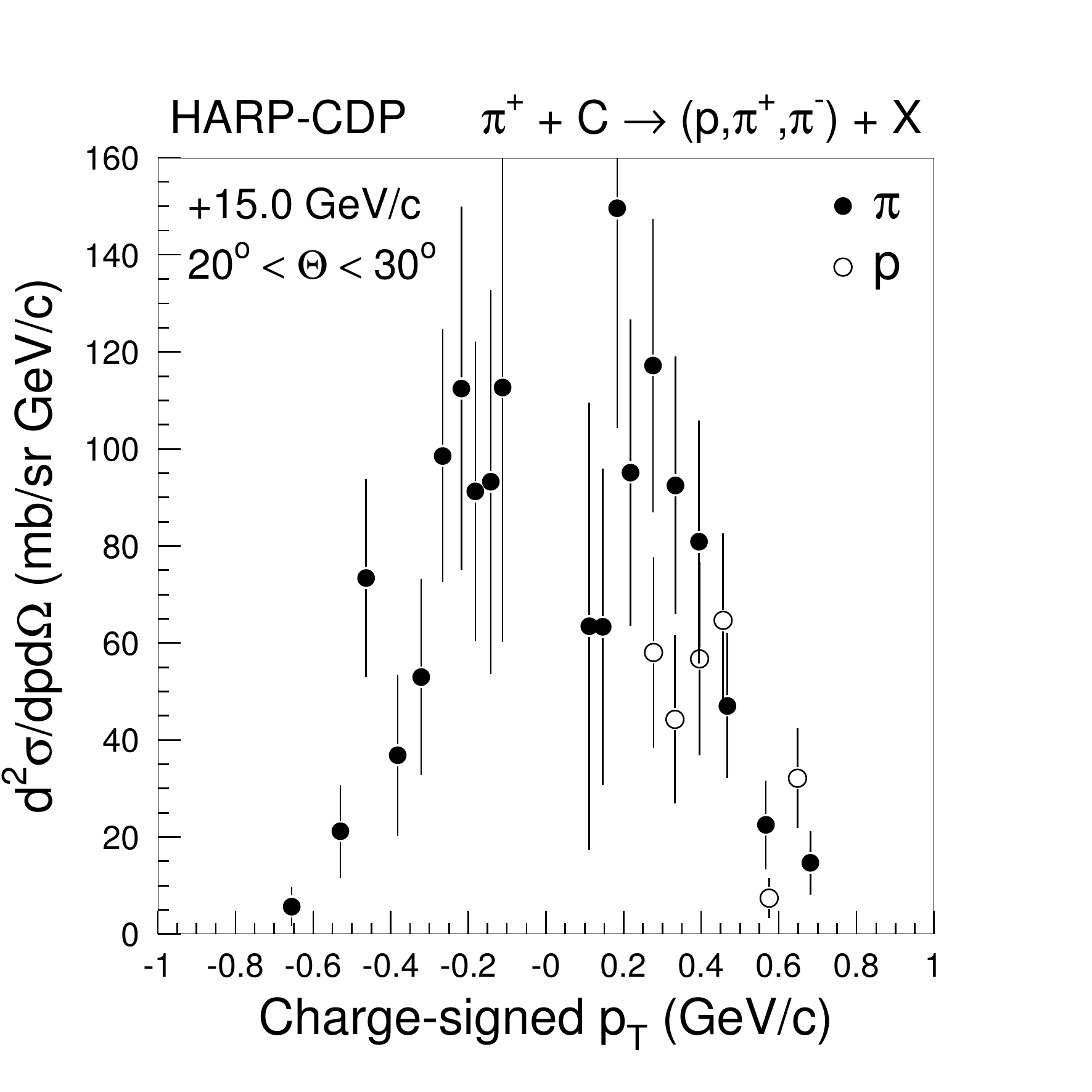} &  \\
\end{tabular}
\caption{Inclusive cross-sections of the production of secondary protons, $\pi^+$'s, and $\pi^-$'s, by $\pi^+$'s on carbon nuclei, in the polar-angle range $20^\circ < \theta < 30^\circ$, for different $\pi^+$ beam momenta, as a function of the charge-signed $p_{\rm T}$ of the secondaries; the shown errors are total errors.}  
\label{xsvsmompip}
\end{center}
\end{figure*}

\begin{figure*}[h]
\begin{center}
\begin{tabular}{cc}
\includegraphics[height=0.30\textheight]{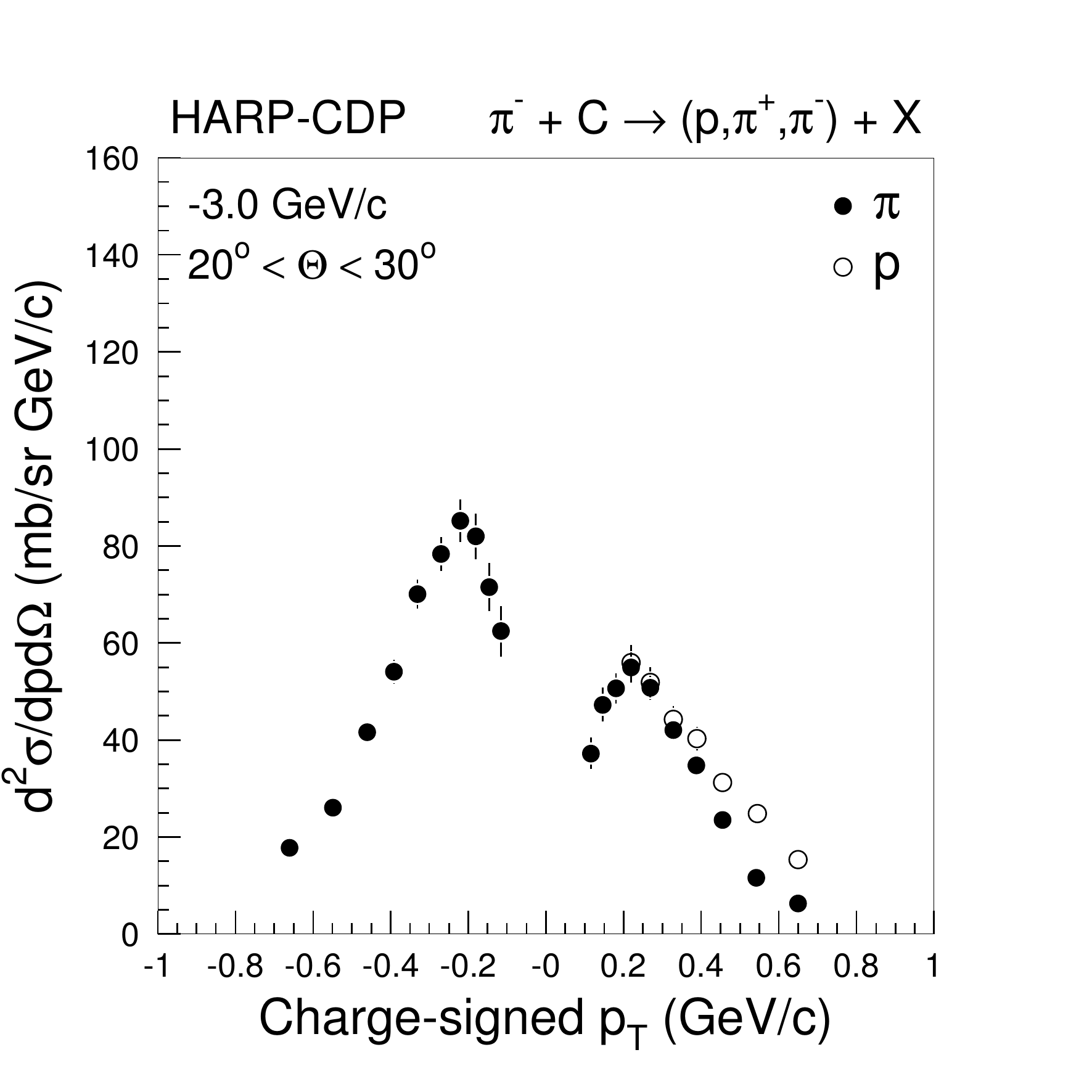} &
\includegraphics[height=0.30\textheight]{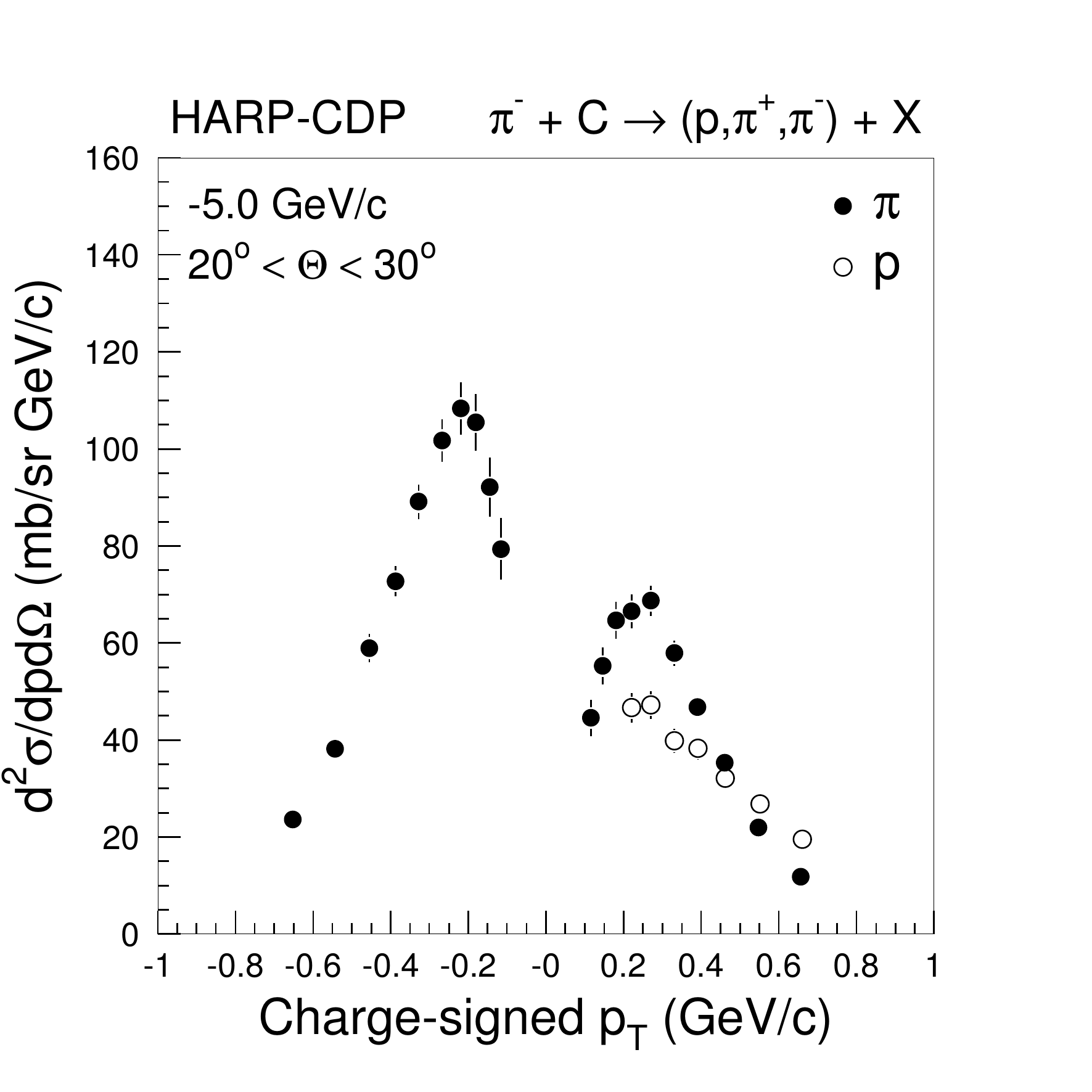} \\
\includegraphics[height=0.30\textheight]{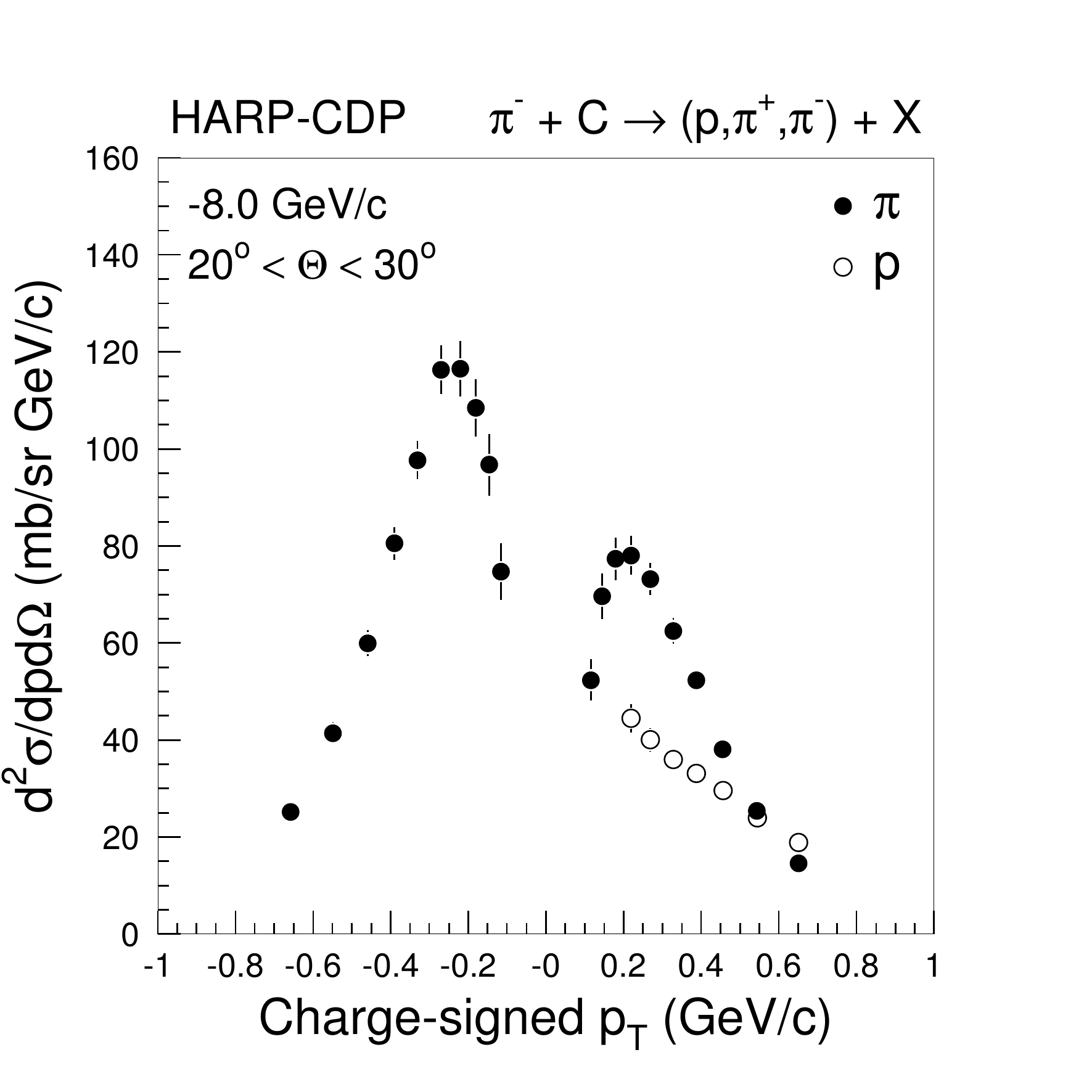} &
\includegraphics[height=0.30\textheight]{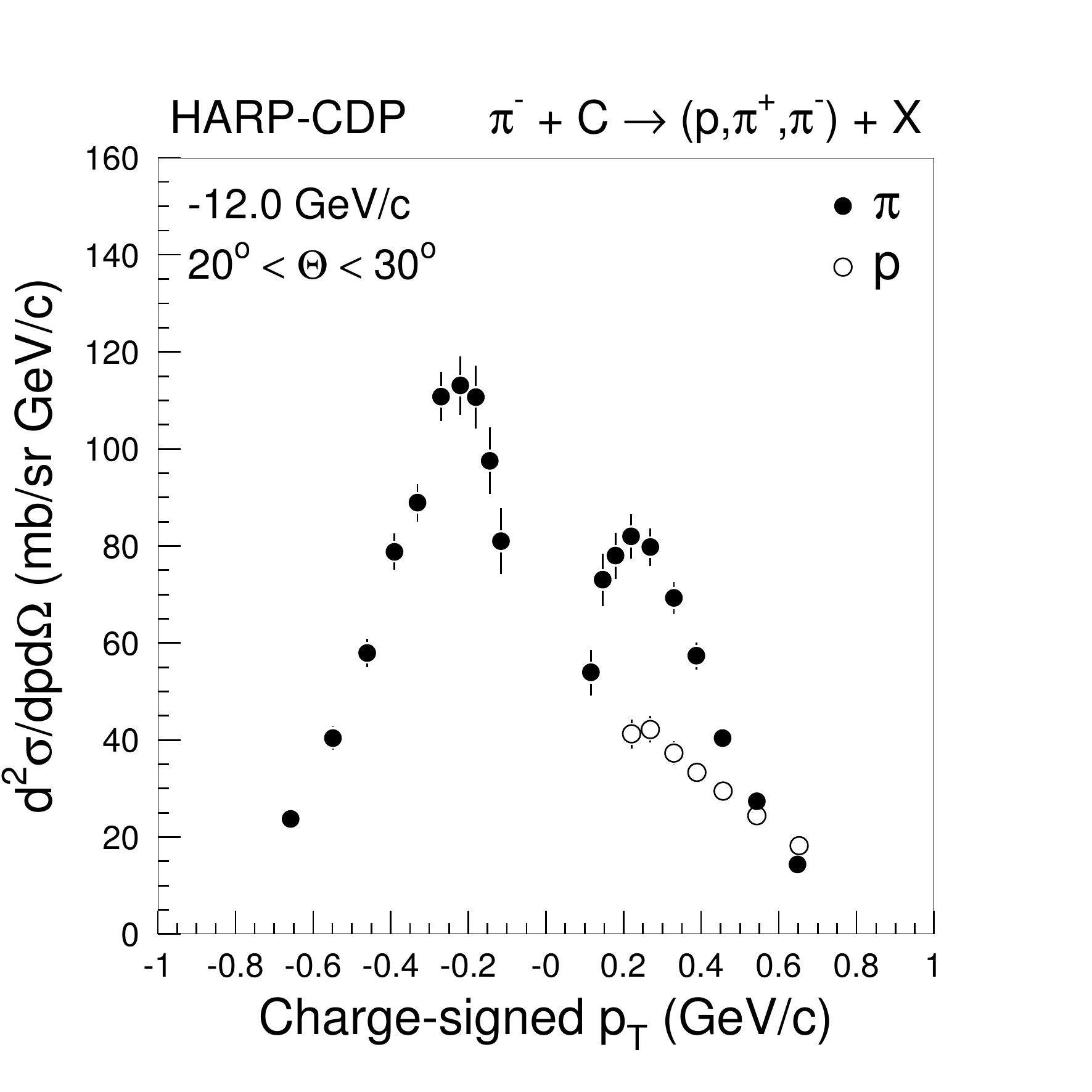} \\
\includegraphics[height=0.30\textheight]{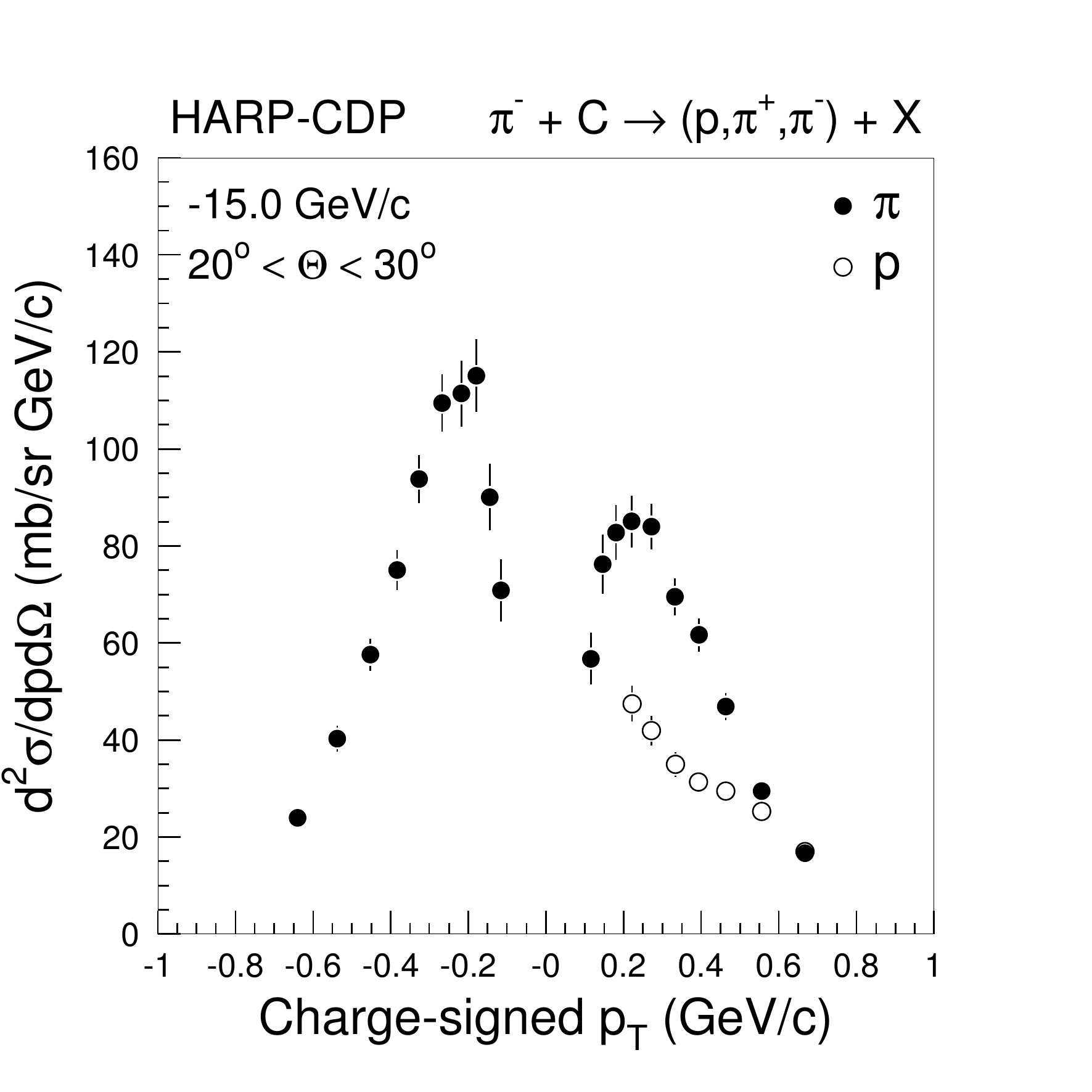} &  \\
\end{tabular}
\caption{Inclusive cross-sections of the production of secondary protons, $\pi^+$'s, and $\pi^-$'s, by $\pi^-$'s on carbon nuclei, in the polar-angle range $20^\circ < \theta < 30^\circ$, for different $\pi^-$ beam momenta, as a function of the charge-signed $p_{\rm T}$ of the secondaries; the shown errors are total errors.} 
\label{xsvsmompim}
\end{center}
\end{figure*}

\begin{figure*}[h]
\begin{center}
\begin{tabular}{cc}
\includegraphics[height=0.30\textheight]{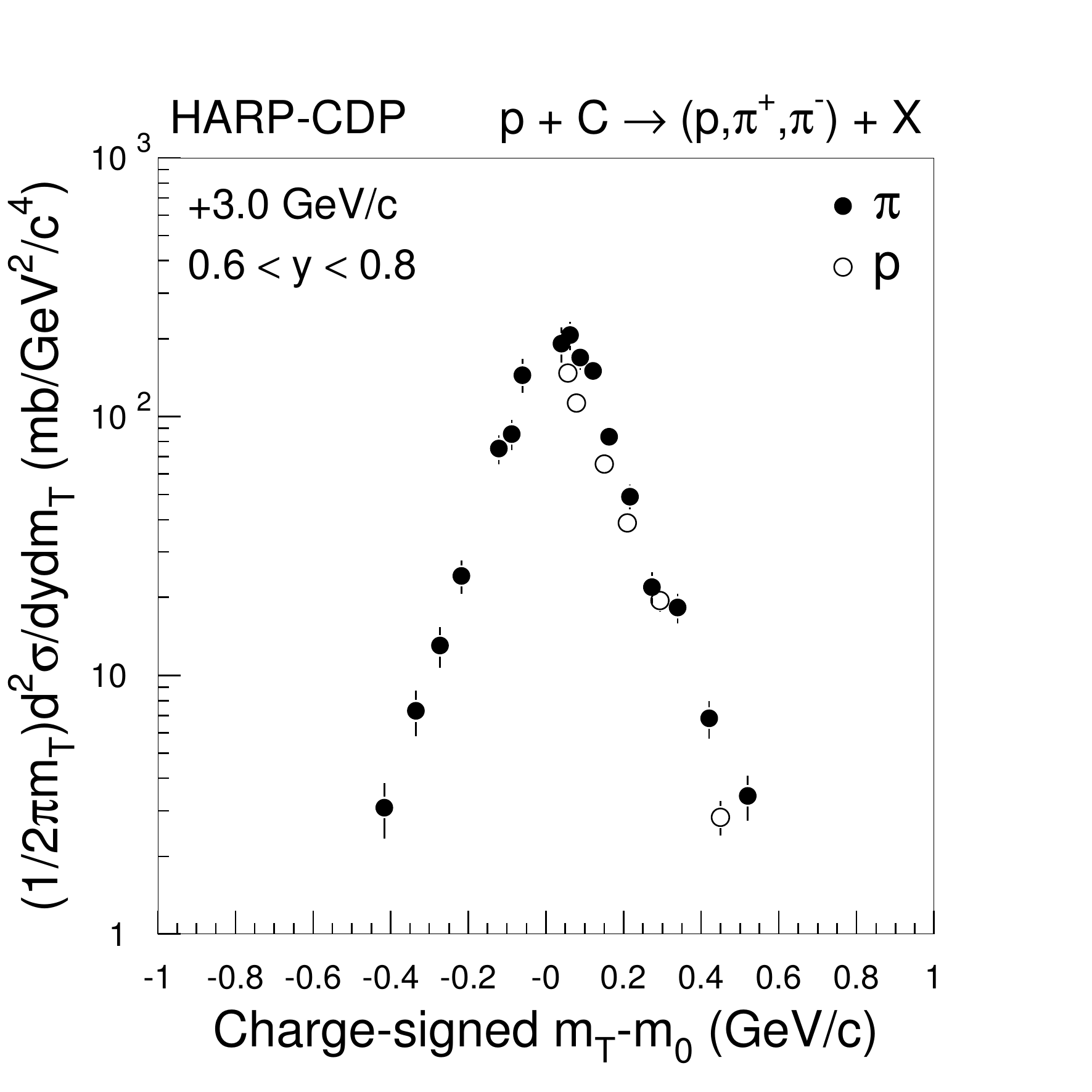} &
\includegraphics[height=0.30\textheight]{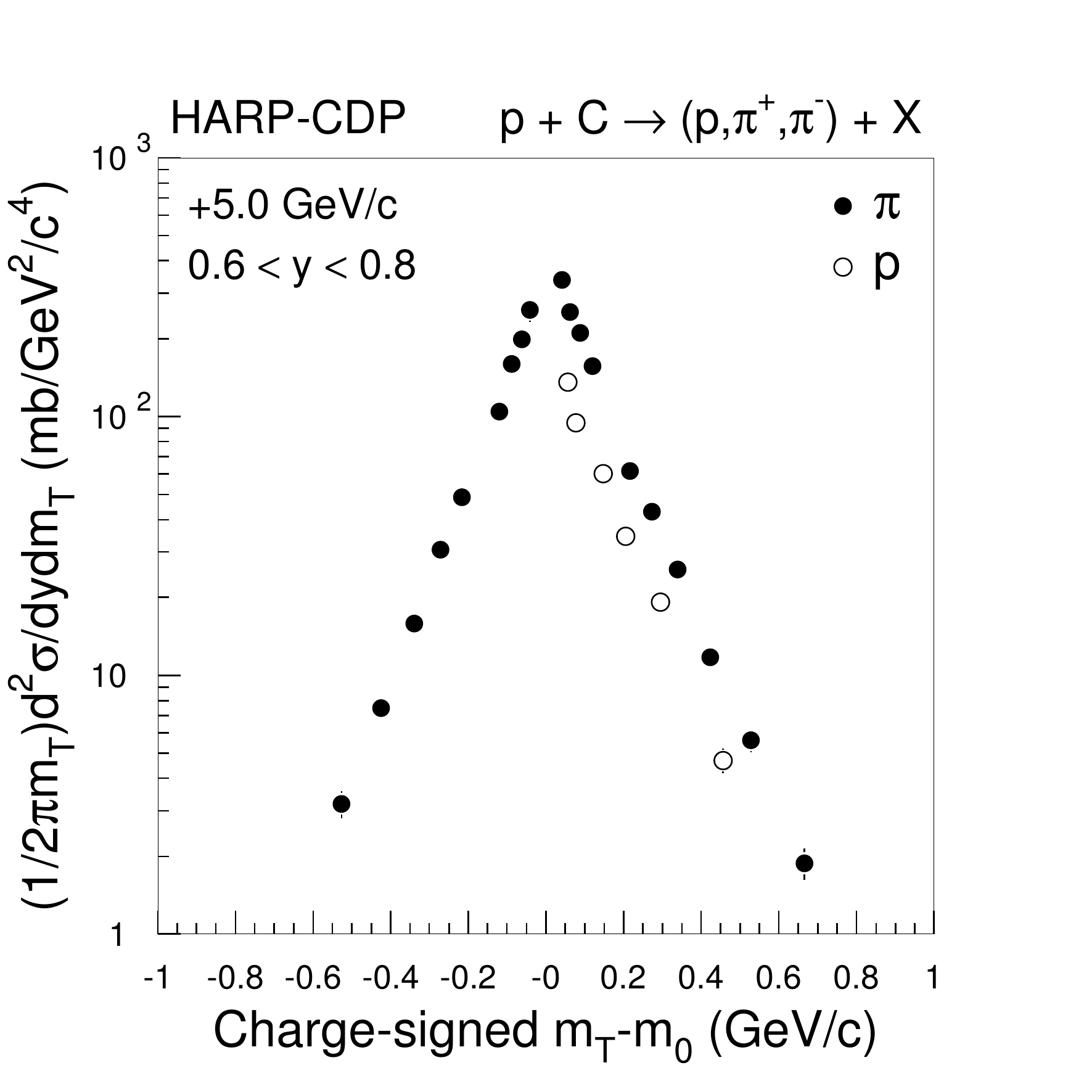} \\
\includegraphics[height=0.30\textheight]{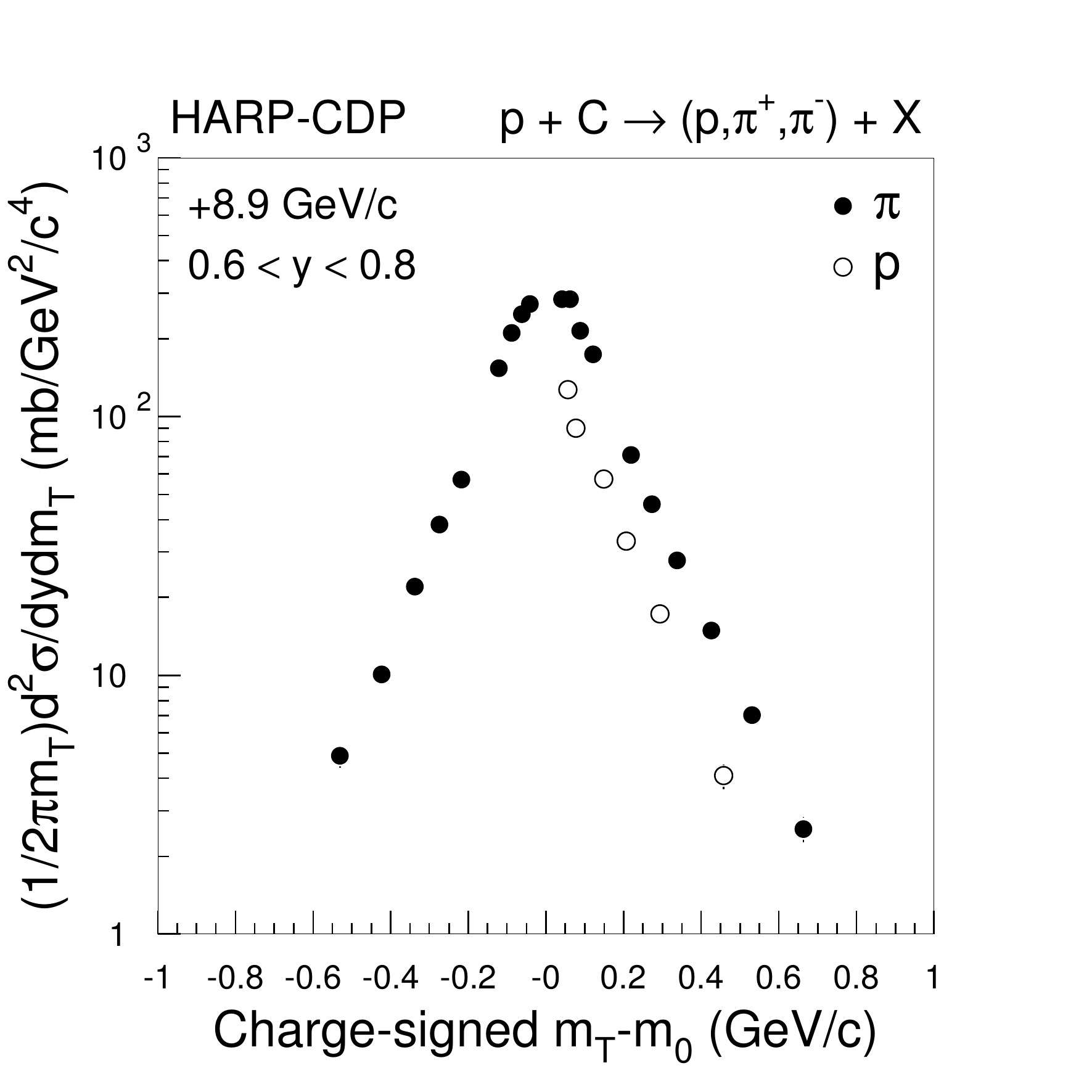} &
\includegraphics[height=0.30\textheight]{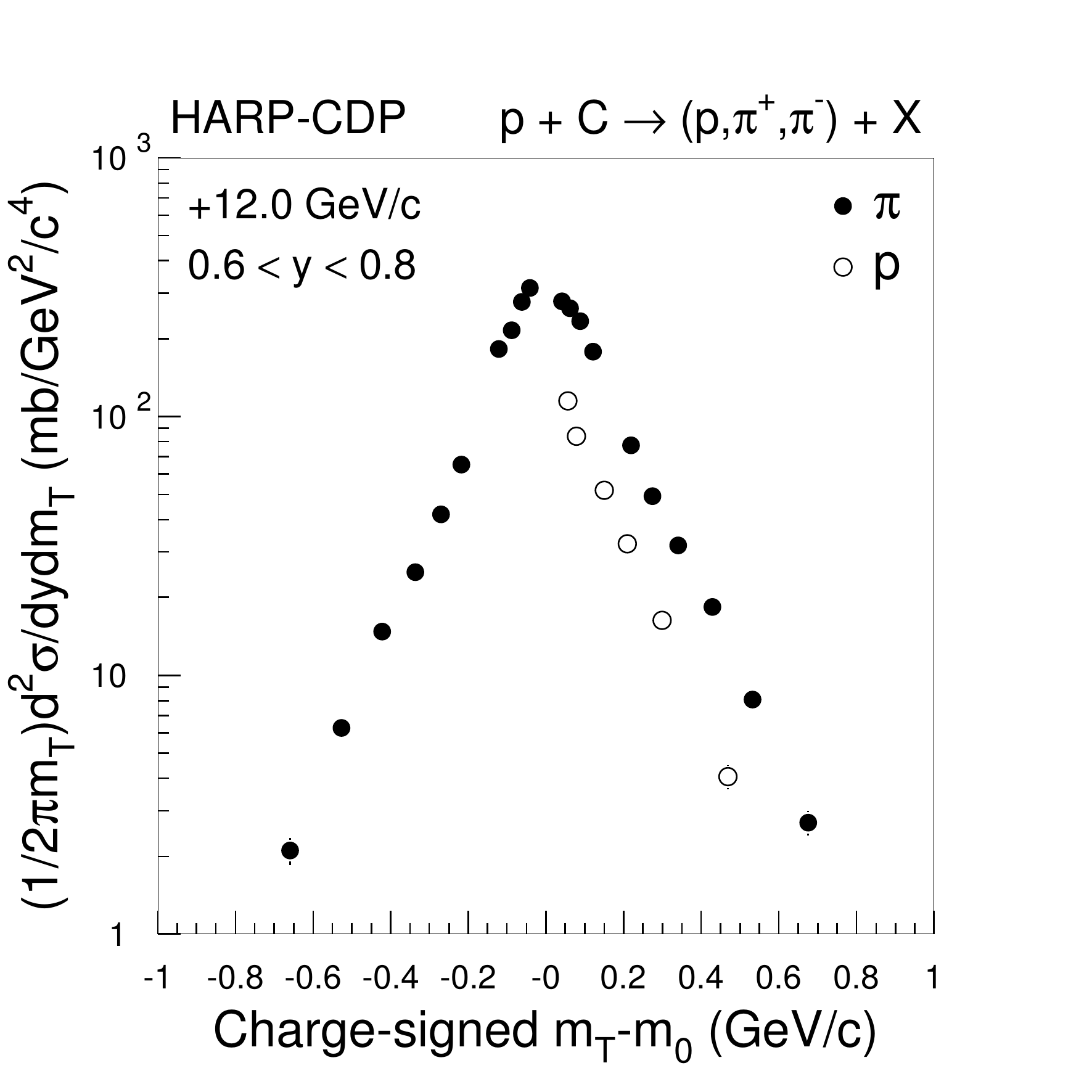} \\
\includegraphics[height=0.30\textheight]{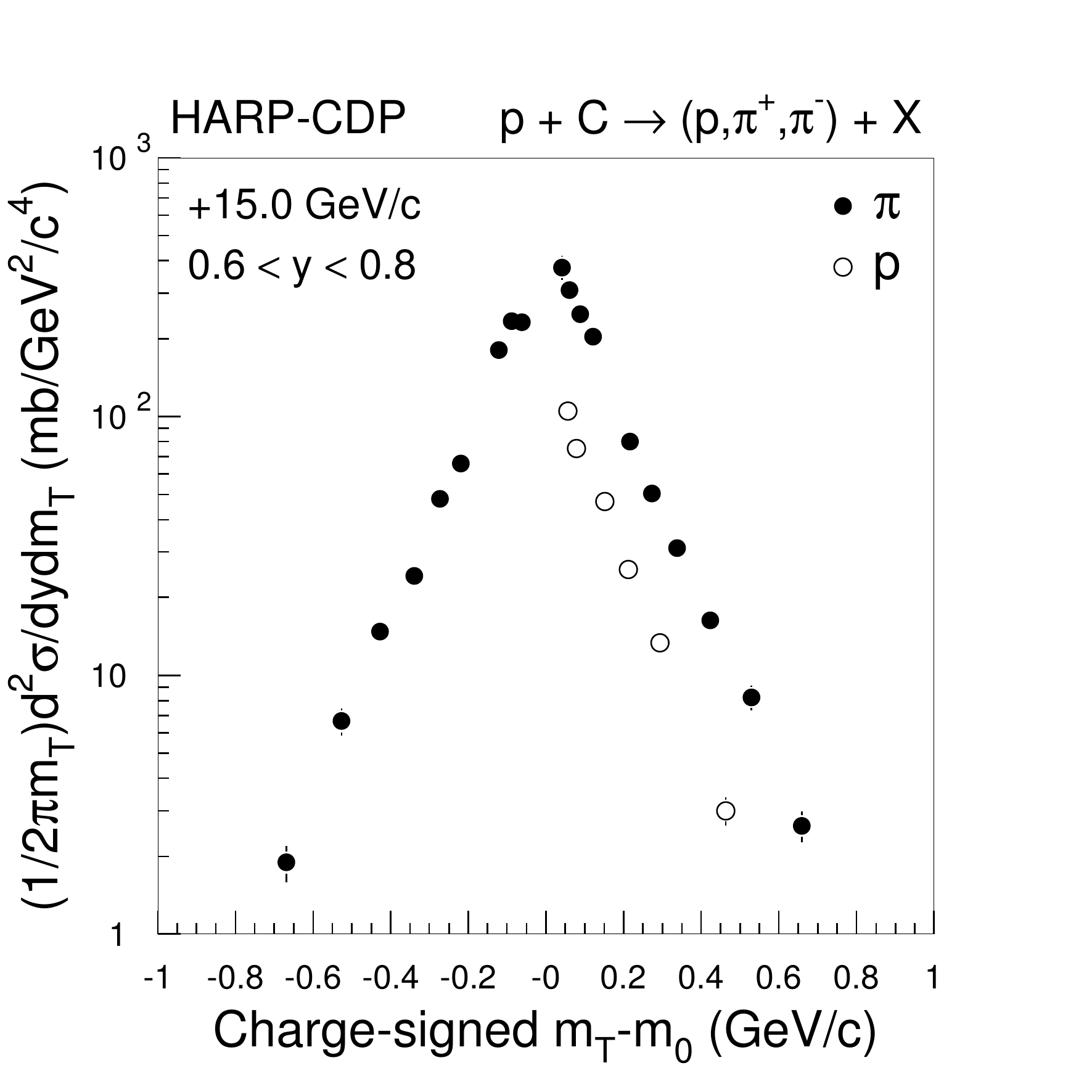} &  \\
\end{tabular}
\caption{Inclusive Lorentz-invariant  
cross-sections of the production of protons, $\pi^+$'s and $\pi^-$'s, by incoming
protons between 3~GeV/{\it c} and 15~GeV/{\it c} momentum, in the rapidity 
range $0.6 < y < 0.8$,
as a function of the charge-signed reduced transverse particle mass, $m_{\rm T} - m_0$,
where $m_0$ is the rest mass of the respective particle;
the shown errors are total errors.} 
\label{xsvsmTpro}
\end{center}
\end{figure*}

\begin{figure*}[h]
\begin{center}
\begin{tabular}{cc}
\includegraphics[height=0.30\textheight]{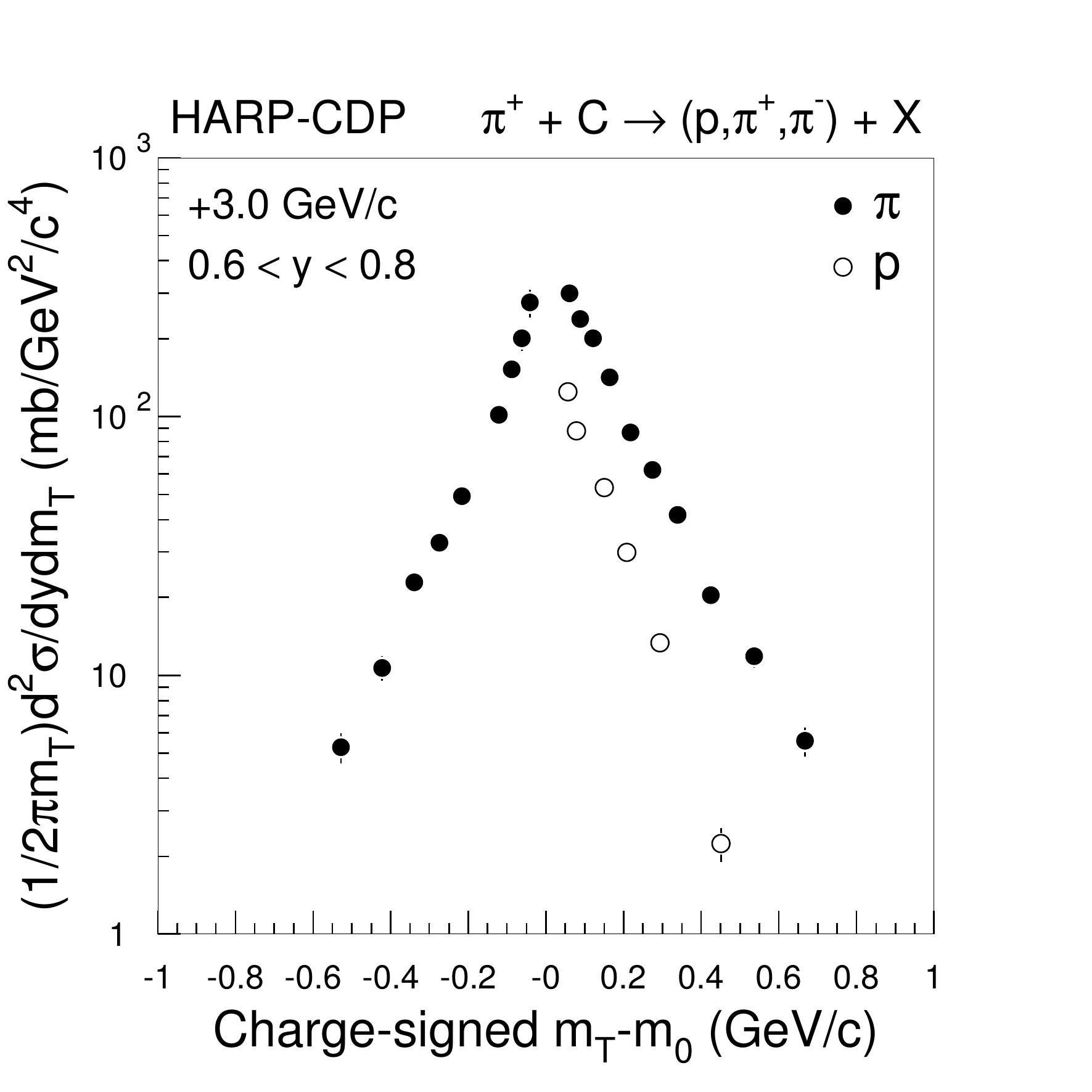} &
\includegraphics[height=0.30\textheight]{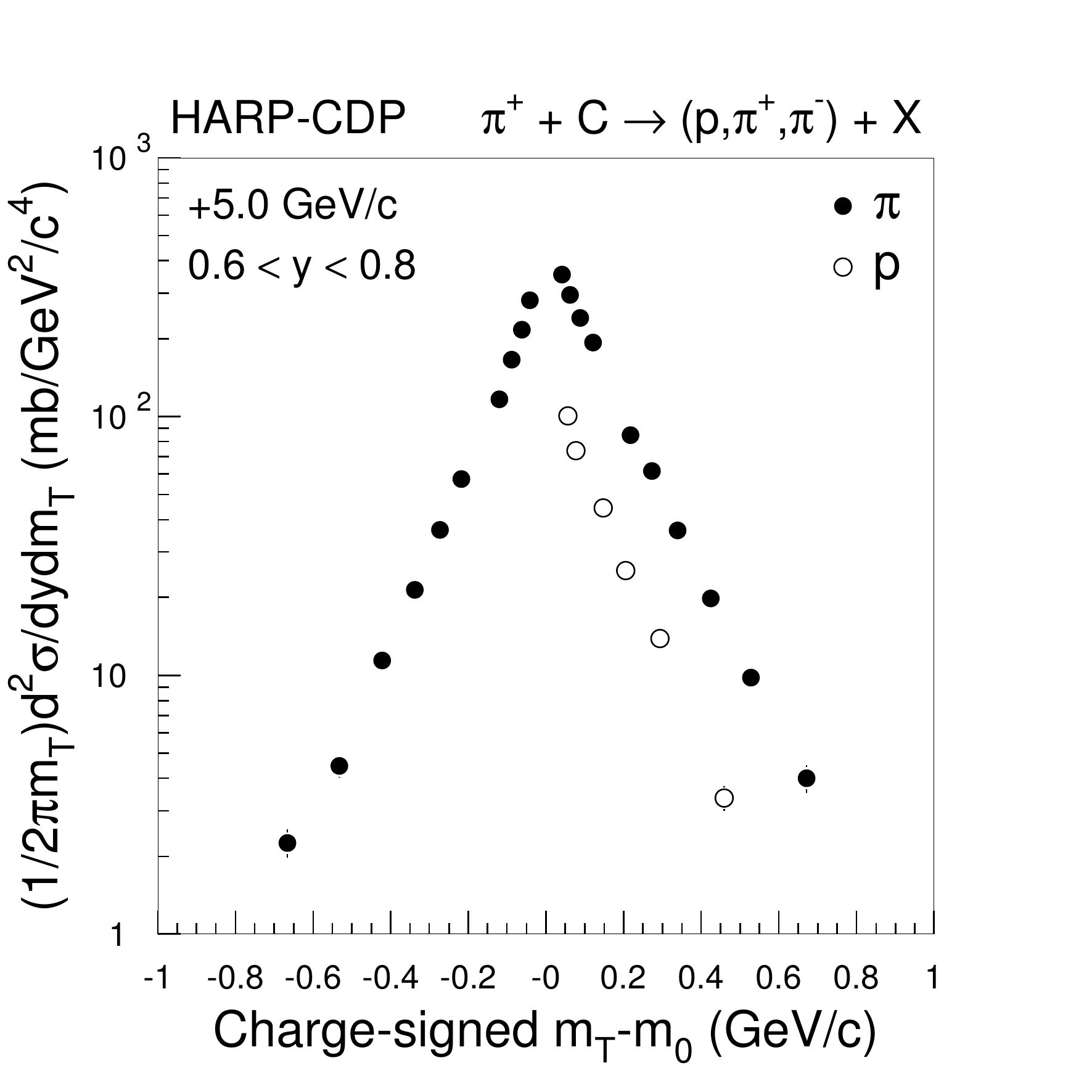} \\
\includegraphics[height=0.30\textheight]{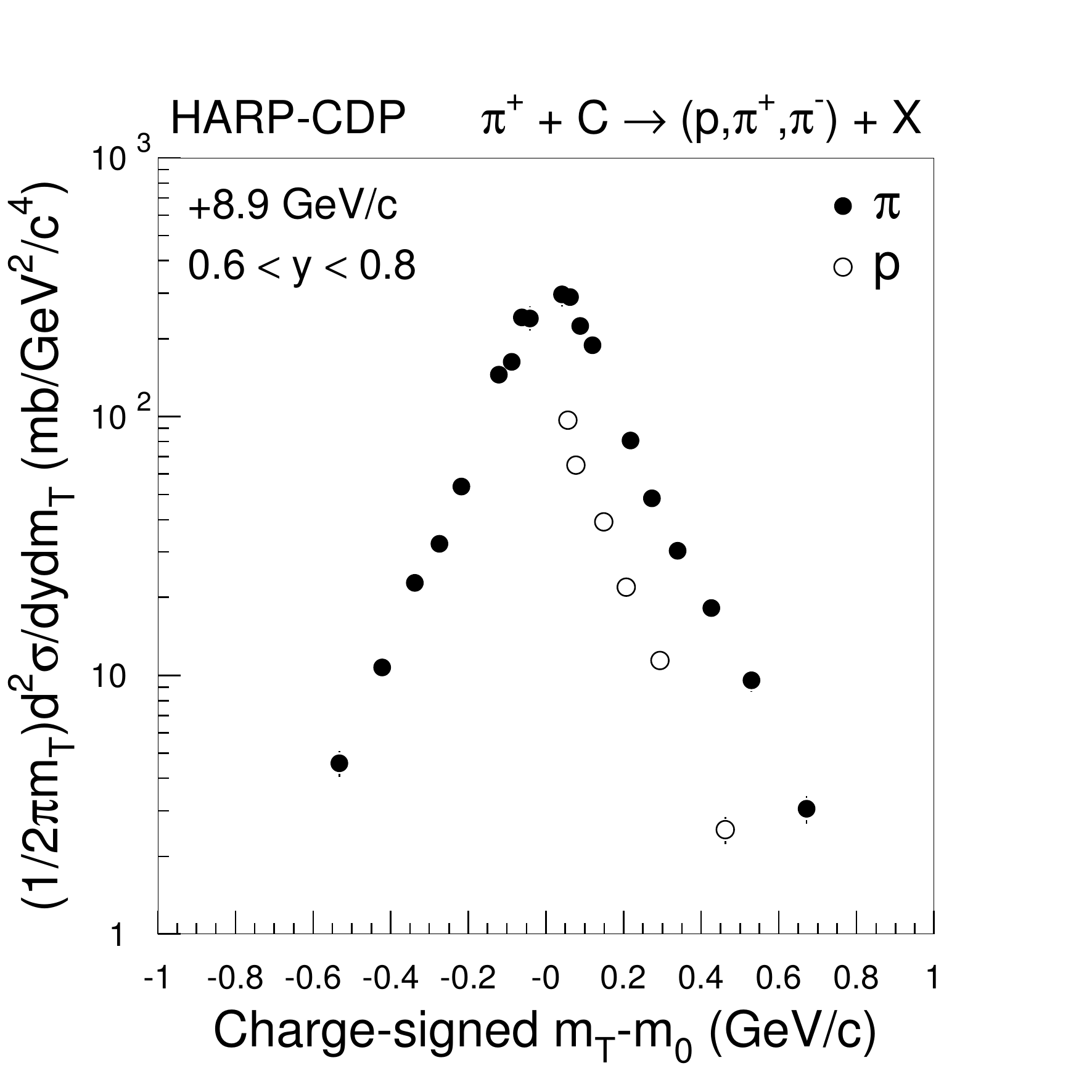} &
\includegraphics[height=0.30\textheight]{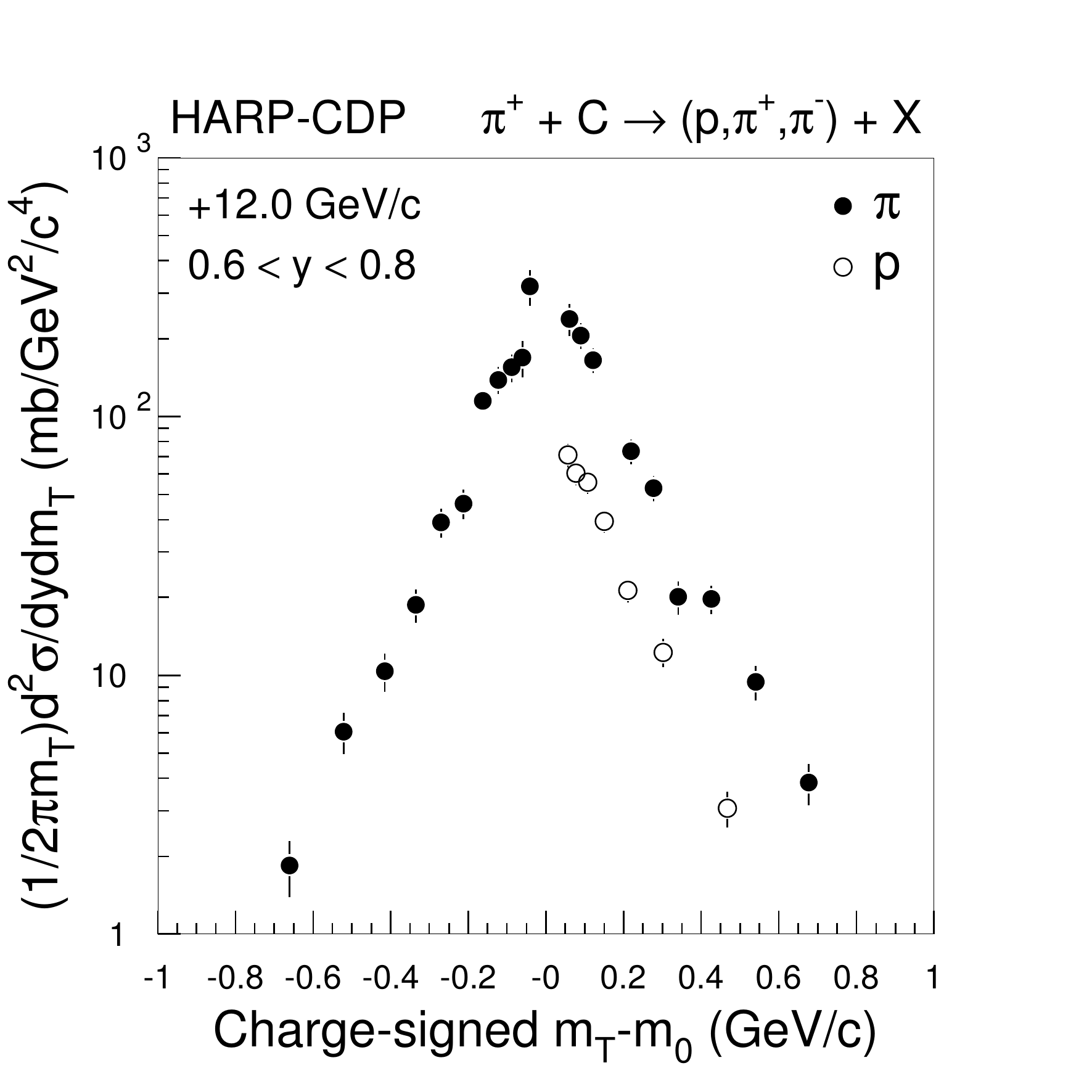} \\
\includegraphics[height=0.30\textheight]{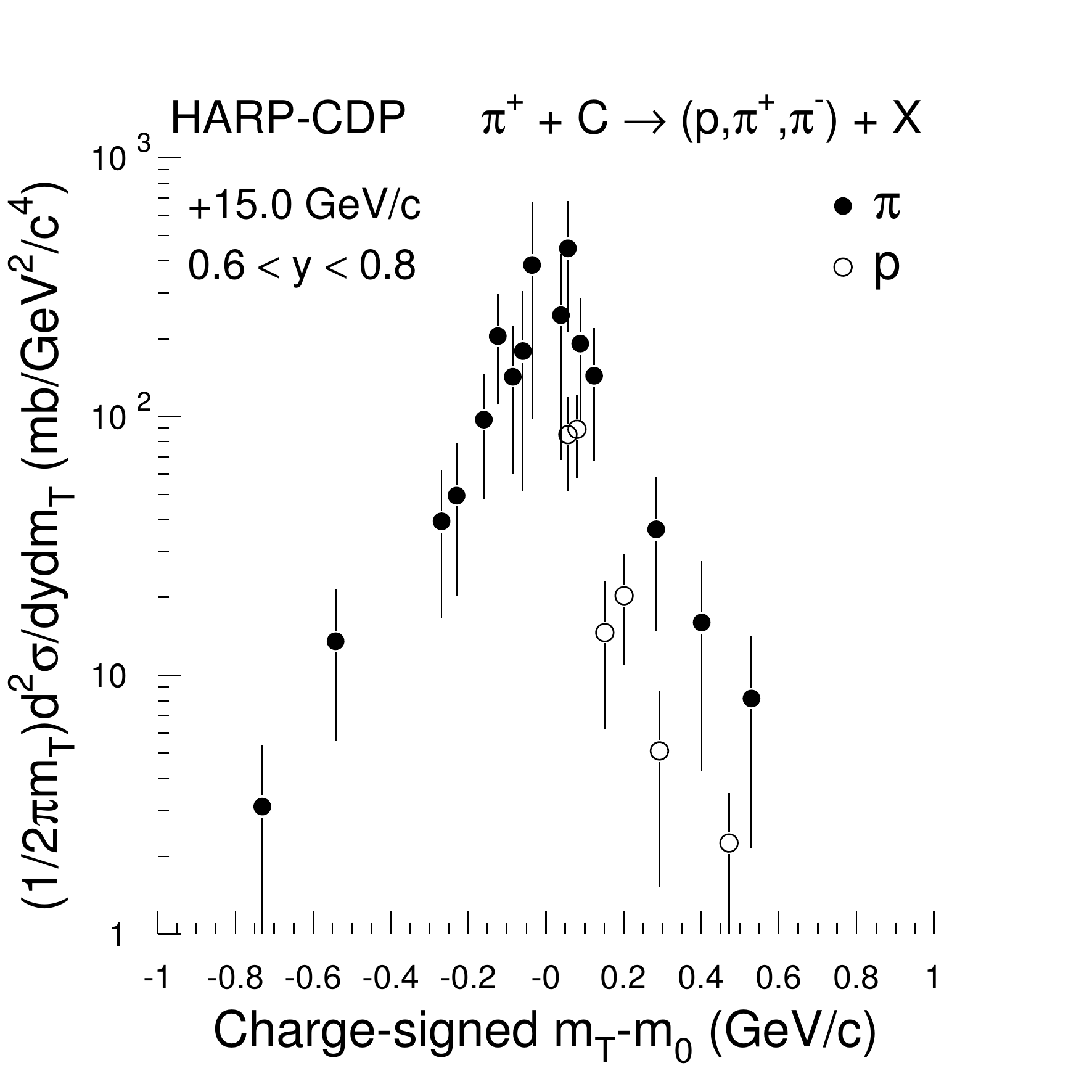} &  \\
\end{tabular}
\caption{Inclusive Lorentz-invariant  
cross-sections of the production of protons, $\pi^+$'s and $\pi^-$'s, by incoming
$\pi^+$'s between 3~GeV/{\it c} and 15~GeV/{\it c} momentum, in the rapidity 
range $0.6 < y < 0.8$,
as a function of the charge-signed reduced transverse pion mass, $m_{\rm T} - m_0$,
where $m_0$ is the rest mass of the respective particle;
the shown errors are total errors.}
\label{xsvsmTpip}
\end{center}
\end{figure*}

\begin{figure*}[h]
\begin{center}
\begin{tabular}{cc}
\includegraphics[height=0.30\textheight]{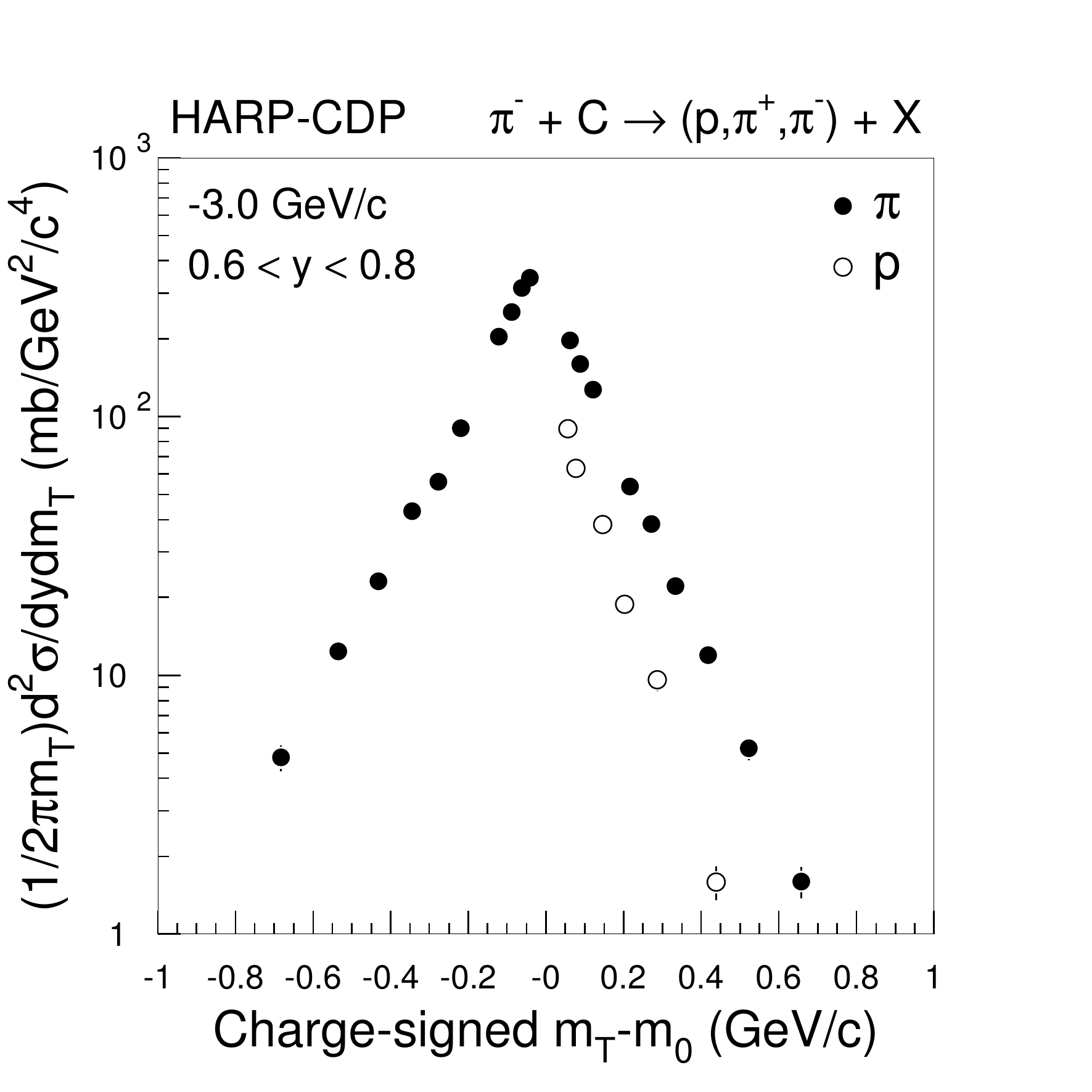} &
\includegraphics[height=0.30\textheight]{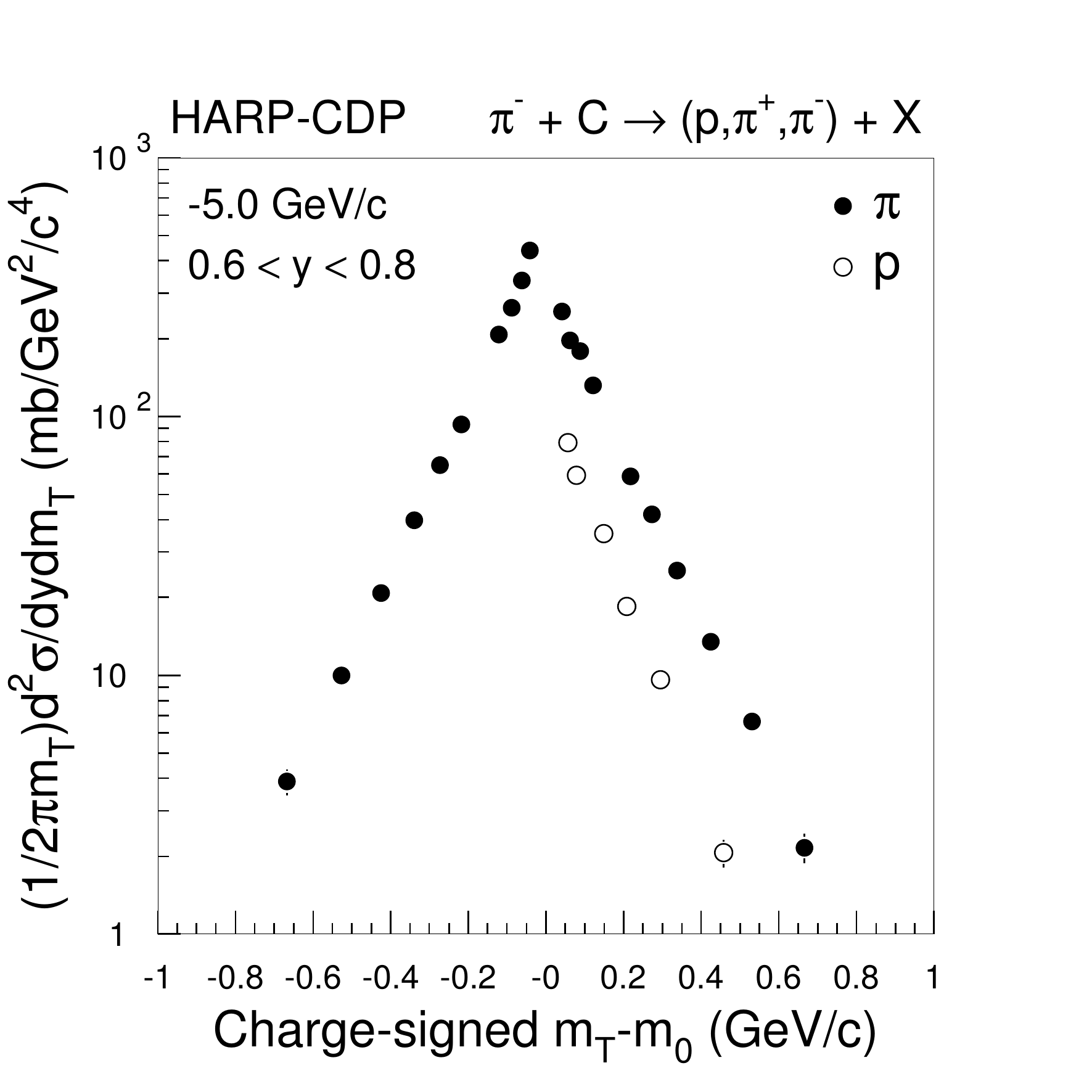} \\
\includegraphics[height=0.30\textheight]{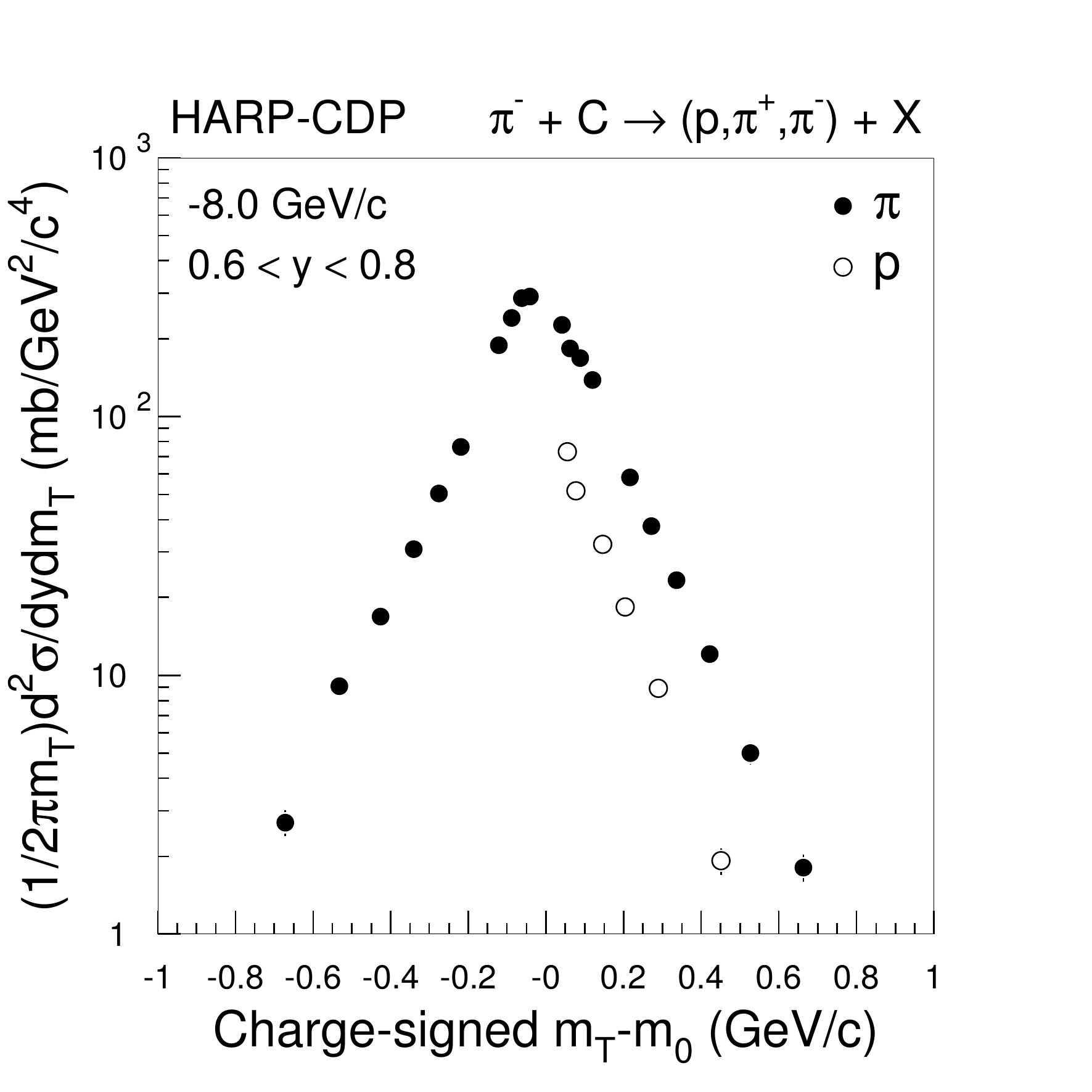} &
\includegraphics[height=0.30\textheight]{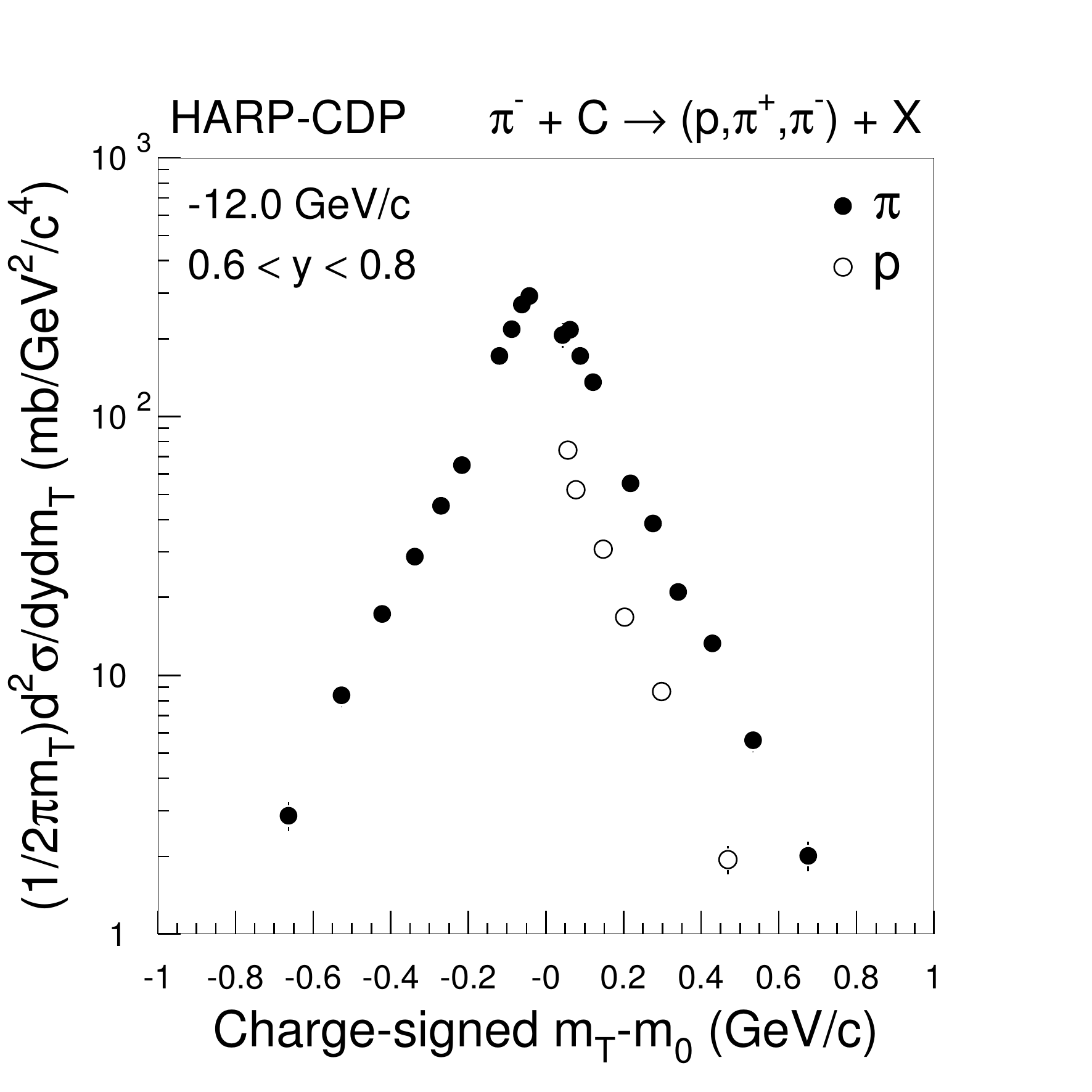} \\
\includegraphics[height=0.30\textheight]{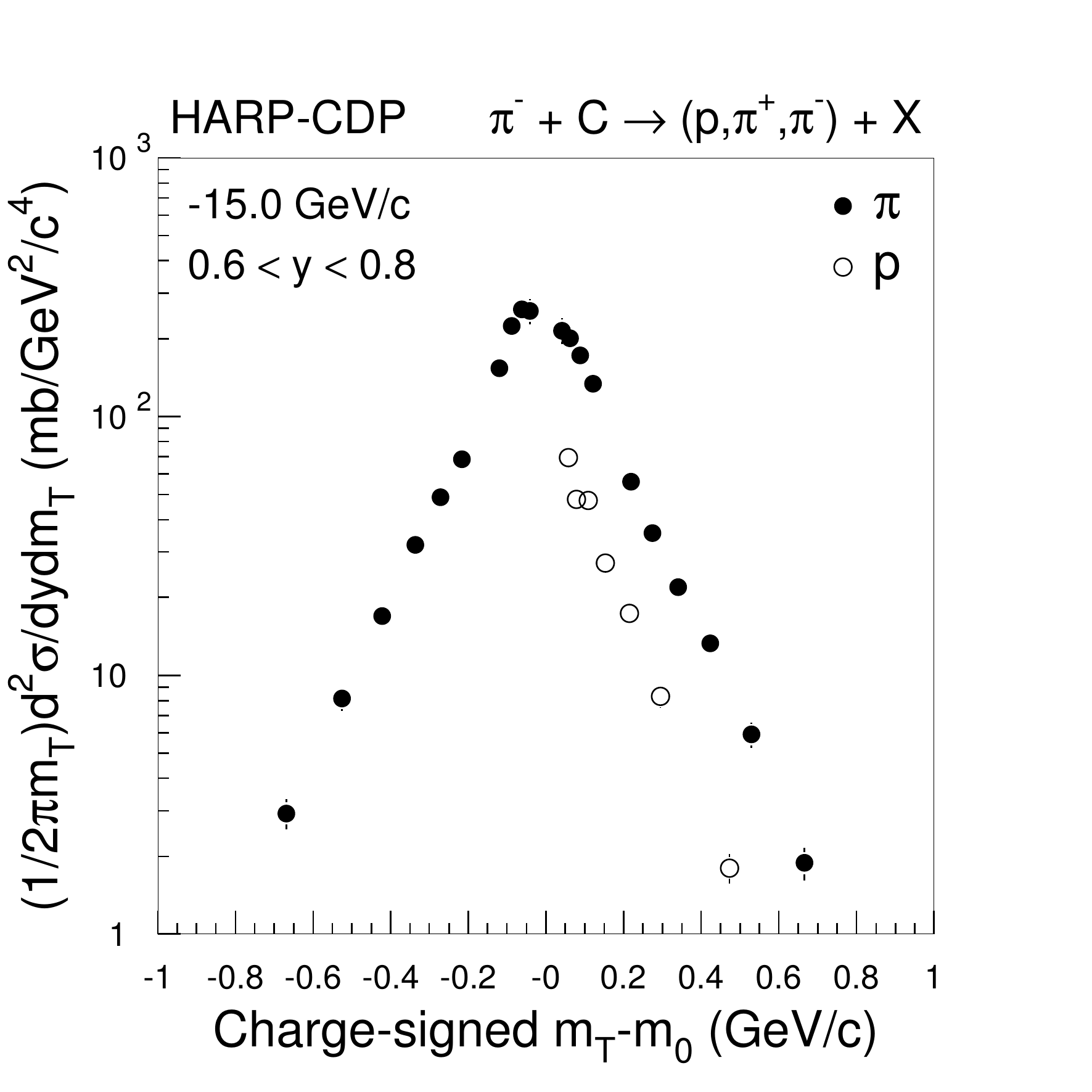} &  \\
\end{tabular}
\caption{Inclusive Lorentz-invariant  
cross-sections of the production of protons, $\pi^+$'s and $\pi^-$'s, by incoming
$\pi^-$'s between 3~GeV/{\it c} and 15~GeV/{\it c} momentum, in the rapidity 
range $0.6 < y < 0.8$,
as a function of the charge-signed reduced transverse pion mass, $m_{\rm T} - m_0$,
where $m_0$ is the rest mass of the respective particle;
the shown errors are total errors.}
\label{xsvsmTpim}
\end{center}
\end{figure*}

\begin{figure*}[h]
\begin{center}
\begin{tabular}{cc}
\includegraphics[height=0.30\textheight]{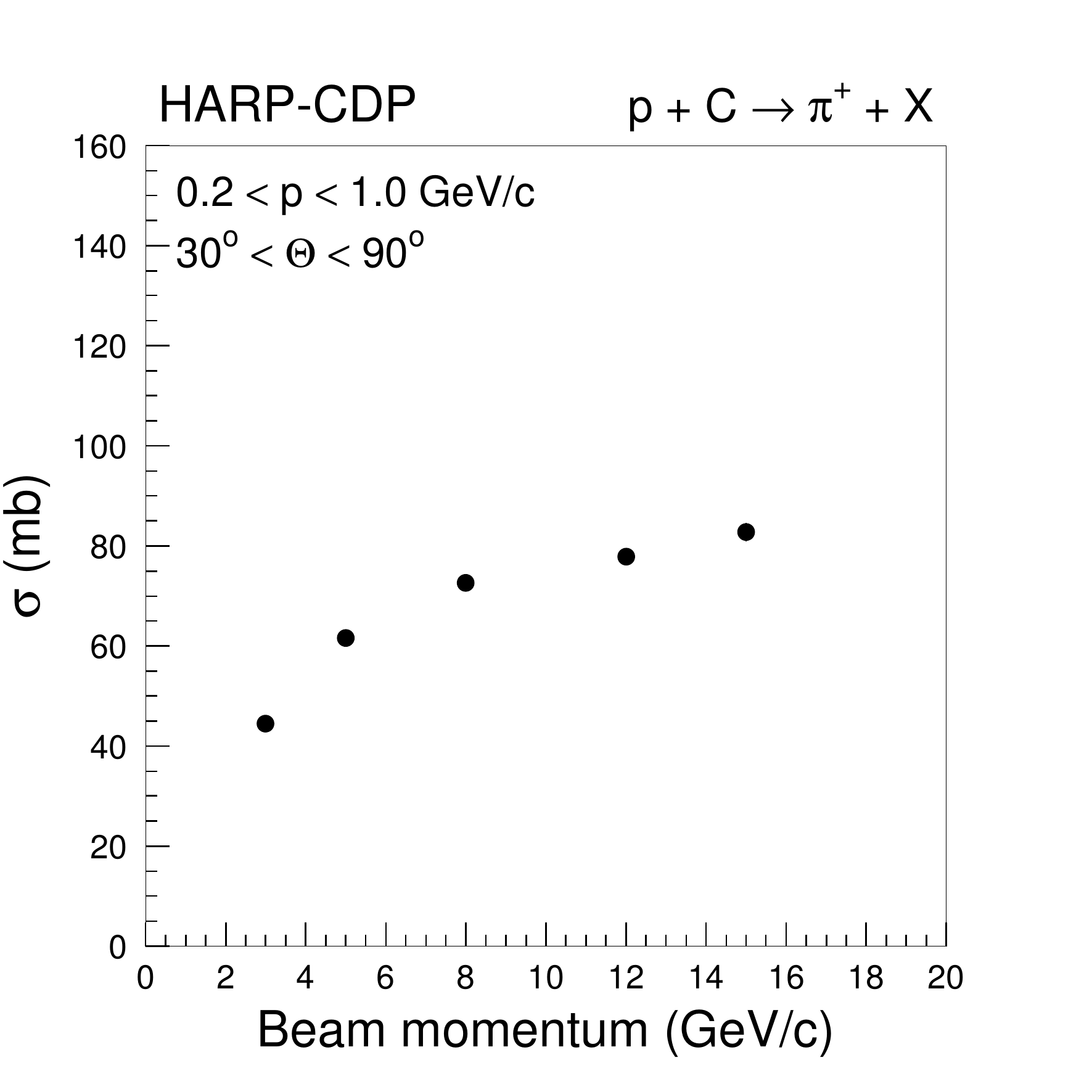} &
\includegraphics[height=0.30\textheight]{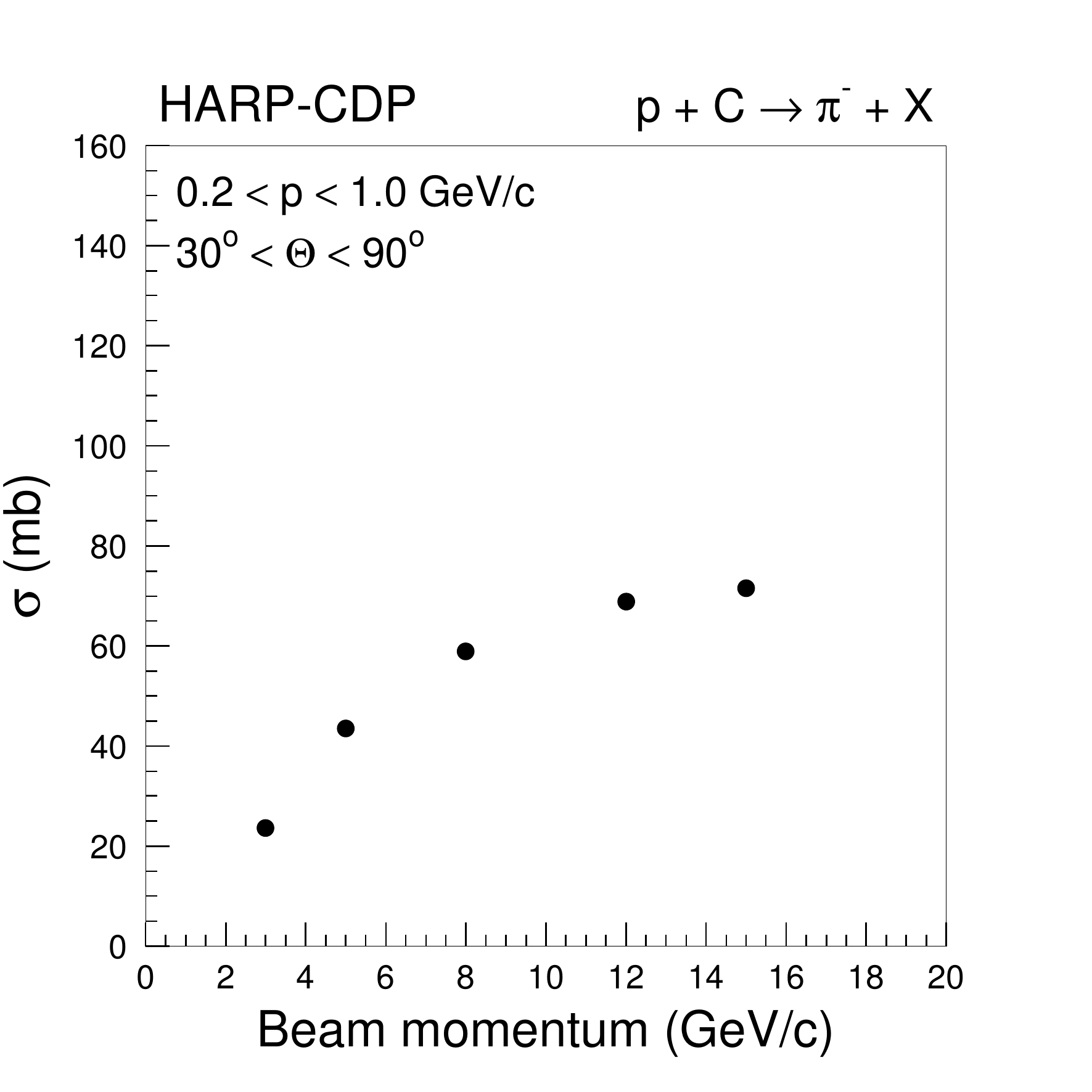} \\
\includegraphics[height=0.30\textheight]{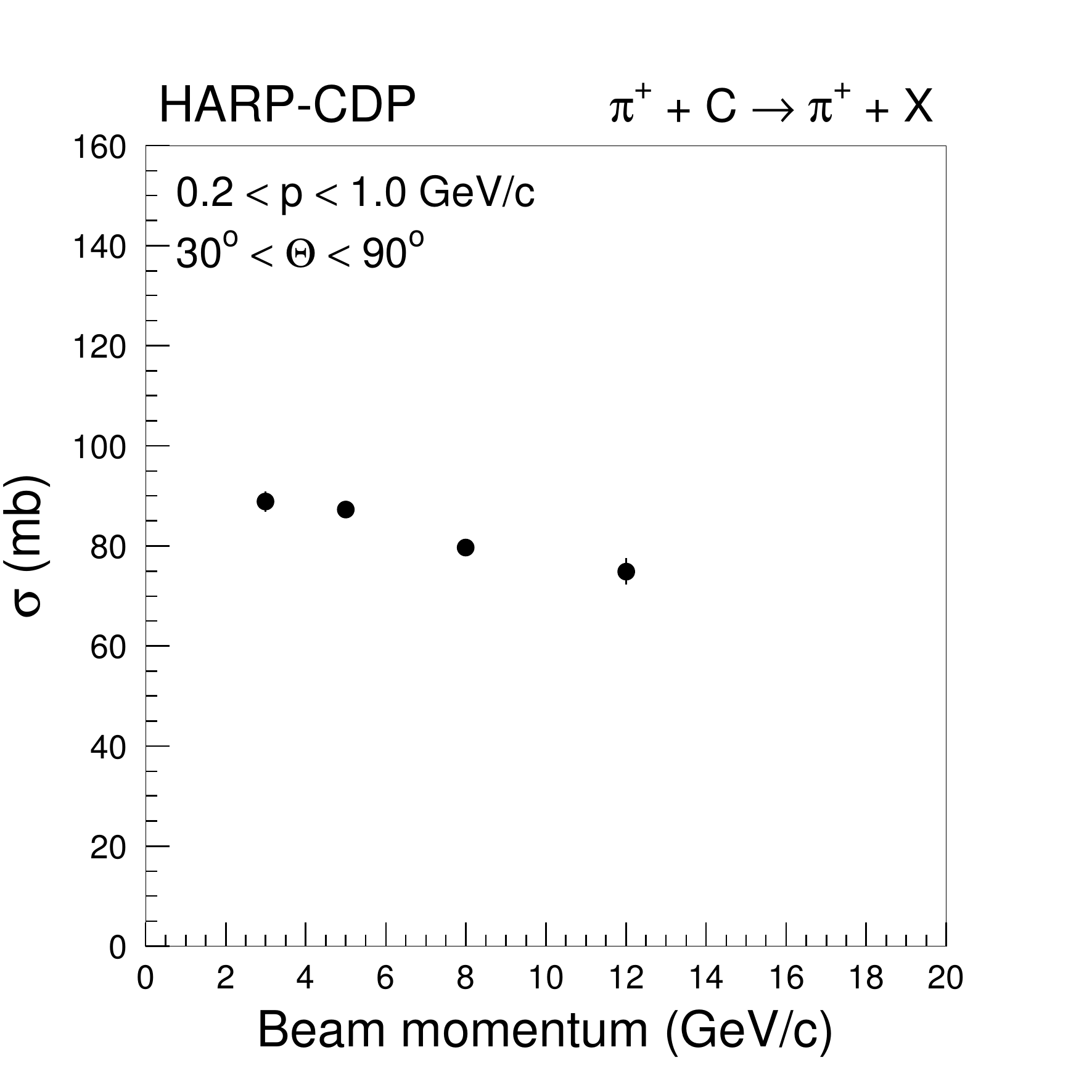} &
\includegraphics[height=0.30\textheight]{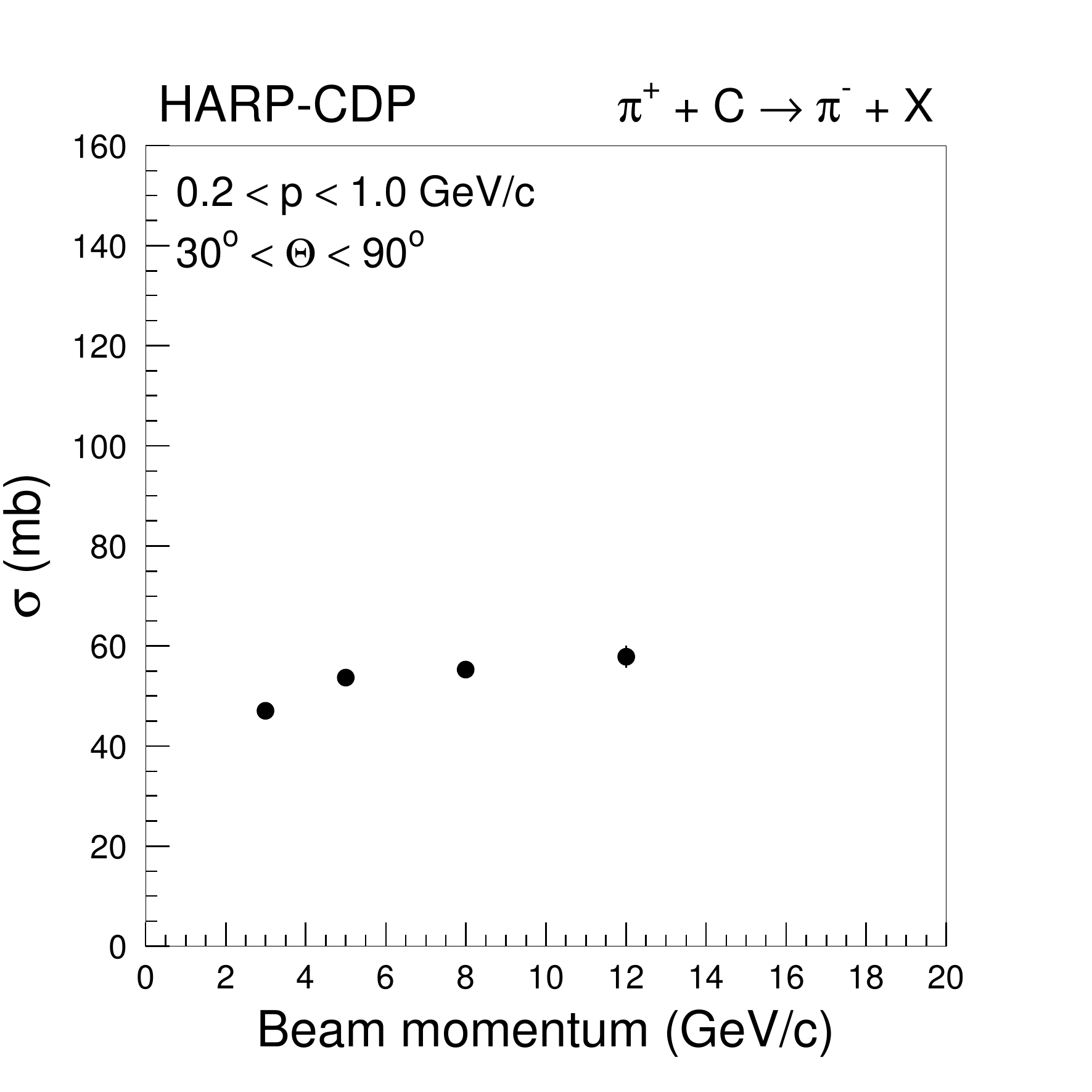} \\
\includegraphics[height=0.30\textheight]{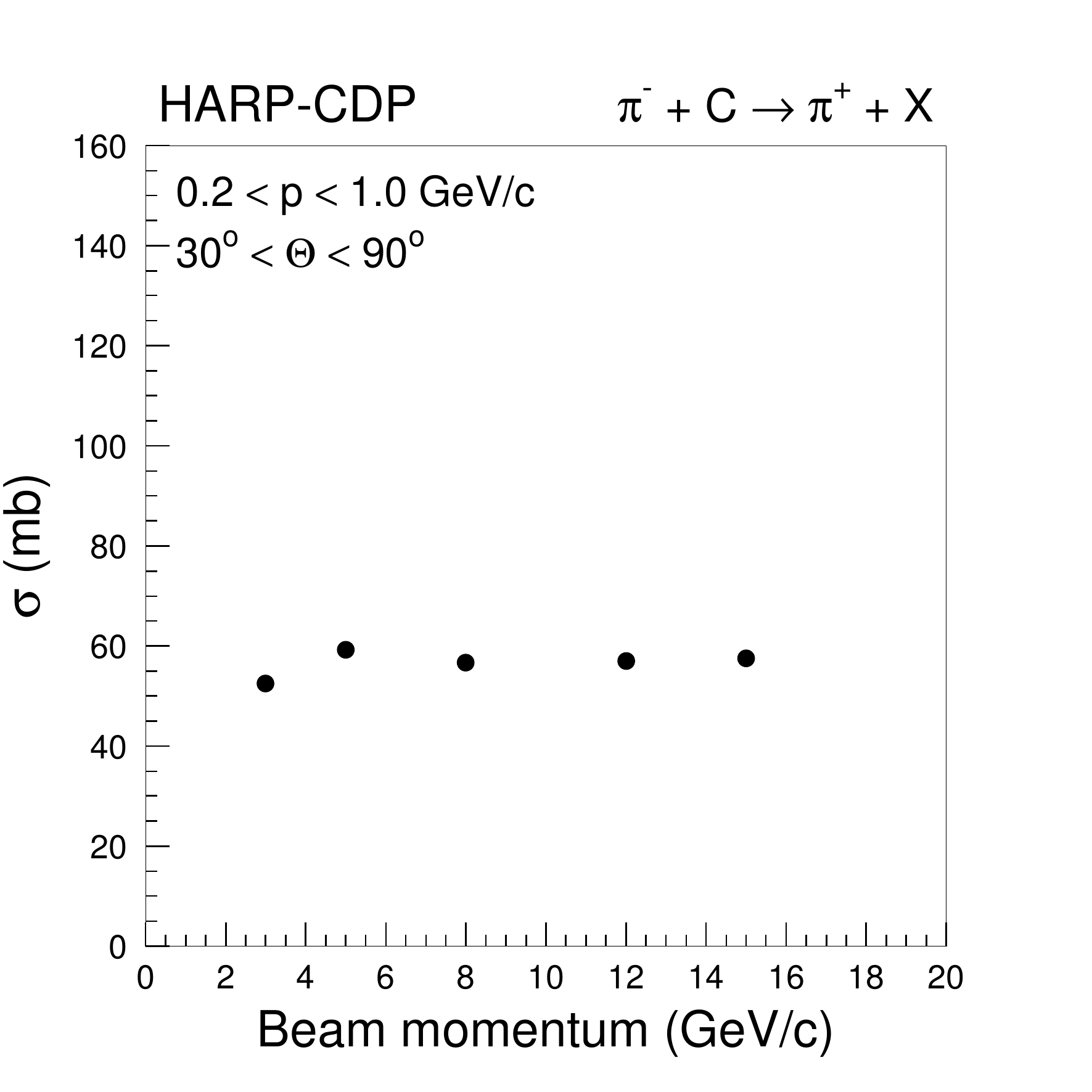} &  
\includegraphics[height=0.30\textheight]{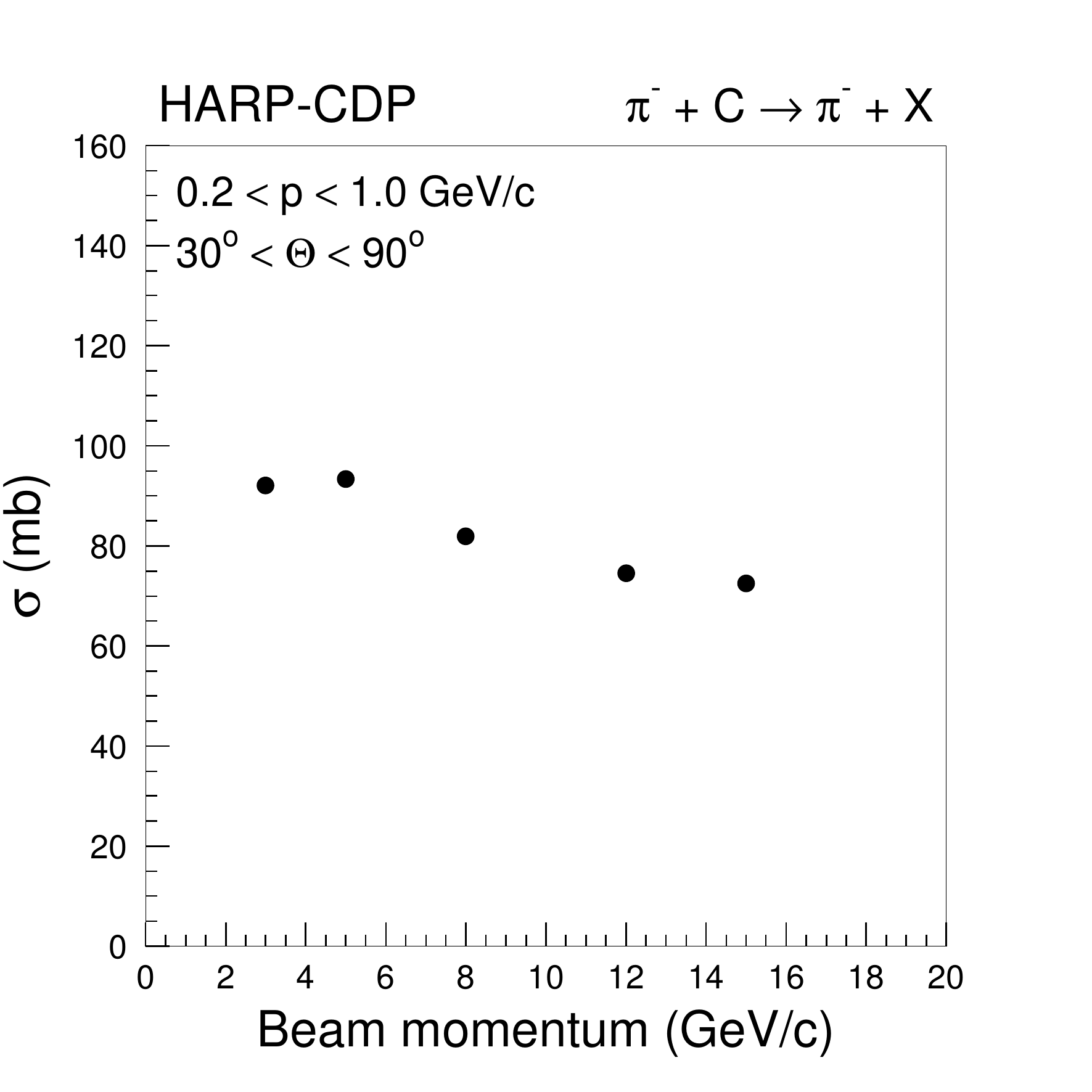} \\
\end{tabular}
\caption{Inclusive cross-sections of the production of secondary $\pi^+$'s and $\pi^-$'s, integrated over the momentum range $0.2 < p < 1.0$~GeV/{\it c} and the polar-angle range $30^\circ < \theta < 90^\circ$, from the interactions on carbon nuclei of protons (top row), $\pi^+$'s (middle row), and $\pi^-$'s (bottom row), as a function of the beam momentum; the shown errors are total errors and mostly smaller than the symbol size.} 
\label{fxsc}
\end{center}
\end{figure*}

We present in Figs.~\ref{xsvsmompro} to \ref{fxsc} what we consider salient features of our cross-sections.

Figure~\ref{xsvsmompro} shows the inclusive cross-sections of the production of protons, $\pi^+$'s, and $\pi^-$'s, from incoming protons between 3~GeV/{\it c} and 15~GeV/{\it c} momentum, as a function of their charge-signed $p_{\rm T}$. The data refer to the polar-angle range $20^\circ < \theta < 30^\circ$. Figures~\ref{xsvsmompip} and \ref{xsvsmompim} show the same for 
incoming $\pi^+$'s and $\pi^-$'s.

Figure~\ref{xsvsmTpro} shows inclusive Lorentz-invariant cross-sections of the production of protons, $\pi^+$'s and $\pi^-$'s, by incoming protons between 3~GeV/{\it c} and 15~GeV/{\it c} momentum, in the rapidity range $0.6 < y < 0.8$, as a function of the charge-signed reduced transverse particle mass, $m_{\rm T} - m_0$,
where $m_0$ is the rest mass of the respective particle. Figures~\ref{xsvsmTpip} and \ref{xsvsmTpim} show the same for incoming $\pi^+$'s and $\pi^-$'s. We note the good representation of particle production by an exponential  falloff with increasing reduced transverse mass.

In Fig.~\ref{fxsc}, we present the inclusive cross-sections of the production of secondary $\pi^+$'s and $\pi^-$'s, integrated over the momentum range $0.2 < p < 1.0$~GeV/{\it c} and the polar-angle range $30^\circ < \theta < 90^\circ$ in the forward hemisphere, as a function of the beam momentum. 

%\clearpage

\clearpage

\section{Comparison of our results with results from 
the HARP Collaboration}

\begin{figure}[ht]
\begin{center}
\begin{tabular}{cc}
\includegraphics[width=0.45\textwidth]{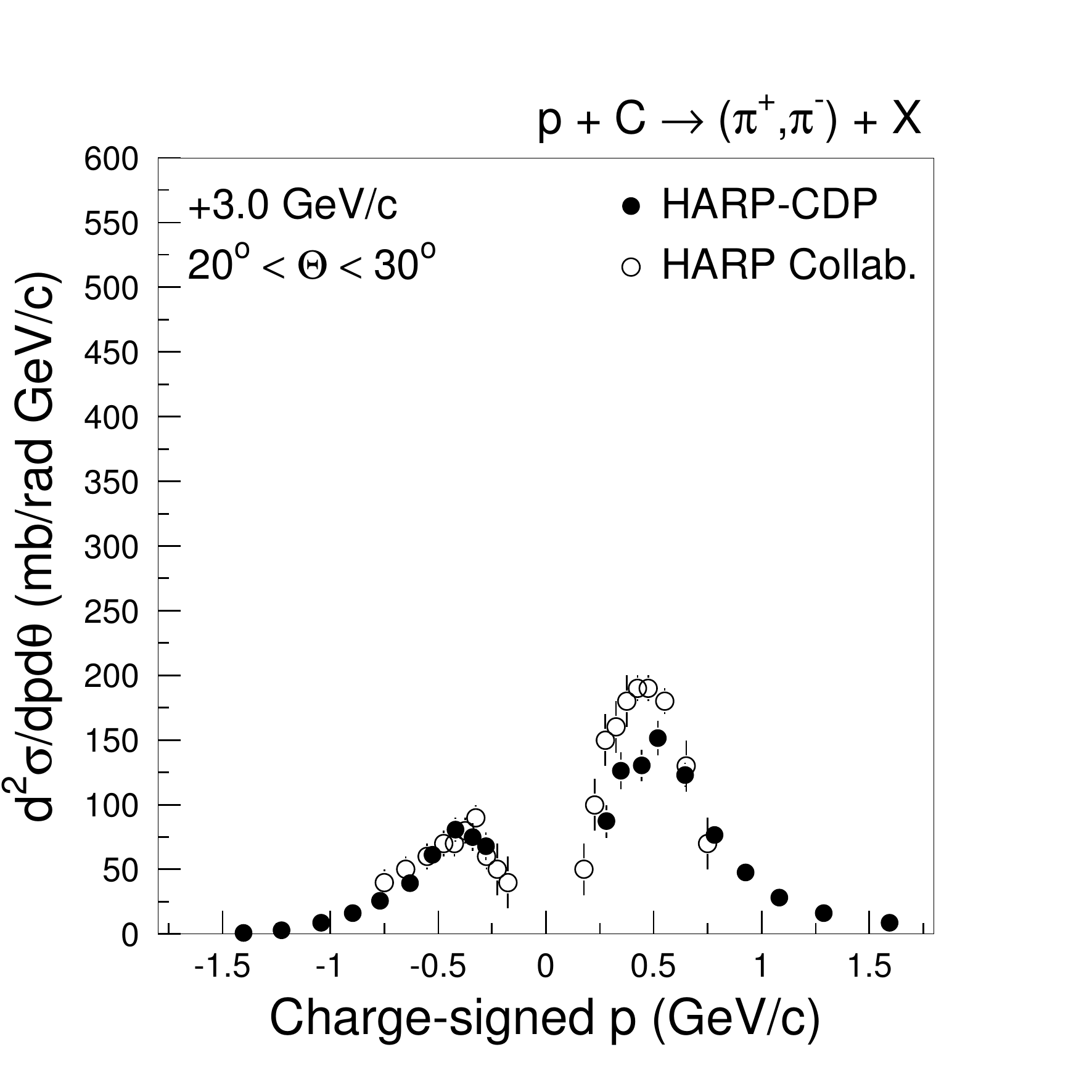}  &
\includegraphics[width=0.45\textwidth]{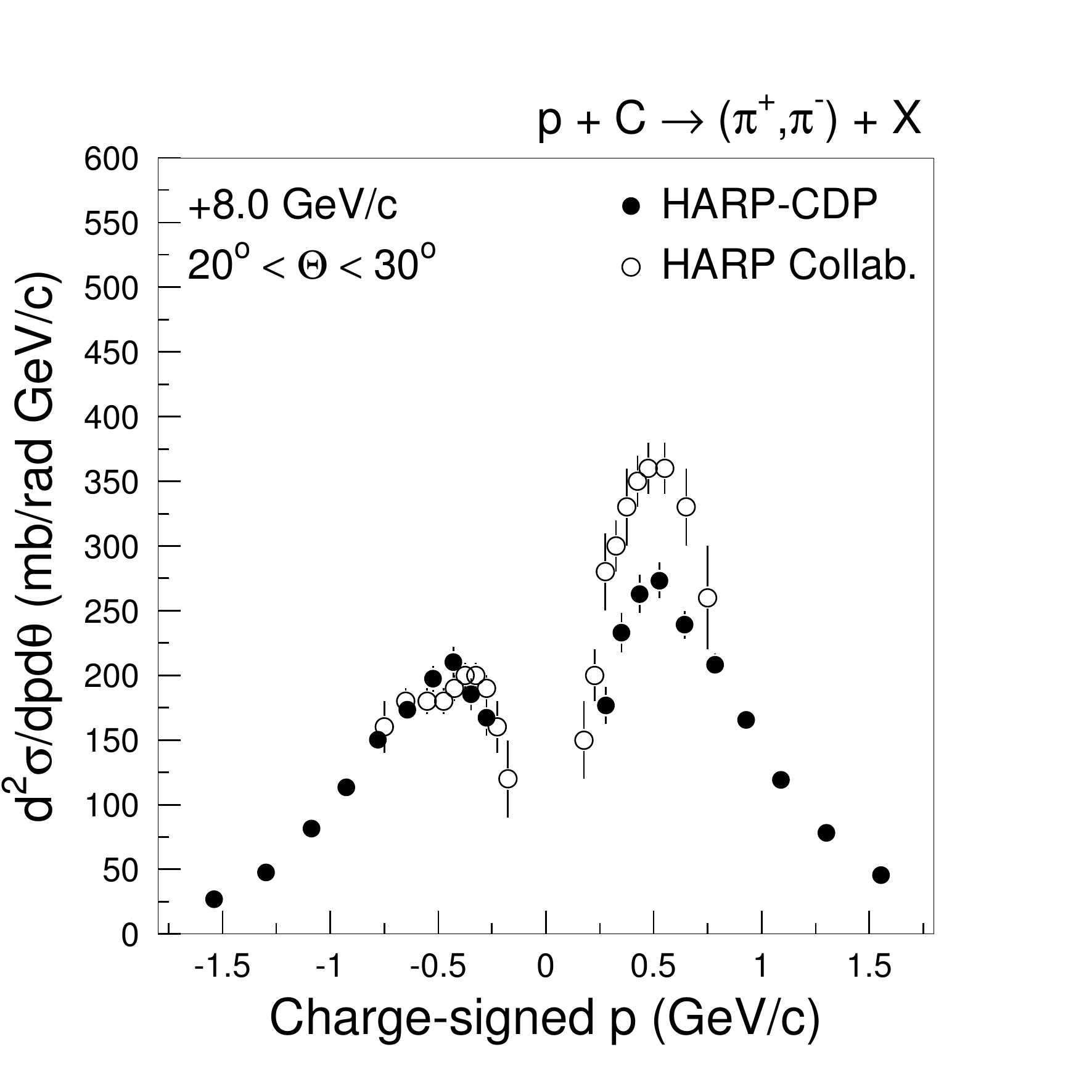}  \\
\includegraphics[width=0.45\textwidth]{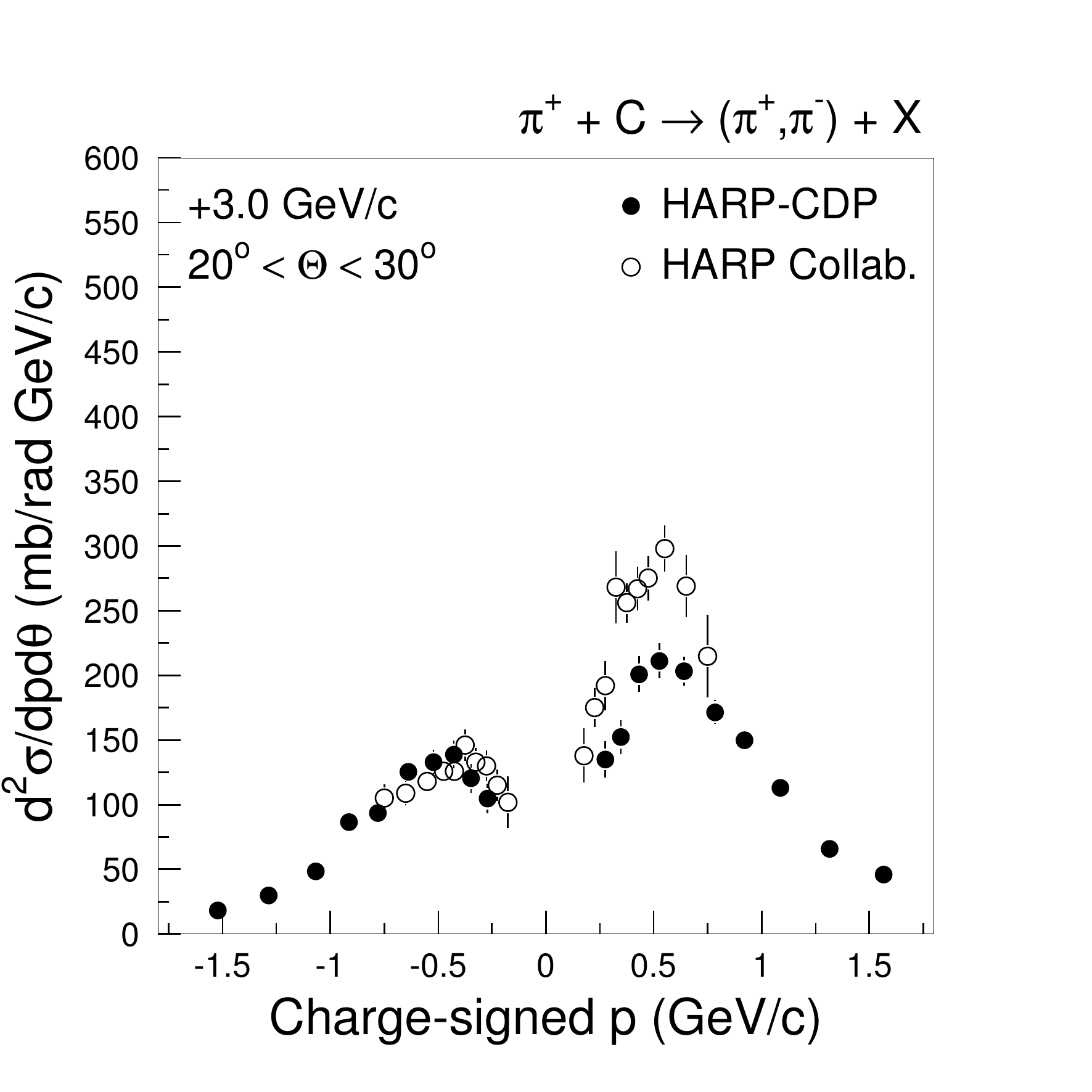}  &
\includegraphics[width=0.45\textwidth]{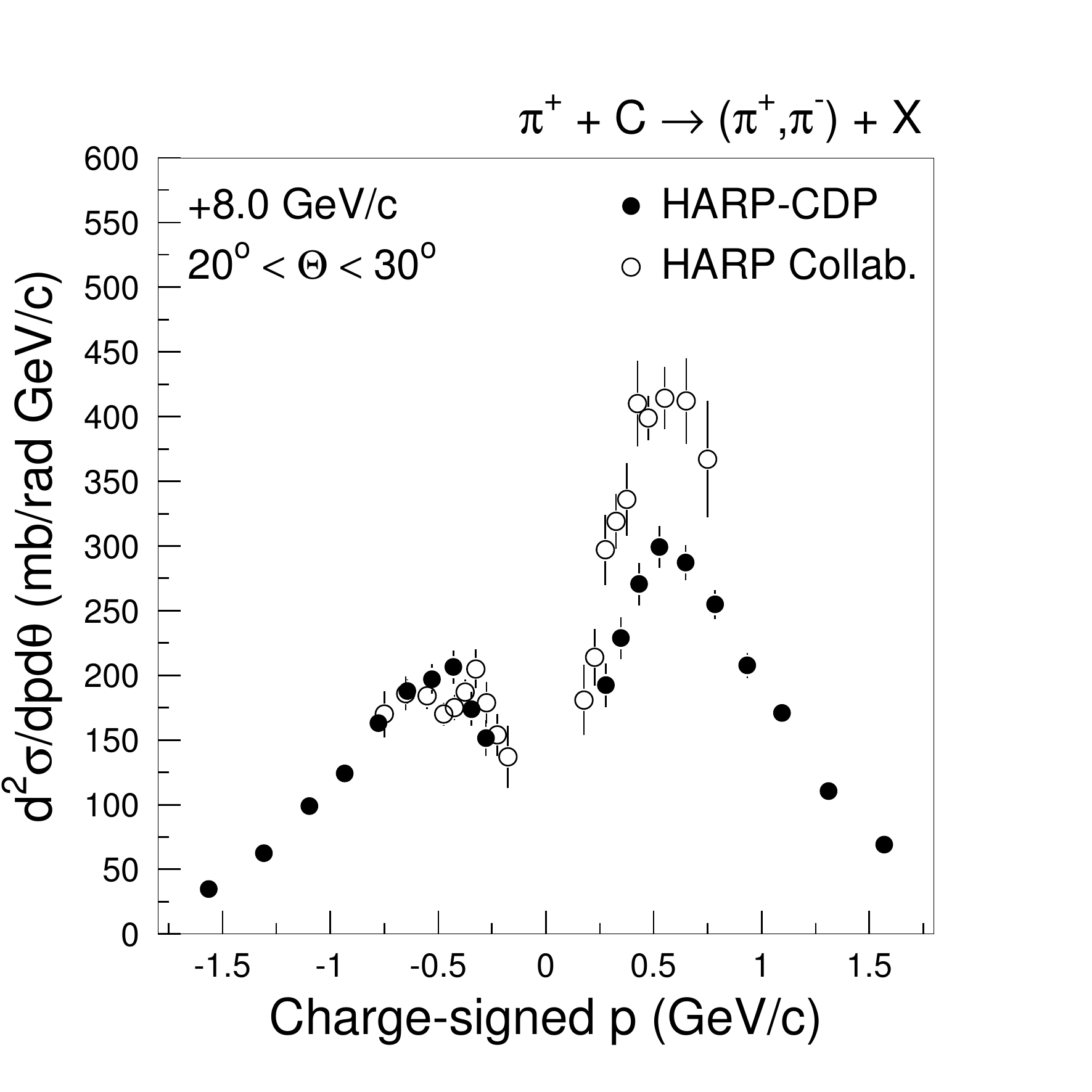}  \\
\includegraphics[width=0.45\textwidth]{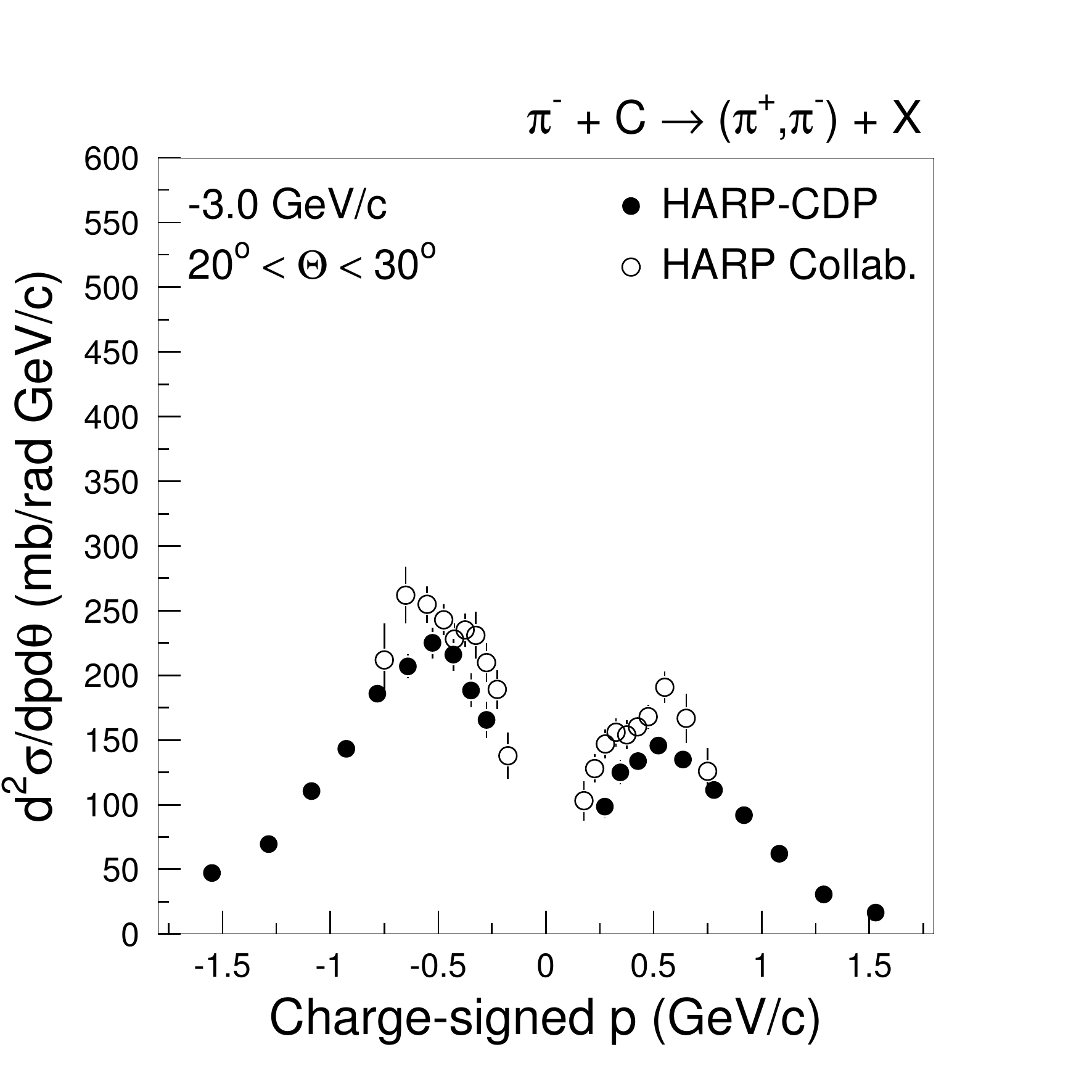} &
\includegraphics[width=0.45\textwidth]{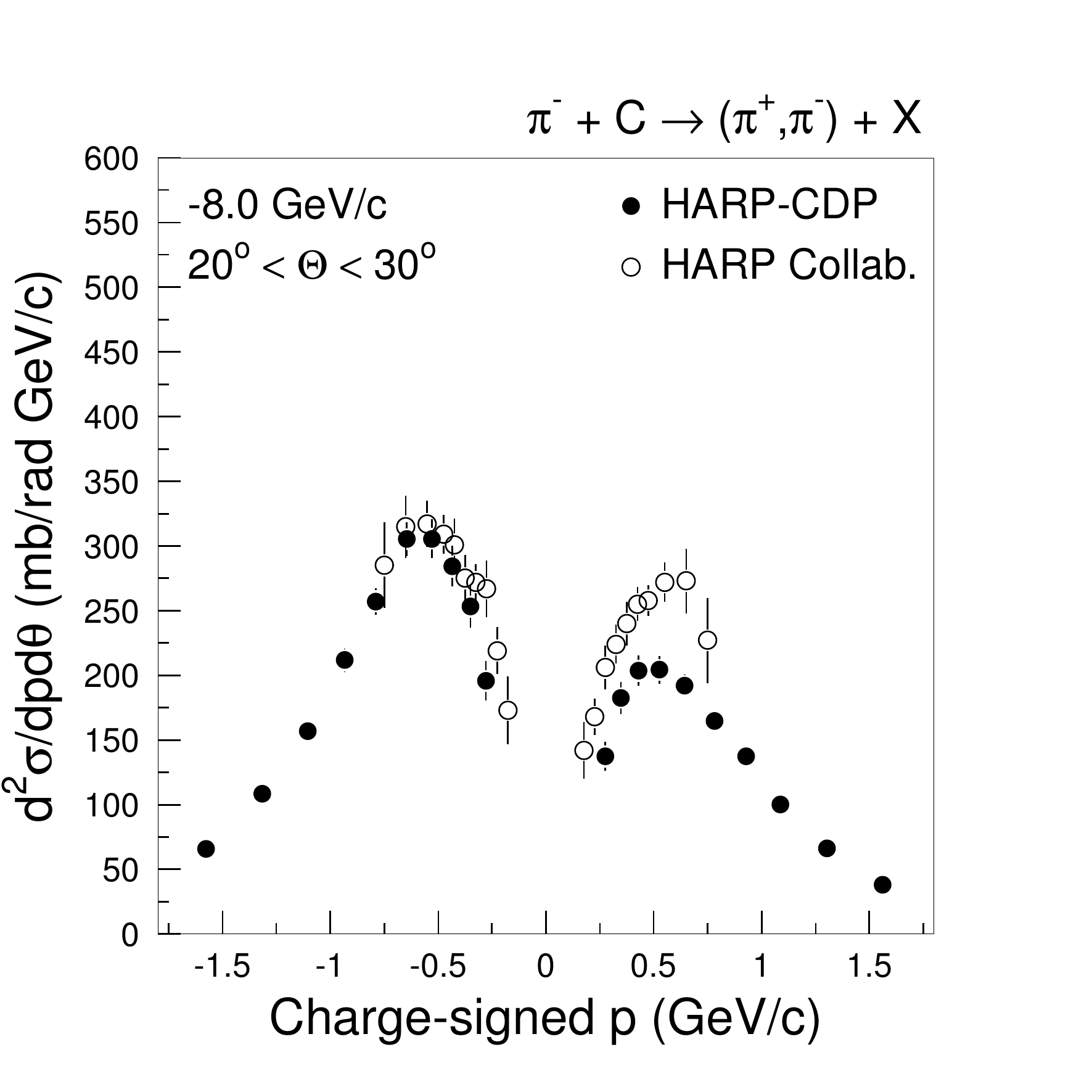} \\
\end{tabular}
\caption{Comparison of HARP--CDP cross-sections (full circles) of $\pi^\pm$ production by protons, $\pi^+$'s  and $\pi^-$'s of 3.0~GeV/{\it c} (left panels) and 8.0~GeV/{\it c} 
momentum (right panels), off carbon nuclei, with the cross-sections published by the HARP Collaboration (open circles).} 
\label{C3and8ComparisonWithOH}
\end{center}
\end{figure}

Figure~\ref{C3and8ComparisonWithOH} shows the comparison of our cross-sections of $\pi^\pm$ production by protons, $\pi^+$'s and $\pi^-$'s of 3.0~GeV/{\it c} and 8.0~GeV/{\it c} momentum, off carbon nuclei, with the ones published by the HARP Collaboration~\cite{OffLAprotonpaper,OffLApionpaper}, in the 
polar-angle range $20^\circ < \theta < 30^\circ$. The latter cross-sections are plotted as published, while we expressed our cross-sections in the unit used by the HARP Collaboration. The errors shown are the published total errors.

The discrepancy between our results and those published by the HARP Collaboration is evident. It shows the same pattern as observed in inclusive cross-sections off other targets that we analyzed and compared to the results of the HARP Collaboration~\cite{Beryllium1,Beryllium2,Copper,Tantalum,Lead}.
We hold that the discrepancy is caused by problems in the HARP Collaboration's data analysis, discussed in detail in Refs~\cite{JINSTpub,EPJCpub,WhiteBook1,WhiteBook2,WhiteBook3}, and summarized in the Appendix of Ref.~\cite{Beryllium1}.

%\clearpage

\section{Comparison of charged-pion production on beryllium, carbon, copper, tantalum and lead}

\begin{figure*}[h]
\begin{center}
\begin{tabular}{cc}
\includegraphics[height=0.30\textheight]{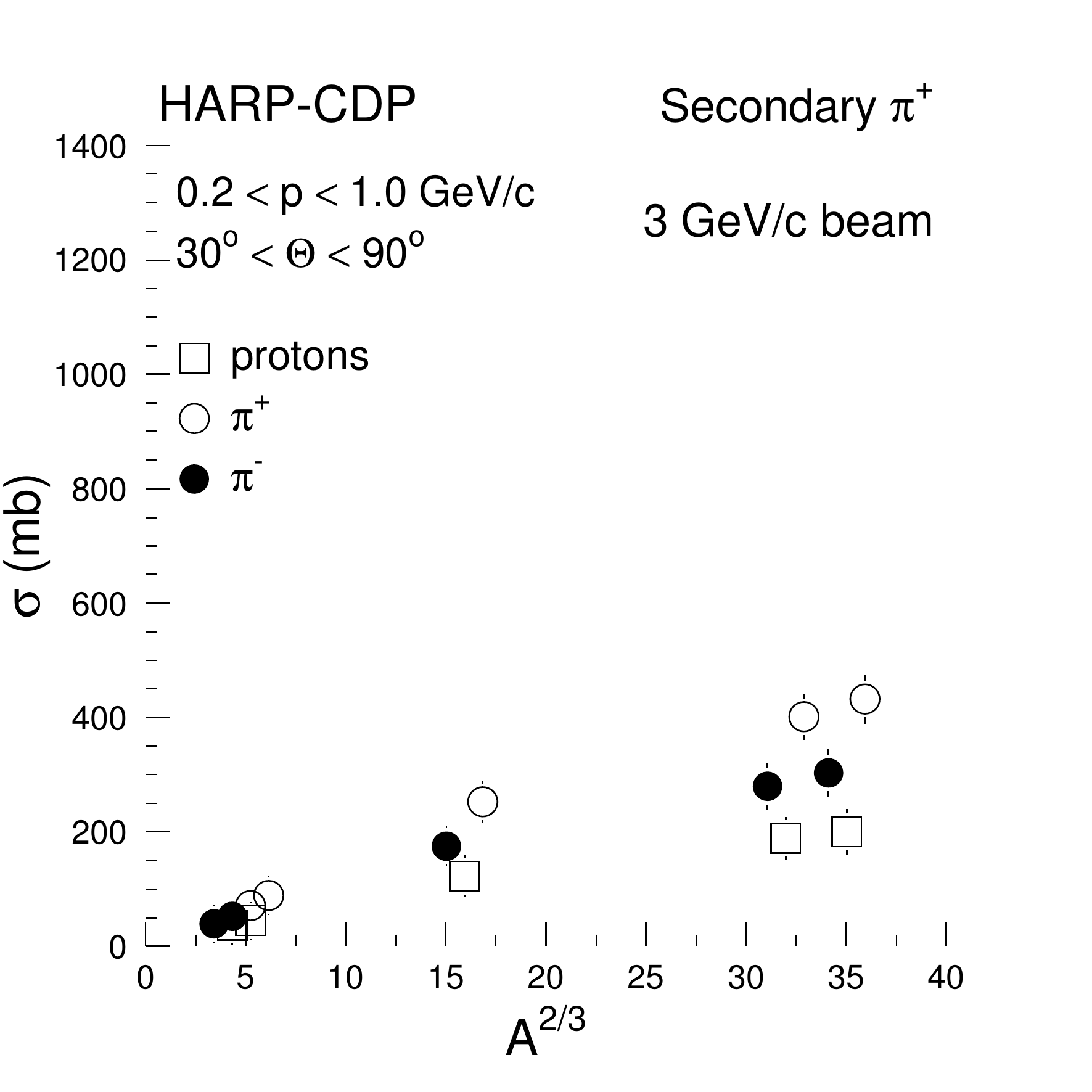} &
\includegraphics[height=0.30\textheight]{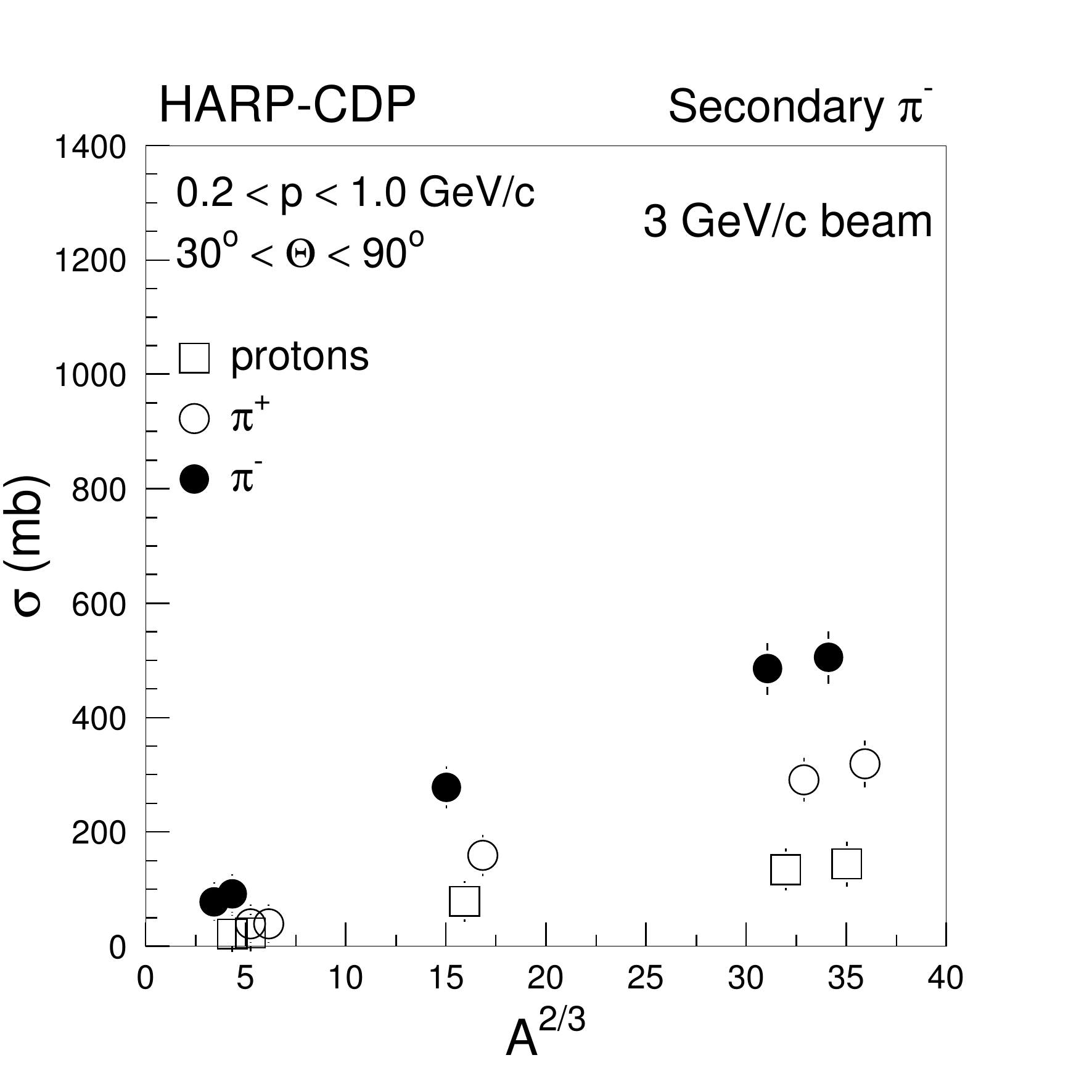} \\
\includegraphics[height=0.30\textheight]{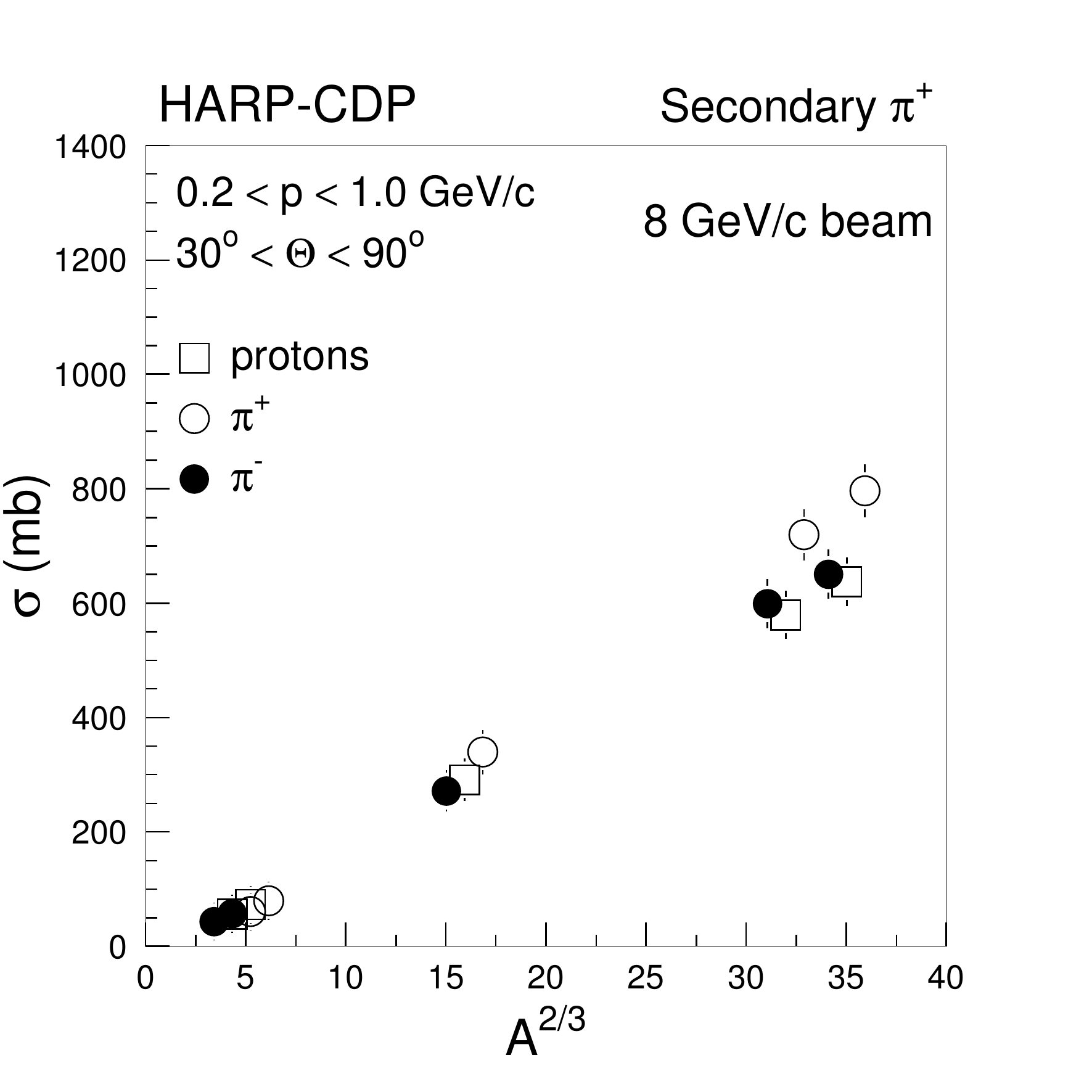} &
\includegraphics[height=0.30\textheight]{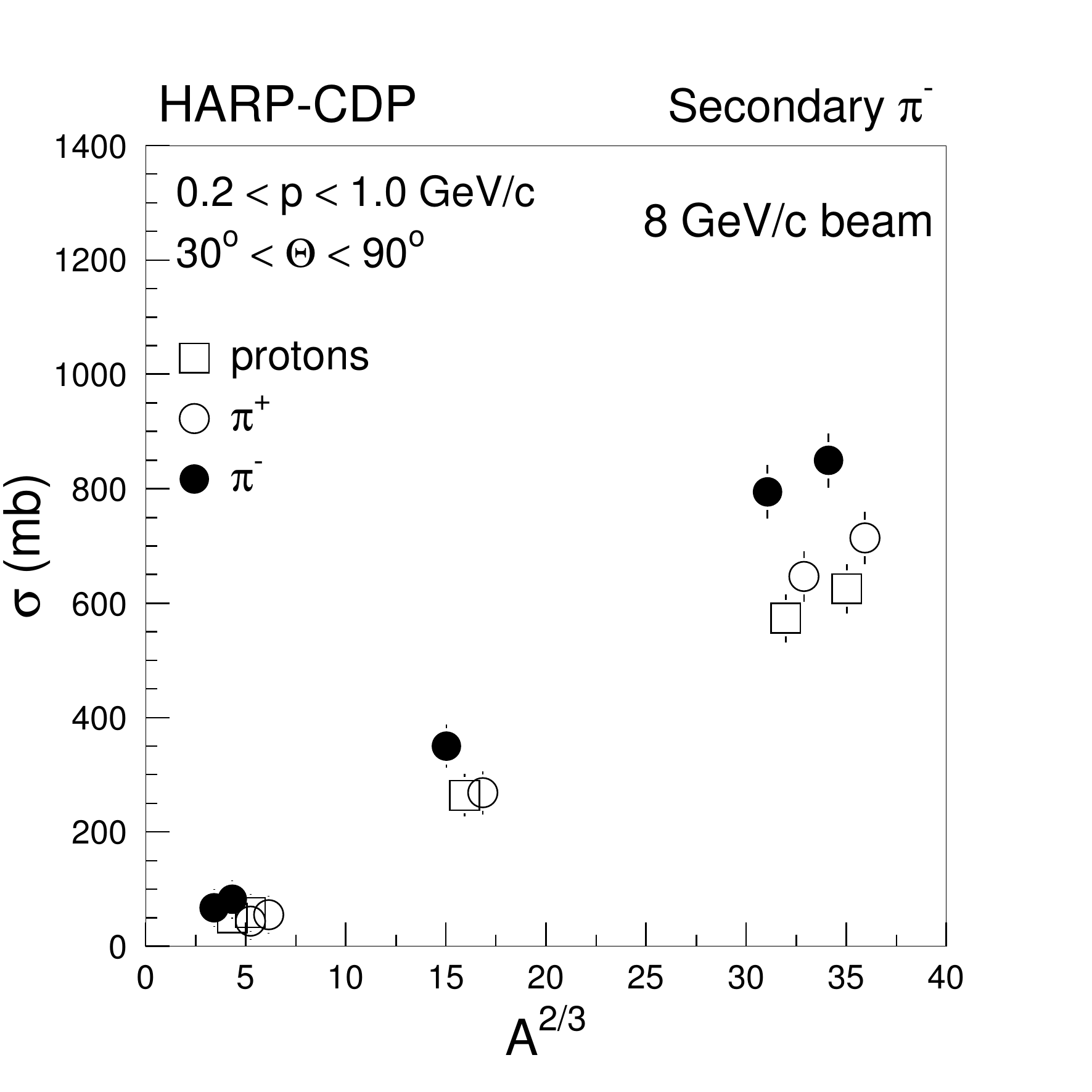} \\
\includegraphics[height=0.30\textheight]{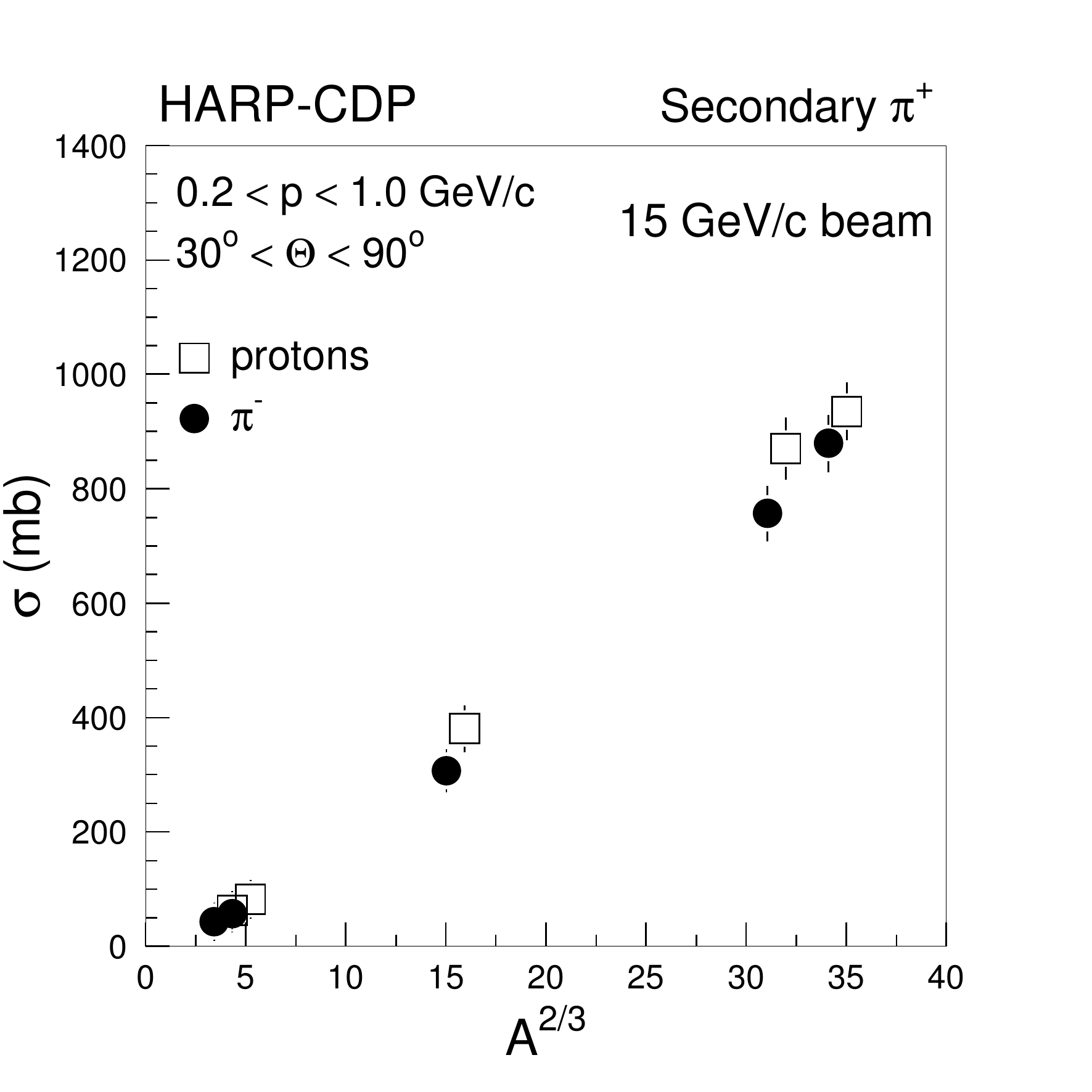} &  
\includegraphics[height=0.30\textheight]{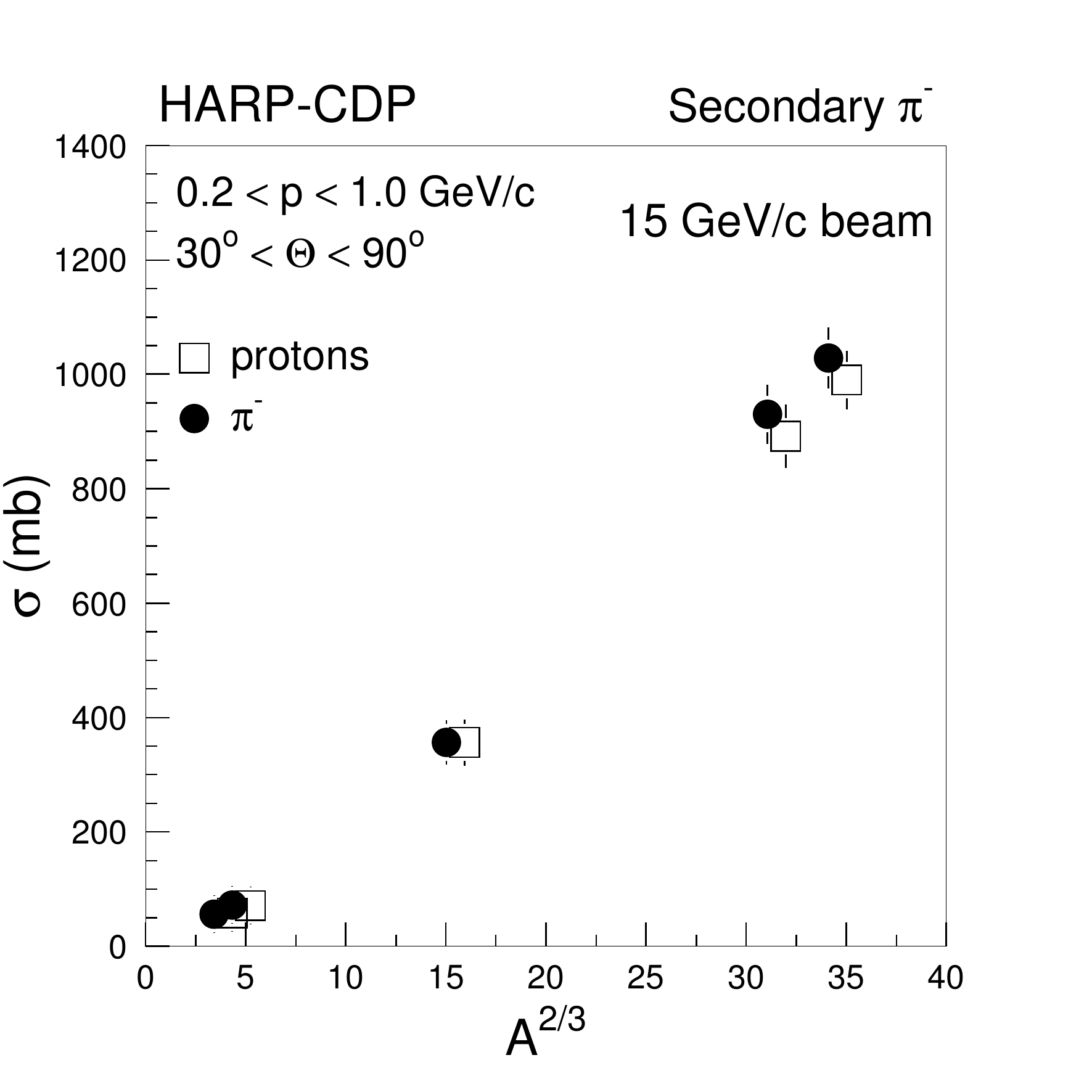} \\
\end{tabular}
\caption{Inclusive cross-sections of $\pi^+$ and $\pi^-$ production by protons (open squares), $\pi^+$'s (open circles), and $\pi^-$'s (black circles), as a function of $A^{2/3}$ for, from left to right, beryllium, carbon, copper, tantalum, and lead nuclei; the cross-sections are integrated over the momentum range $0.2 < p < 1.0$~GeV/{\it c} and the polar-angle range $30^\circ < \theta < 90^\circ$; the shown errors are total errors and often smaller than the symbol size.} 
\label{ComparisonxsecBeCCuTaPb}
\end{center}
\end{figure*}

Figure~\ref{ComparisonxsecBeCCuTaPb} presents a comparison between the inclusive cross-sections of $\pi^+$ and $\pi^-$ production, integrated over the secondaries' momentum range 
$0.2 < p < 1.0$~GeV/{\it c} and polar-angle range $30^\circ < \theta < 90^\circ$, in the interactions of protons, $\pi^+$ and $\pi^-$, with beryllium (A~=~9.01), carbon (A~=~12.01), copper (A~=~63.55), tantalum (A~=~181.0), and lead (A~=~207.2) nuclei\footnote{The beryllium data with $+8.9$~GeV/{\it c} beam momentum~\cite{Beryllium1,Beryllium2} have been scaled, by interpolation, to a beam momentum of $+8.0$~GeV/{\it c}.}. The comparison employs the scaling variable $A^{2/3}$ where $A$ is the atomic number of the respective nucleus. We note the approximately linear dependence on this scaling variable. At low beam momentum, the slope exhibits a strong dependence on beam particle type, which tends to disappear with higher beam momentum. 

\begin{figure*}[h]
\begin{center}
\begin{tabular}{cc}
\includegraphics[height=0.30\textheight]{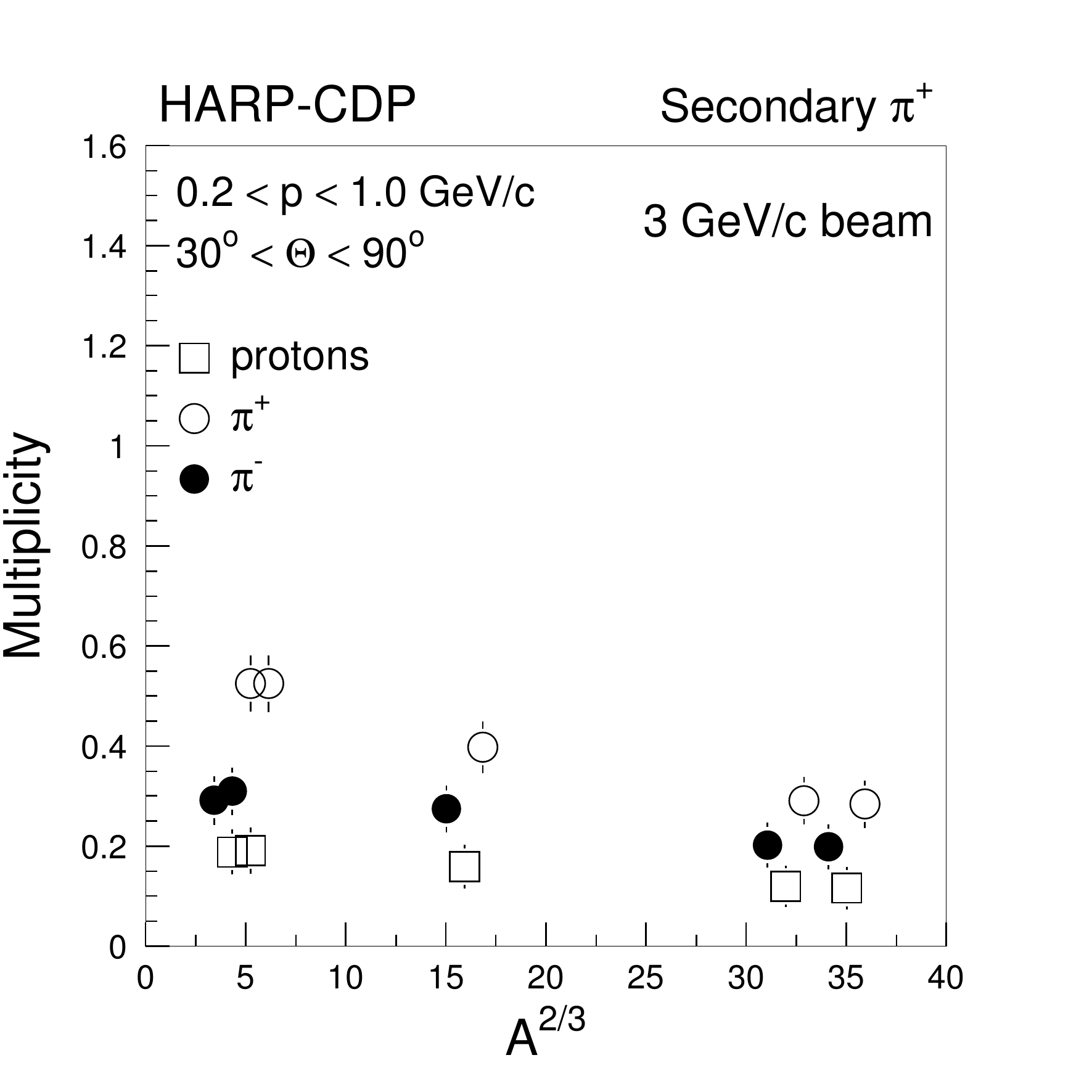} &
\includegraphics[height=0.30\textheight]{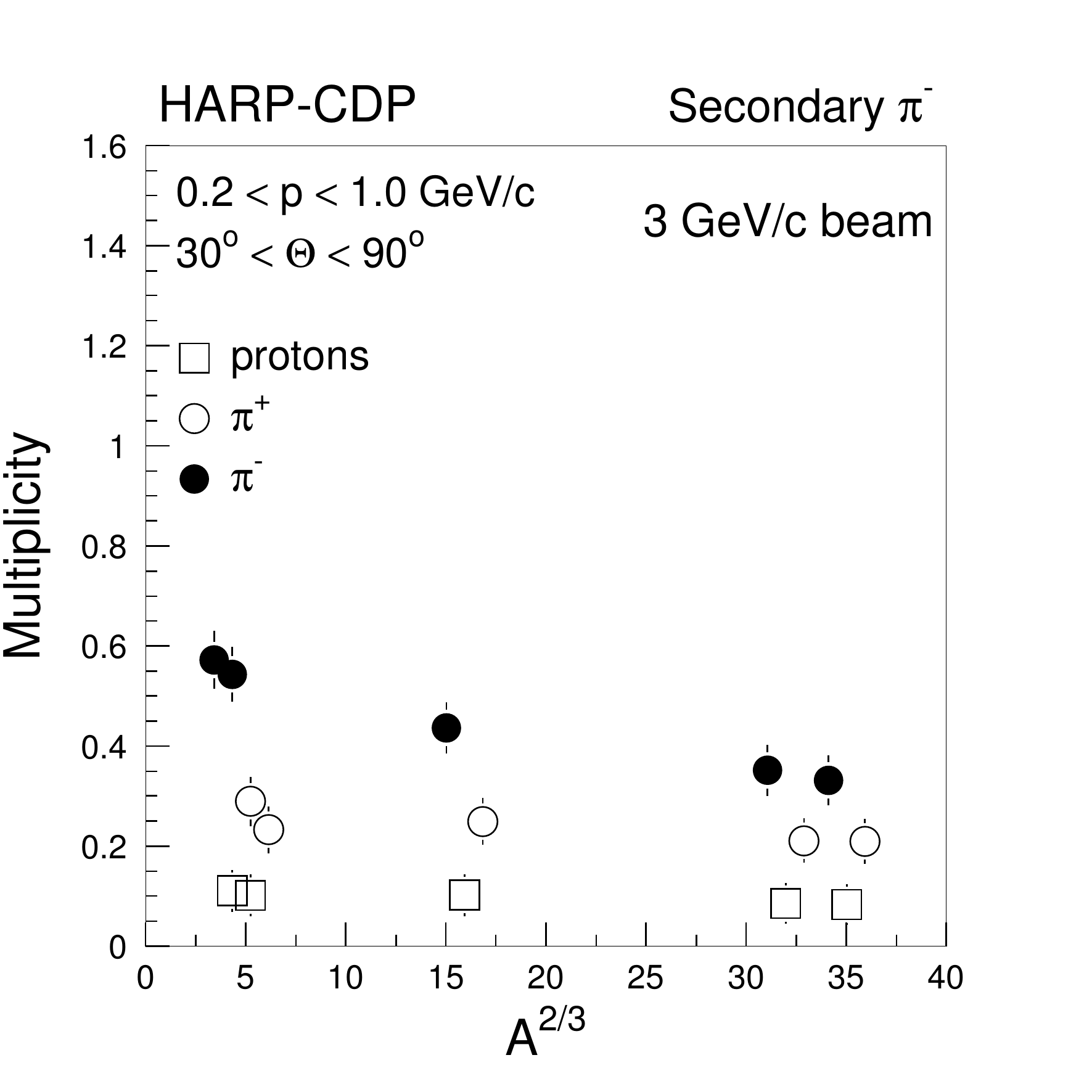} \\
\includegraphics[height=0.30\textheight]{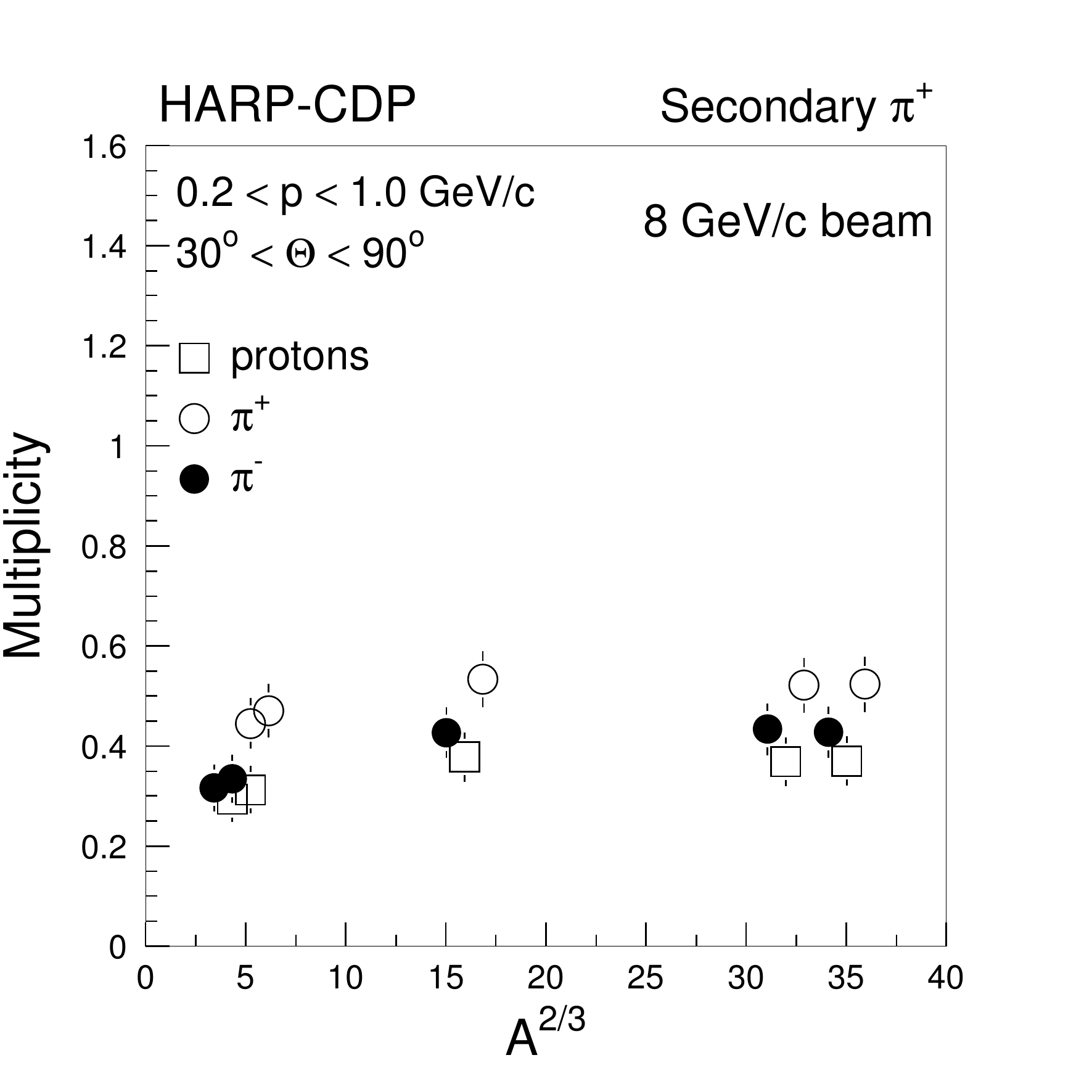} &
\includegraphics[height=0.30\textheight]{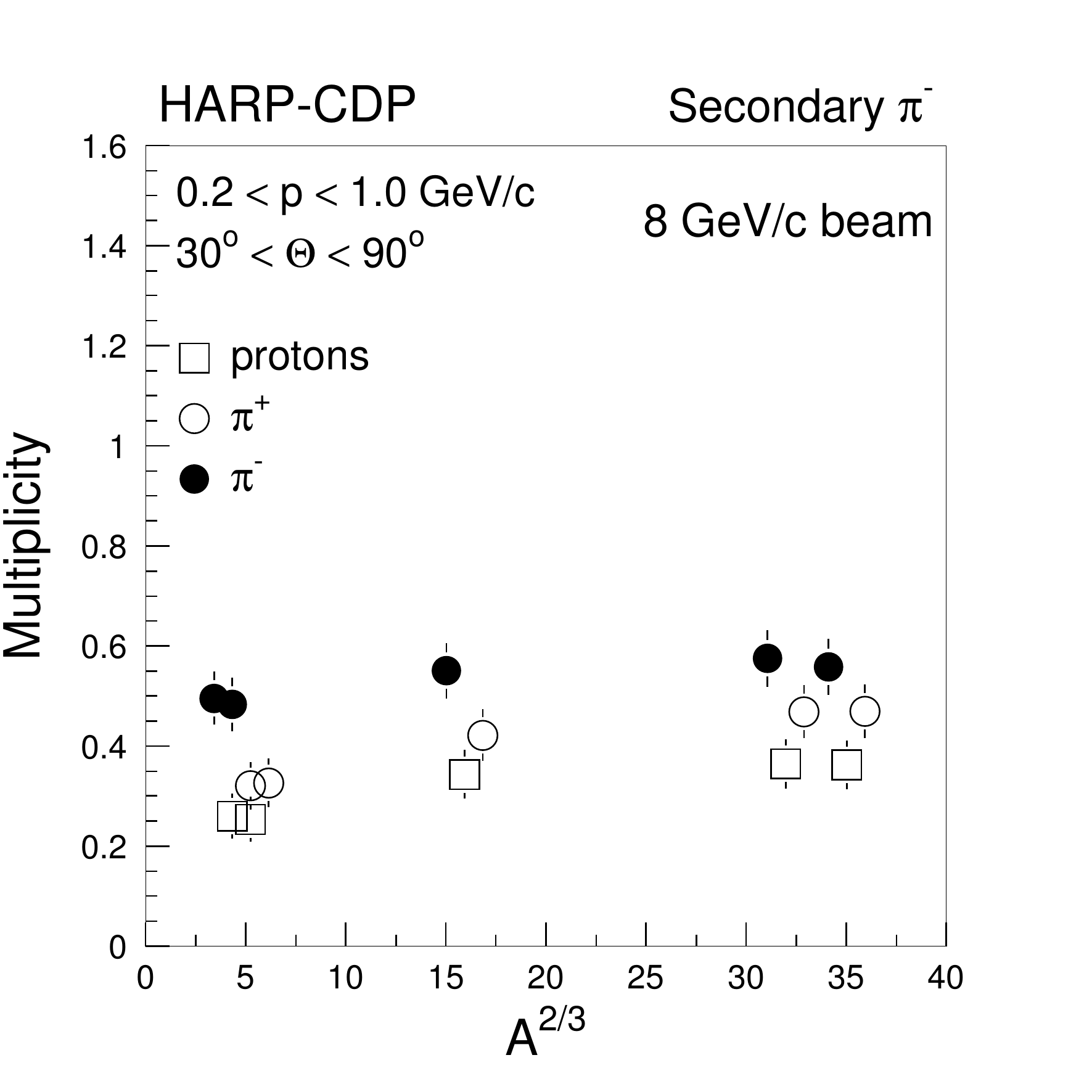} \\
\includegraphics[height=0.30\textheight]{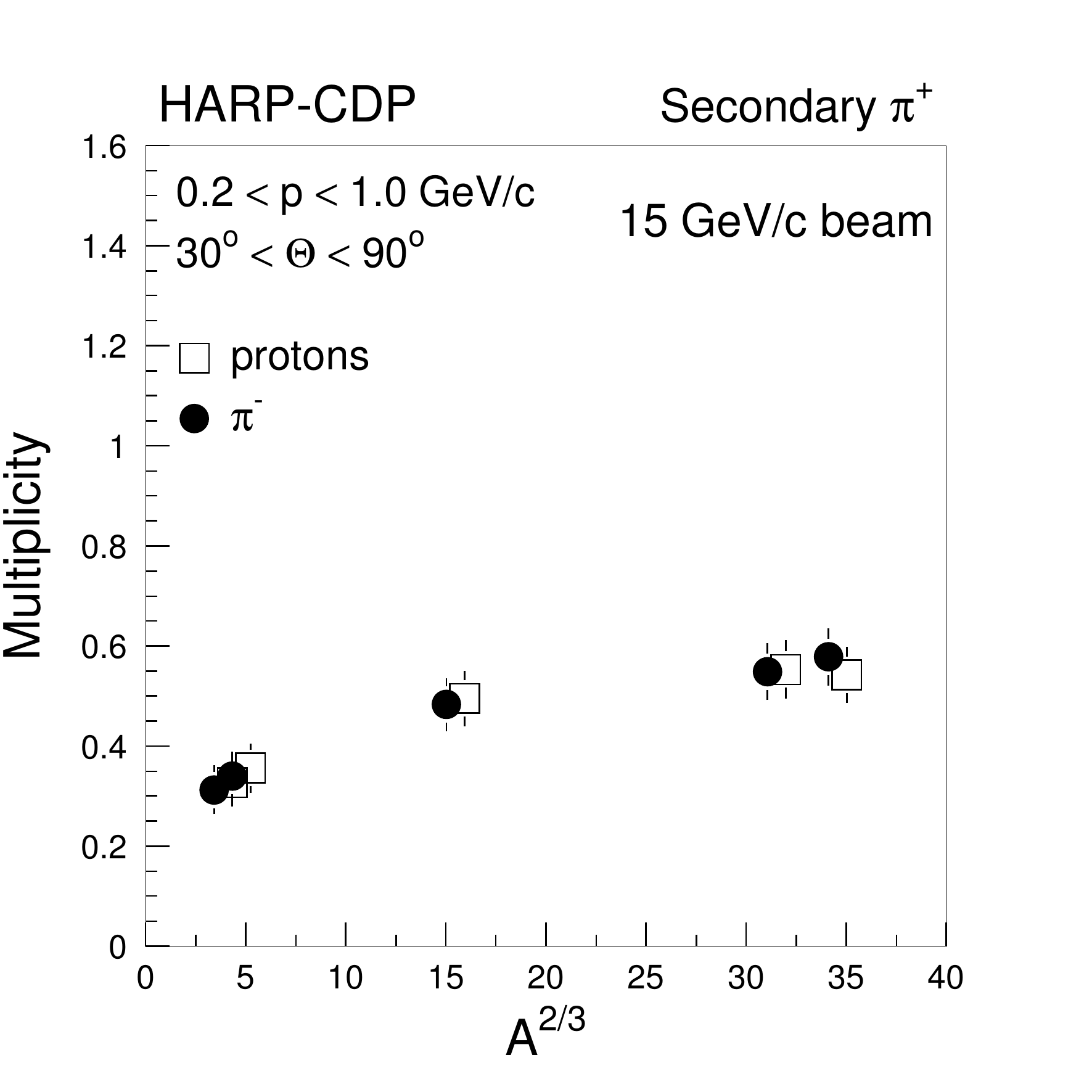} &  
\includegraphics[height=0.30\textheight]{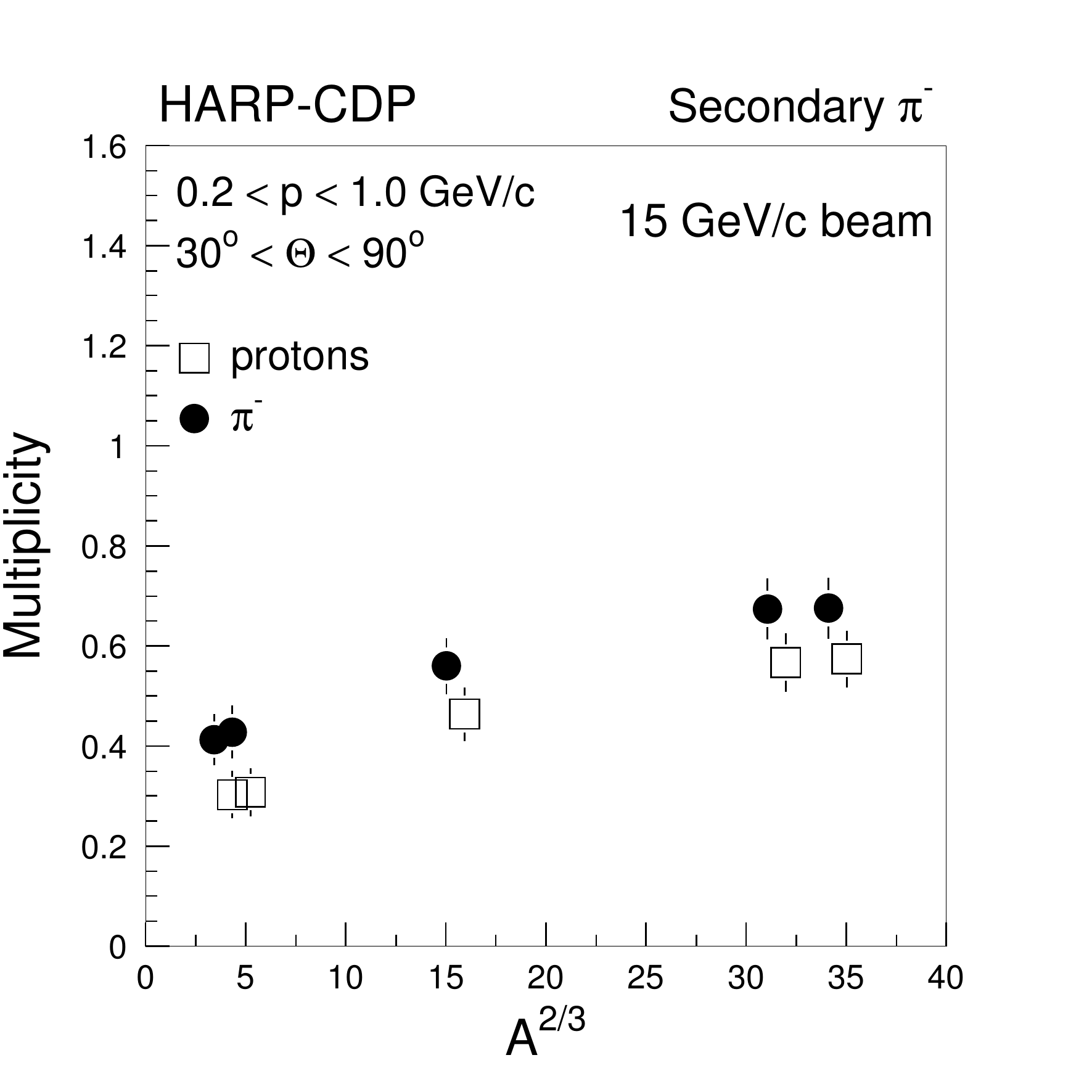} \\
\end{tabular}
\caption{Forward multiplicity of $\pi^+$'s and $\pi^-$'s produced by protons (open squares), $\pi^+$'s (open circles), and $\pi^-$'s (black circles), as a function of $A^{2/3}$ for, from left to right, beryllium, carbon, copper, tantalum, and lead nuclei; the forward multiplicity refers to the momentum range $0.2 < p < 1.0$~GeV/{\it c} and the polar-angle range $30^\circ < \theta < 90^\circ$ of secondary pions.} 
\label{ComparisonmultBeCCuTaPb}
\end{center}
\end{figure*}

Figure~\ref{ComparisonmultBeCCuTaPb} compares the `forward multiplicity' of secondary $\pi^+$'s and $\pi^-$'s in the interaction of protons and pions with beryllium, carbon, copper, tantalum, and lead target nuclei. The forward multiplicities are averaged over the momentum range 
$0.2 < p < 1.0$~GeV/{\it c} and the polar-angle range $30^\circ < \theta < 90^\circ$. They have been obtained by dividing the measured inclusive cross-section by the total cross-section inferred from the nuclear interaction lengths and pion interaction lengths, respectively, as published by the Particle Data Group~\cite{WebsitePDG2009} and reproduced in Table~\ref{interactionlengths}. The errors of the forward multiplicities are dominated by a 3\% systematic uncertainty.
\begin{table}[h]
\caption{Nuclear and pion interactions lengths used for the calculation of pion forward multiplicities.}
\label{interactionlengths}
\begin{center}
\begin{tabular}{|l||c|c|}
\hline
Nucleus   & $\lambda^{\rm nucl}_{\rm int}$ [g cm$^{-2}$] 
                 & $\lambda^{\rm pion}_{\rm int}$ [g cm$^{-2}$]  \\
\hline
\hline
Beryllium    	& 77.8 	&   109.9  \\
Carbon 		& 85.8	& 117.8   \\
Copper   		& 137.3 	&  165.9  \\
Tantalum    	& 191.0	&  217.7  \\
Lead   		&  199.6 	&  226.2  \\
\hline
\end{tabular}
\end{center}
\end{table}

The forward multiplicities display a 'leading particle effect' that mirrors the incoming beam particle. It is also interesting that the forward multiplicity decreases with the nuclear mass at low beam momentum but increases at high beam momentum. We interpret this as the effect of the nuclear medium on secondary pions from the primary interaction of the incoming beam particle. At low beam momentum, the secondary pions have low momentum and tend to fall below the 0.2~GeV/{\it c} threshold imposed in our analysis if there is more nuclear medium to be traversed before escape. At high beam momentum, the secondary pions have high enough momentum such that tertiary pions from the re-interaction of secondary pions in the nuclear medium tend to pass the 0.2~GeV/{\it c} threshold.

\begin{figure}[ht]
\begin{center}
\includegraphics[width=1.0\textwidth]{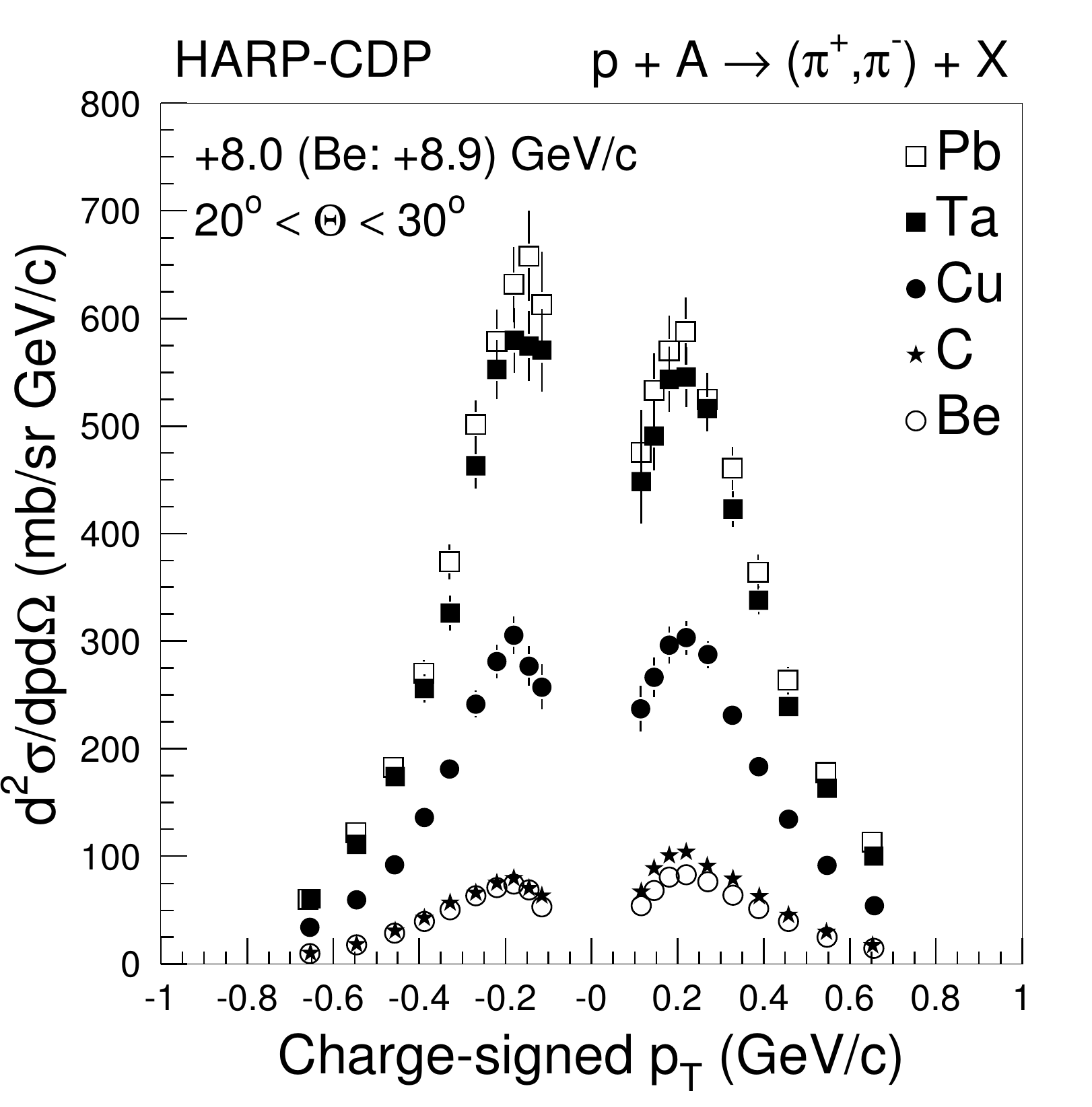} 
\caption{Comparison of inclusive pion production cross-sections in the forward region between beryllium, carbon, copper, tantalum, and lead target nuclei, as a function of the pion momentum.}
\label{pippim20to30proBeCCuTaPb}
\end{center}
\end{figure}

Figure~\ref{pippim20to30proBeCCuTaPb} shows the increase of the inclusive cross-sections of $\pi^+$'s and $\pi^-$'s production by incoming protons of  8.0~GeV/{\it c} (in the case of beryllium target nuclei: +8.9~GeV/{\it c}) from the light beryllium nucleus to the heavy lead nucleus, for pions in the polar angle range $20^\circ < \theta < 30^\circ$. It is interesting to note that $\pi^-$ production is slightly favoured on heavy nuclei, while $\pi^+$ production is slightly favoured on light nuclei.

Comparing the cross-sections for carbon and beryllium targets shown in Fig.~\ref{pippim20to30proBeCCuTaPb} we note that $\pi^-$ production on carbon nuclei is nearly the same as on beryllium nuclei while the production of $\pi^+$'s scales in approximate agreement with the $A^{2/3}$ law. A similar effect was observed many years ago in an experiment at JINR Dubna~\cite{Meshkovski1,Meshkovski2} with a 660 MeV/{\it c} proton beam\footnote{An even stronger effect was observed with a 600 MeV/{\it c} neutron beam~\cite{Oganesian}.}.
The author of Ref.\cite{Oganesian} interprets the effect in terms of a model in which the beryllium nucleus consists of two alpha particles and a nearly free neutron, while the carbon nucleus consists of three alpha particles. The inelastic cross-section on a free neutron is expected to be much higher than on the neutrons which are tightly bound within the alpha particles. This explains the anomalously large yield of negative pions on beryllium nuclei.

\clearpage

\section{Deuteron production}

Besides pions and protons, also deuterons are produced in sizeable quantities on carbon nuclei. Up to momenta of about 1~GeV/{\it c}, deuterons are easily separated from protons by \dedx . 

Table~\ref{deuteronsbypropippim} gives the deuteron-to-proton production ratio as a function of the momentum at the vertex, for 8~GeV/{\it c} beam protons, $\pi^+$'s, and $\pi^-$'s\footnote{We observe no appreciable dependence of the deuteron-to-proton production ratio on beam momentum.}. Cross-section ratios are not given if the data are scarce and the statistical error becomes comparable with the ratio itself---which is the case for deuterons at the high-momentum end of the spectrum.

  
 \begin{table}[h]
 \caption{Ratio of deuterons to protons produced by beam protons, $\pi^+$'s and $\pi^-$'s of 8~GeV/{\it c} momentum, as a function of the particle momentum {\it p} [GeV/{\it c}] at the vertex, for differents bins of polar angle $\theta$.}
 \label{deuteronsbypropippim}
 \begin{center}
 \begin{tabular}{|c|c||rcr|rcr|rcr|}
 \hline
 Polar angle & &  \multicolumn{3}{c|}{Beam p }  & \multicolumn{3}{c|}{Beam $\pi^+$} & \multicolumn{3}{c|}{Beam $\pi^-$}  \\
 \cline{2-11}
  $\theta$ &  {\it p} &  \multicolumn{3}{c|}{d/p} &  \multicolumn{3}{c|}{d/p} &  \multicolumn{3}{c|}{d/p} \\
 \hline
$20^\circ - 30^\circ$  & 0.663 & 0.146 & $\!\!\!\pm\!\!\!$ & 0.072 & 0.116 & $\!\!\!\pm\!\!\!$ & 0.035 & 0.107 & $\!\!\!\pm\!\!\!$ & 0.027 \\
                       & 0.715 & 0.127 & $\!\!\!\pm\!\!\!$ & 0.034 & 0.095 & $\!\!\!\pm\!\!\!$ & 0.031 & 0.136 & $\!\!\!\pm\!\!\!$ & 0.035 \\
                       & 0.778 & 0.135 & $\!\!\!\pm\!\!\!$ & 0.029 & 0.064 & $\!\!\!\pm\!\!\!$ & 0.021 & 0.147 & $\!\!\!\pm\!\!\!$ & 0.035 \\
                       & 0.848 & 0.095 & $\!\!\!\pm\!\!\!$ & 0.018 & 0.118 & $\!\!\!\pm\!\!\!$ & 0.035 & 0.098 & $\!\!\!\pm\!\!\!$ & 0.019 \\
                       & 0.926 & 0.156 & $\!\!\!\pm\!\!\!$ & 0.048 & 0.132 & $\!\!\!\pm\!\!\!$ & 0.049 & 0.161 & $\!\!\!\pm\!\!\!$ & 0.061 \\
                       & 1.011 & 0.088 & $\!\!\!\pm\!\!\!$ & 0.017 & 0.082 & $\!\!\!\pm\!\!\!$ & 0.028 & 0.088 & $\!\!\!\pm\!\!\!$ & 0.027 \\
\hline
 \end{tabular}
 \end{center}
 \end{table}

\begin{figure*}[ht]
\begin{center}
\includegraphics[height=0.55\textheight]{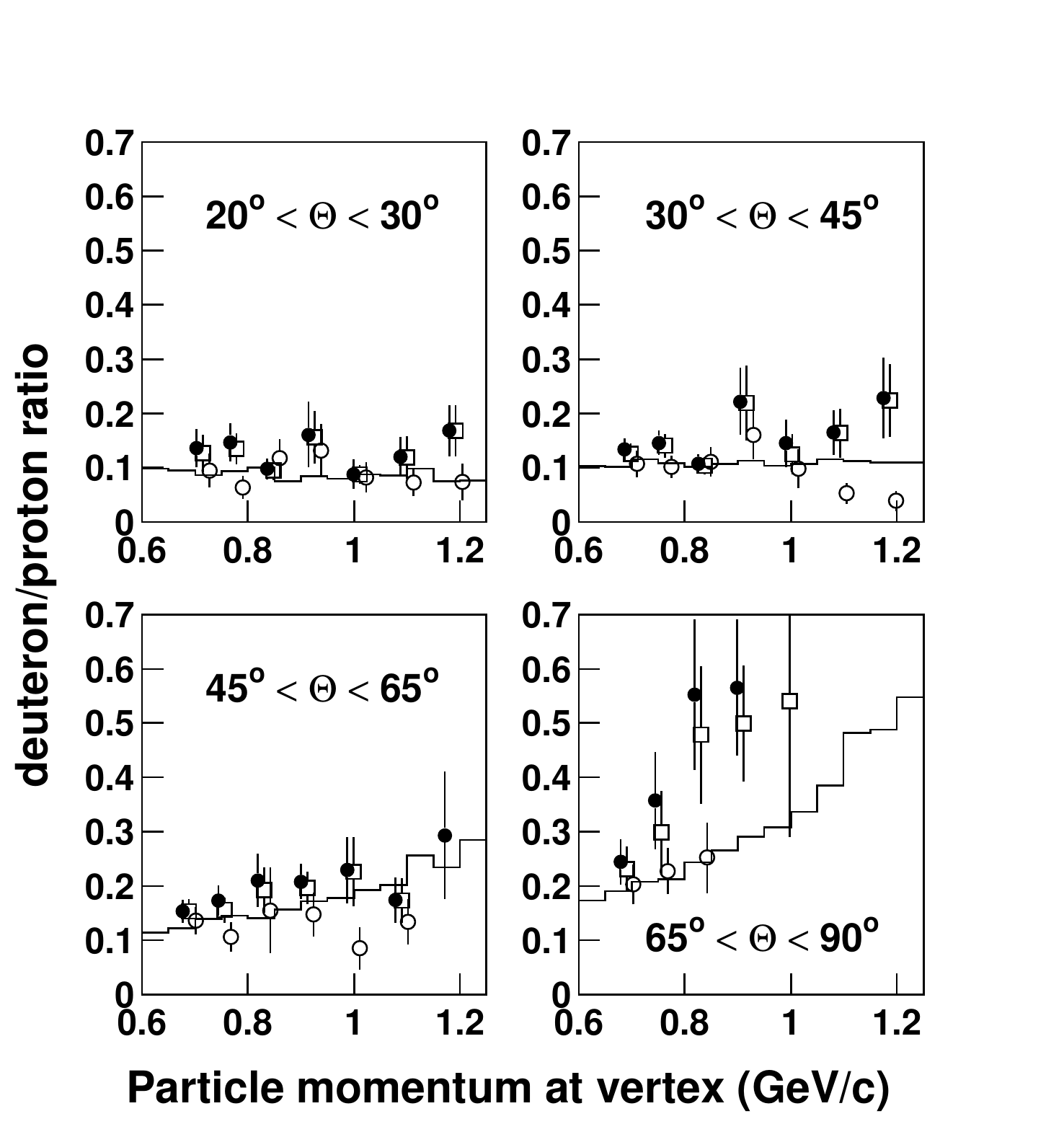} 
\caption{Deuteron-to-proton production ratios for 8~GeV/{\it c} beam particles on carbon nuclei, as a function of the momentum at the vertex, for four polar-angle regions; open squares denote beam protons, open circles beam $\pi^+$'s, and full circles beam $\pi^-$'s; the full lines denotes predictions of Geant4's FRITIOF model for $\pi^+$ beam particles.} 
\label{dtopratio}
\end{center}
\end{figure*}

The measured deuteron-to-proton production ratios are illustrated in Fig.~\ref{dtopratio}, and compared with the predictions of Geant4's  FRITIOF model. FRITIOF's predictions are shown for 
$\pi^+$ beam particles\footnote{There is virtually no difference between its predictions for incoming protons, $\pi^+$'s and $\pi^-$'s.}. FRITIOF reproduces deuteron production reasonably
well, except perhaps at large polar angles.

%

%How does the deuteron-to-proton ratio vary with the mass of the target nucleus? Figure~\ref{dtopratiofornuclei} shows for the polar-angle region $30^\circ < \theta < 45^\circ$ these ratios for 8~GeV/{\it c} beam protons on beryllium, carbon, copper, tantalum and lead nuclei.

\begin{figure*}[h]
\begin{center}
\includegraphics[height=0.5\textheight]{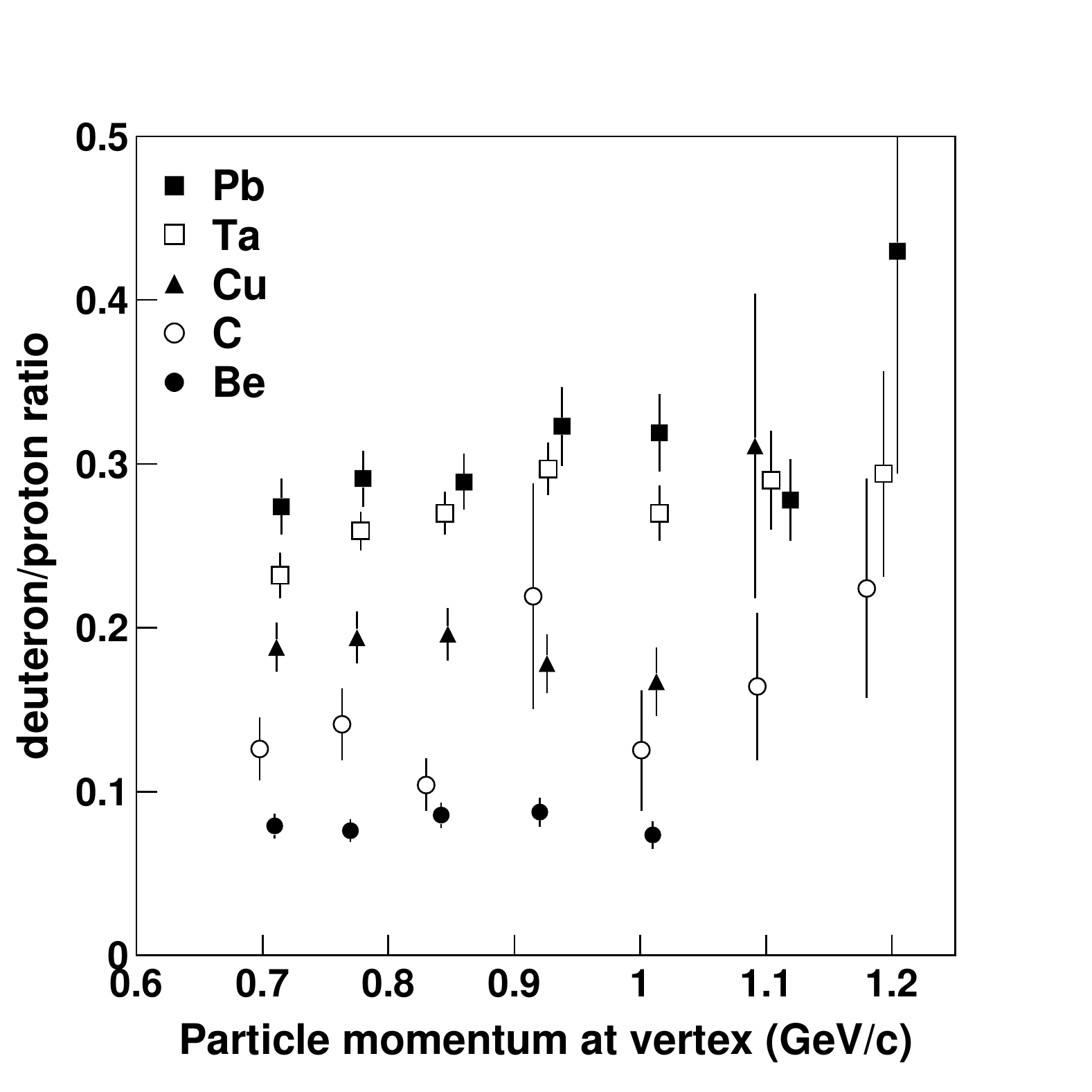} 
\caption{Deuteron-to-proton production ratios for 8~GeV/{\it c} beam protons on beryllium,
carbon, copper, tantalum and lead nuclei, as a function of the momentum at the vertex, for
the polar-angle region $30^\circ < \theta < 45^\circ$.}
\label{dtopratiofornuclei}
\end{center}
\end{figure*}

In Fig.~\ref{dtopratiofornuclei} we show, for the polar-angle region $30^\circ < \theta < 45^\circ$, how the deuteron-to-proton ratio varies with the mass of the target nucleus. The ratios are for 8~GeV/{\it c} beam protons on beryllium, carbon, copper, tantalum and lead nuclei.

\clearpage

\section{Summary}

From the analysis of data from the HARP large-angle spectrometer (polar angle $\theta$ in the range $20^\circ < \theta < 125^\circ$), double-differential cross-sections ${\rm d}^2 \sigma / {\rm d}p {\rm d}\Omega$ of the production of secondary protons, $\pi^+$'s, and $\pi^-$'s, and of deuterons, have been obtained. The incoming beam particles were protons and pions with momenta from $\pm 3$ to $\pm 15$~GeV/{\it c}, impinging on a 5\% $\lambda_{\rm int}$ thick stationary carbon target.

In the same way as for the other target nuclei which we have analyzed, our cross-sections for $\pi^+$ and $\pi^-$ production disagree with results of the HARP Collaboration that were obtained from the same raw data.

We have compared the inclusive carbon  $\pi^+$ and $\pi^-$ production cross-sections with those on beryllium, copper, tantalum, and lead and find an approximately linear dependence on the scaling variable $A^{2/3}$.

We also observe a sizeable production of deuterons off carbon nuclei that we compared to the deuteron production on beryllium, copper, tantalum, and lead.

\section*{Acknowledgements}

We are greatly indebted to many technical collaborators whose 
diligent and hard work made the HARP detector a well-functioning 
instrument. We thank all HARP colleagues who devoted time and 
effort to the design and construction of the detector, to data taking, 
and to setting up the computing and software infrastructure. 
We express our sincere gratitude to HARP's funding agencies 
for their support.  

%\clearpage

\clearpage

\appendix

\section{Cross-section Tables}

%%%%%%%%%%%%%%%%%%%%%%%%%%%%%%%%%%%%%%
% Here start the 3 GeV/c tables
%%%%%%%%%%%%%%%%%%%%%%%%%%%%%%%%%%%%%%

  
 \begin{table}[h]
 \begin{scriptsize}
 \caption{Double-differential inclusive cross-section ${\rm d}^2 \sigma /{\rm d}p{\rm d}\Omega$
  [mb/(GeV/{\it c} sr)] of the production of protons in p + C $\rightarrow$ p + X interactions with $+3.0$~GeV/{\it c} beam momentum; the first error is statistical, the second systematic; $p_{\rm T}$ in GeV/{\it c}, polar angle $\theta$ in degrees.}
 \label{pro.proc3}
 \begin{center}
 \begin{tabular}{|c||c|c|rcrcr||c|c|rcrcr|}
 \hline
   & \multicolumn{7}{c||}{$20<\theta<30$} & \multicolumn{7}{c|}{$30<\theta<40$} \\
 \hline
 $p_{\rm T}$ & $\langle p_{\rm T} \rangle$ & $\langle \theta \rangle$ & \multicolumn{5}{c||}{${\rm d}^2 \sigma /{\rm d}p{\rm d}\Omega$} &$\langle p_{\rm T} \rangle$ & $\langle \theta \rangle$ & \multicolumn{5}{c|}{${\rm d}^2 \sigma /{\rm d}p{\rm d}\Omega$} \\
 \hline
0.20--0.24 & 0.222 &  24.7 &   92.39 & $\!\!\pm\!\!$ &   5.08 & $\!\!\pm\!\!$ &   5.12 &  &  &  \multicolumn{5}{c|}{ } \\
0.24--0.30 & 0.271 &  24.9 &   83.64 & $\!\!\pm\!\!$ &   3.94 & $\!\!\pm\!\!$ &   4.31 & 0.272 &  34.7 &   71.69 & $\!\!\pm\!\!$ &   3.49 & $\!\!\pm\!\!$ &   3.42 \\
0.30--0.36 & 0.332 &  24.8 &   74.65 & $\!\!\pm\!\!$ &   3.76 & $\!\!\pm\!\!$ &   3.61 & 0.329 &  35.1 &   78.59 & $\!\!\pm\!\!$ &   3.76 & $\!\!\pm\!\!$ &   3.43 \\
0.36--0.42 & 0.391 &  25.1 &   72.47 & $\!\!\pm\!\!$ &   3.76 & $\!\!\pm\!\!$ &   3.32 & 0.391 &  35.0 &   63.24 & $\!\!\pm\!\!$ &   3.47 & $\!\!\pm\!\!$ &   2.90 \\
0.42--0.50 & 0.461 &  24.9 &   71.82 & $\!\!\pm\!\!$ &   3.15 & $\!\!\pm\!\!$ &   2.97 & 0.459 &  35.2 &   53.49 & $\!\!\pm\!\!$ &   2.81 & $\!\!\pm\!\!$ &   2.40 \\
0.50--0.60 & 0.552 &  24.9 &   54.18 & $\!\!\pm\!\!$ &   2.48 & $\!\!\pm\!\!$ &   2.23 & 0.551 &  34.8 &   45.29 & $\!\!\pm\!\!$ &   2.32 & $\!\!\pm\!\!$ &   2.11 \\
\hline
 \end{tabular}
 \end{center}
 \end{scriptsize}
 \end{table}
  
 \begin{table}[h]
 \begin{scriptsize}
 \caption{Double-differential inclusive cross-section ${\rm d}^2 \sigma /{\rm d}p{\rm d}\Omega$
  [mb/(GeV/{\it c} sr)] of the production of $\pi^+$'s in p + C $\rightarrow$ $\pi^+$ + X interactions with $+3.0$~GeV/{\it c} beam momentum; the first error is statistical, the second systematic; $p_{\rm T}$ in GeV/{\it c}, polar angle $\theta$ in degrees.}
 \label{pip.proc3}
 \begin{center}
 \begin{tabular}{|c||c|c|rcrcr||c|c|rcrcr|}
 \hline
   & \multicolumn{7}{c||}{$20<\theta<30$} & \multicolumn{7}{c|}{$30<\theta<40$} \\
 \hline
 $p_{\rm T}$ & $\langle p_{\rm T} \rangle$ & $\langle \theta \rangle$ & \multicolumn{5}{c||}{${\rm d}^2 \sigma /{\rm d}p{\rm d}\Omega$} &$\langle p_{\rm T} \rangle$ & $\langle \theta \rangle$ & \multicolumn{5}{c|}{${\rm d}^2 \sigma /{\rm d}p{\rm d}\Omega$} \\
 \hline
0.10--0.13 & 0.117 &  24.7 &   33.25 & $\!\!\pm\!\!$ &   4.06 & $\!\!\pm\!\!$ &   2.56 & 0.114 &  36.0 &   30.10 & $\!\!\pm\!\!$ &   3.94 & $\!\!\pm\!\!$ &   2.33 \\
0.13--0.16 & 0.146 &  24.9 &   47.78 & $\!\!\pm\!\!$ &   4.53 & $\!\!\pm\!\!$ &   3.00 & 0.145 &  35.2 &   45.34 & $\!\!\pm\!\!$ &   4.29 & $\!\!\pm\!\!$ &   2.91 \\
0.16--0.20 & 0.182 &  24.1 &   50.68 & $\!\!\pm\!\!$ &   3.86 & $\!\!\pm\!\!$ &   2.70 & 0.181 &  34.8 &   40.10 & $\!\!\pm\!\!$ &   3.34 & $\!\!\pm\!\!$ &   2.19 \\
0.20--0.24 & 0.221 &  25.2 &   56.63 & $\!\!\pm\!\!$ &   4.09 & $\!\!\pm\!\!$ &   2.71 & 0.220 &  34.8 &   44.13 & $\!\!\pm\!\!$ &   3.44 & $\!\!\pm\!\!$ &   2.15 \\
0.24--0.30 & 0.269 &  24.7 &   46.97 & $\!\!\pm\!\!$ &   2.93 & $\!\!\pm\!\!$ &   2.01 & 0.269 &  34.9 &   37.37 & $\!\!\pm\!\!$ &   2.61 & $\!\!\pm\!\!$ &   1.61 \\
0.30--0.36 & 0.330 &  25.0 &   28.80 & $\!\!\pm\!\!$ &   2.33 & $\!\!\pm\!\!$ &   1.20 & 0.329 &  35.1 &   26.17 & $\!\!\pm\!\!$ &   2.18 & $\!\!\pm\!\!$ &   1.09 \\
\hline
 \end{tabular}
 \end{center}
 \end{scriptsize}
 \end{table}
  
 \begin{table}[h]
 \begin{scriptsize}
 \caption{Double-differential inclusive cross-section ${\rm d}^2 \sigma /{\rm d}p{\rm d}\Omega$
  [mb/(GeV/{\it c} sr)] of the production of $\pi^-$'s in p + C $\rightarrow$ $\pi^-$ + X interactions with $+3.0$~GeV/{\it c} beam momentum; the first error is statistical, the second systematic; $p_{\rm T}$ in GeV/{\it c}, polar angle $\theta$ in degrees.}
 \label{pim.proc3}
 \begin{center}
 \begin{tabular}{|c||c|c|rcrcr||c|c|rcrcr|}
 \hline
   & \multicolumn{7}{c||}{$20<\theta<30$} & \multicolumn{7}{c|}{$30<\theta<40$} \\
 \hline
 $p_{\rm T}$ & $\langle p_{\rm T} \rangle$ & $\langle \theta \rangle$ & \multicolumn{5}{c||}{${\rm d}^2 \sigma /{\rm d}p{\rm d}\Omega$} &$\langle p_{\rm T} \rangle$ & $\langle \theta \rangle$ & \multicolumn{5}{c|}{${\rm d}^2 \sigma /{\rm d}p{\rm d}\Omega$} \\
 \hline
0.10--0.13 & 0.115 &  24.5 &   26.18 & $\!\!\pm\!\!$ &   3.58 & $\!\!\pm\!\!$ &   2.06 & 0.116 &  35.3 &   25.40 & $\!\!\pm\!\!$ &   3.47 & $\!\!\pm\!\!$ &   2.08 \\
0.13--0.16 & 0.144 &  25.0 &   28.28 & $\!\!\pm\!\!$ &   3.49 & $\!\!\pm\!\!$ &   1.97 & 0.144 &  35.1 &   20.25 & $\!\!\pm\!\!$ &   2.79 & $\!\!\pm\!\!$ &   1.43 \\
0.16--0.20 & 0.181 &  25.5 &   29.85 & $\!\!\pm\!\!$ &   2.95 & $\!\!\pm\!\!$ &   1.72 & 0.180 &  34.7 &   25.31 & $\!\!\pm\!\!$ &   2.65 & $\!\!\pm\!\!$ &   1.51 \\
0.20--0.24 & 0.221 &  24.8 &   23.19 & $\!\!\pm\!\!$ &   2.53 & $\!\!\pm\!\!$ &   1.22 & 0.217 &  35.0 &   21.56 & $\!\!\pm\!\!$ &   2.39 & $\!\!\pm\!\!$ &   1.17 \\
0.24--0.30 & 0.268 &  25.2 &   14.68 & $\!\!\pm\!\!$ &   1.65 & $\!\!\pm\!\!$ &   0.70 & 0.267 &  35.0 &   18.47 & $\!\!\pm\!\!$ &   1.81 & $\!\!\pm\!\!$ &   0.88 \\
0.30--0.36 & 0.319 &  24.5 &    9.80 & $\!\!\pm\!\!$ &   1.33 & $\!\!\pm\!\!$ &   0.49 & 0.329 &  35.2 &   12.76 & $\!\!\pm\!\!$ &   1.51 & $\!\!\pm\!\!$ &   0.62 \\
\hline
 \end{tabular}
 \end{center}
 \end{scriptsize}
 \end{table}
  
 \begin{table}[h]
 \begin{scriptsize}
 \caption{Double-differential inclusive cross-section ${\rm d}^2 \sigma /{\rm d}p{\rm d}\Omega$
  [mb/(GeV/{\it c} sr)] of the production of protons in $\pi^+$ + C $\rightarrow$ p + X interactions with $+3.0$~GeV/{\it c} beam momentum; the first error is statistical, the second systematic; $p_{\rm T}$ in GeV/{\it c}, polar angle $\theta$ in degrees.}
 \label{pro.pipc3}
 \begin{center}
 \begin{tabular}{|c||c|c|rcrcr||c|c|rcrcr|}
 \hline
   & \multicolumn{7}{c||}{$20<\theta<30$} & \multicolumn{7}{c|}{$30<\theta<40$} \\
 \hline
 $p_{\rm T}$ & $\langle p_{\rm T} \rangle$ & $\langle \theta \rangle$ & \multicolumn{5}{c||}{${\rm d}^2 \sigma /{\rm d}p{\rm d}\Omega$} &$\langle p_{\rm T} \rangle$ & $\langle \theta \rangle$ & \multicolumn{5}{c|}{${\rm d}^2 \sigma /{\rm d}p{\rm d}\Omega$} \\
 \hline
0.20--0.24 & 0.221 &  24.6 &   76.55 & $\!\!\pm\!\!$ &   3.37 & $\!\!\pm\!\!$ &   4.27 &  &  &  \multicolumn{5}{c|}{ } \\
0.24--0.30 & 0.271 &  25.1 &   67.20 & $\!\!\pm\!\!$ &   2.59 & $\!\!\pm\!\!$ &   3.50 & 0.271 &  34.8 &   65.27 & $\!\!\pm\!\!$ &   2.44 & $\!\!\pm\!\!$ &   3.14 \\
0.30--0.36 & 0.331 &  25.2 &   61.82 & $\!\!\pm\!\!$ &   2.52 & $\!\!\pm\!\!$ &   3.05 & 0.330 &  34.9 &   63.49 & $\!\!\pm\!\!$ &   2.47 & $\!\!\pm\!\!$ &   2.80 \\
0.36--0.42 & 0.391 &  24.9 &   57.73 & $\!\!\pm\!\!$ &   2.42 & $\!\!\pm\!\!$ &   2.73 & 0.390 &  35.2 &   59.08 & $\!\!\pm\!\!$ &   2.45 & $\!\!\pm\!\!$ &   2.76 \\
0.42--0.50 & 0.460 &  24.9 &   52.03 & $\!\!\pm\!\!$ &   1.93 & $\!\!\pm\!\!$ &   2.22 & 0.461 &  34.9 &   48.66 & $\!\!\pm\!\!$ &   1.94 & $\!\!\pm\!\!$ &   2.22 \\
0.50--0.60 & 0.550 &  25.0 &   40.92 & $\!\!\pm\!\!$ &   1.54 & $\!\!\pm\!\!$ &   1.75 & 0.551 &  34.9 &   36.67 & $\!\!\pm\!\!$ &   1.50 & $\!\!\pm\!\!$ &   1.73 \\
\hline
 \end{tabular}
 \end{center}
 \end{scriptsize}
 \end{table}
  
 \begin{table}[h]
 \begin{scriptsize}
 \caption{Double-differential inclusive cross-section ${\rm d}^2 \sigma /{\rm d}p{\rm d}\Omega$
  [mb/(GeV/{\it c} sr)] of the production of $\pi^+$'s in $\pi^+$ + C $\rightarrow$ $\pi^+$ + X interactions with $+3.0$~GeV/{\it c} beam momentum; the first error is statistical, the second systematic; $p_{\rm T}$ in GeV/{\it c}, polar angle $\theta$ in degrees.}
 \label{pip.pipc3}
 \begin{center}
 \begin{tabular}{|c||c|c|rcrcr||c|c|rcrcr|}
 \hline
   & \multicolumn{7}{c||}{$20<\theta<30$} & \multicolumn{7}{c|}{$30<\theta<40$} \\
 \hline
 $p_{\rm T}$ & $\langle p_{\rm T} \rangle$ & $\langle \theta \rangle$ & \multicolumn{5}{c||}{${\rm d}^2 \sigma /{\rm d}p{\rm d}\Omega$} &$\langle p_{\rm T} \rangle$ & $\langle \theta \rangle$ & \multicolumn{5}{c|}{${\rm d}^2 \sigma /{\rm d}p{\rm d}\Omega$} \\
 \hline
0.10--0.13 & 0.117 &  25.1 &   50.75 & $\!\!\pm\!\!$ &   3.71 & $\!\!\pm\!\!$ &   3.76 & 0.116 &  34.7 &   47.93 & $\!\!\pm\!\!$ &   3.48 & $\!\!\pm\!\!$ &   3.61 \\
0.13--0.16 & 0.146 &  24.8 &   57.70 & $\!\!\pm\!\!$ &   3.60 & $\!\!\pm\!\!$ &   3.46 & 0.146 &  35.2 &   49.70 & $\!\!\pm\!\!$ &   3.25 & $\!\!\pm\!\!$ &   3.04 \\
0.16--0.20 & 0.181 &  24.7 &   76.43 & $\!\!\pm\!\!$ &   3.48 & $\!\!\pm\!\!$ &   3.98 & 0.181 &  34.8 &   59.58 & $\!\!\pm\!\!$ &   2.95 & $\!\!\pm\!\!$ &   3.12 \\
0.20--0.24 & 0.220 &  24.8 &   80.25 & $\!\!\pm\!\!$ &   3.52 & $\!\!\pm\!\!$ &   3.77 & 0.220 &  34.9 &   67.84 & $\!\!\pm\!\!$ &   3.12 & $\!\!\pm\!\!$ &   3.16 \\
0.24--0.30 & 0.270 &  25.0 &   76.53 & $\!\!\pm\!\!$ &   2.74 & $\!\!\pm\!\!$ &   3.15 & 0.272 &  34.8 &   62.94 & $\!\!\pm\!\!$ &   2.46 & $\!\!\pm\!\!$ &   2.58 \\
0.30--0.36 & 0.330 &  24.9 &   64.95 & $\!\!\pm\!\!$ &   2.53 & $\!\!\pm\!\!$ &   2.48 & 0.331 &  35.0 &   51.07 & $\!\!\pm\!\!$ &   2.21 & $\!\!\pm\!\!$ &   1.94 \\
\hline
 \end{tabular}
 \end{center}
 \end{scriptsize}
 \end{table}
  
 \begin{table}[h]
 \begin{scriptsize}
 \caption{Double-differential inclusive cross-section ${\rm d}^2 \sigma /{\rm d}p{\rm d}\Omega$
  [mb/(GeV/{\it c} sr)] of the production of $\pi^-$'s in $\pi^+$ + C $\rightarrow$ $\pi^-$ + X interactions with $+3.0$~GeV/{\it c} beam momentum; the first error is statistical, the second systematic; $p_{\rm T}$ in GeV/{\it c}, polar angle $\theta$ in degrees.}
 \label{pim.pipc3}
 \begin{center}
 \begin{tabular}{|c||c|c|rcrcr||c|c|rcrcr|}
 \hline
   & \multicolumn{7}{c||}{$20<\theta<30$} & \multicolumn{7}{c|}{$30<\theta<40$} \\
 \hline
 $p_{\rm T}$ & $\langle p_{\rm T} \rangle$ & $\langle \theta \rangle$ & \multicolumn{5}{c||}{${\rm d}^2 \sigma /{\rm d}p{\rm d}\Omega$} &$\langle p_{\rm T} \rangle$ & $\langle \theta \rangle$ & \multicolumn{5}{c|}{${\rm d}^2 \sigma /{\rm d}p{\rm d}\Omega$} \\
 \hline
0.10--0.13 & 0.116 &  25.4 &   38.91 & $\!\!\pm\!\!$ &   3.13 & $\!\!\pm\!\!$ &   2.99 & 0.116 &  35.0 &   32.75 & $\!\!\pm\!\!$ &   2.84 & $\!\!\pm\!\!$ &   2.58 \\
0.13--0.16 & 0.145 &  24.6 &   46.04 & $\!\!\pm\!\!$ &   3.22 & $\!\!\pm\!\!$ &   2.98 & 0.145 &  35.2 &   39.25 & $\!\!\pm\!\!$ &   2.84 & $\!\!\pm\!\!$ &   2.55 \\
0.16--0.20 & 0.181 &  25.0 &   52.31 & $\!\!\pm\!\!$ &   2.79 & $\!\!\pm\!\!$ &   2.78 & 0.181 &  34.9 &   40.10 & $\!\!\pm\!\!$ &   2.43 & $\!\!\pm\!\!$ &   2.20 \\
0.20--0.24 & 0.219 &  24.9 &   50.16 & $\!\!\pm\!\!$ &   2.72 & $\!\!\pm\!\!$ &   2.36 & 0.218 &  34.8 &   41.07 & $\!\!\pm\!\!$ &   2.39 & $\!\!\pm\!\!$ &   1.99 \\
0.24--0.30 & 0.270 &  25.0 &   47.19 & $\!\!\pm\!\!$ &   2.15 & $\!\!\pm\!\!$ &   1.94 & 0.269 &  34.5 &   36.87 & $\!\!\pm\!\!$ &   1.83 & $\!\!\pm\!\!$ &   1.55 \\
0.30--0.36 & 0.328 &  24.9 &   35.47 & $\!\!\pm\!\!$ &   1.85 & $\!\!\pm\!\!$ &   1.37 & 0.328 &  34.7 &   30.27 & $\!\!\pm\!\!$ &   1.66 & $\!\!\pm\!\!$ &   1.20 \\
\hline
 \end{tabular}
 \end{center}
 \end{scriptsize}
 \end{table}
  
 \begin{table}[h]
 \begin{scriptsize}
 \caption{Double-differential inclusive cross-section ${\rm d}^2 \sigma /{\rm d}p{\rm d}\Omega$
  [mb/(GeV/{\it c} sr)] of the production of protons in $\pi^-$ + C $\rightarrow$ p + X interactions with $-3.0$~GeV/{\it c} beam momentum; the first error is statistical, the second systematic; $p_{\rm T}$ in GeV/{\it c}, polar angle $\theta$ in degrees.}
 \label{pro.pimc3}
 \begin{center}
 \begin{tabular}{|c||c|c|rcrcr||c|c|rcrcr|}
 \hline
   & \multicolumn{7}{c||}{$20<\theta<30$} & \multicolumn{7}{c|}{$30<\theta<40$} \\
 \hline
 $p_{\rm T}$ & $\langle p_{\rm T} \rangle$ & $\langle \theta \rangle$ & \multicolumn{5}{c||}{${\rm d}^2 \sigma /{\rm d}p{\rm d}\Omega$} &$\langle p_{\rm T} \rangle$ & $\langle \theta \rangle$ & \multicolumn{5}{c|}{${\rm d}^2 \sigma /{\rm d}p{\rm d}\Omega$} \\
 \hline
0.20--0.24 & 0.219 &  25.0 &   55.99 & $\!\!\pm\!\!$ &   1.70 & $\!\!\pm\!\!$ &   3.18 &  &  &  \multicolumn{5}{c|}{ } \\
0.24--0.30 & 0.269 &  25.1 &   51.89 & $\!\!\pm\!\!$ &   1.36 & $\!\!\pm\!\!$ &   2.82 & 0.270 &  35.0 &   51.35 & $\!\!\pm\!\!$ &   1.29 & $\!\!\pm\!\!$ &   2.49 \\
0.30--0.36 & 0.329 &  25.1 &   44.32 & $\!\!\pm\!\!$ &   1.26 & $\!\!\pm\!\!$ &   2.37 & 0.329 &  35.1 &   43.83 & $\!\!\pm\!\!$ &   1.20 & $\!\!\pm\!\!$ &   2.03 \\
0.36--0.42 & 0.388 &  25.1 &   40.31 & $\!\!\pm\!\!$ &   1.21 & $\!\!\pm\!\!$ &   2.14 & 0.388 &  35.0 &   40.08 & $\!\!\pm\!\!$ &   1.20 & $\!\!\pm\!\!$ &   2.03 \\
0.42--0.50 & 0.455 &  25.1 &   31.30 & $\!\!\pm\!\!$ &   0.89 & $\!\!\pm\!\!$ &   1.67 & 0.456 &  35.1 &   32.43 & $\!\!\pm\!\!$ &   0.95 & $\!\!\pm\!\!$ &   1.68 \\
0.50--0.60 & 0.544 &  25.2 &   24.80 & $\!\!\pm\!\!$ &   0.71 & $\!\!\pm\!\!$ &   1.38 & 0.543 &  34.9 &   25.78 & $\!\!\pm\!\!$ &   0.75 & $\!\!\pm\!\!$ &   1.43 \\
\hline
 \end{tabular}
 \end{center}
 \end{scriptsize}
 \end{table}
  
 \begin{table}[h]
 \begin{scriptsize}
 \caption{Double-differential inclusive cross-section ${\rm d}^2 \sigma /{\rm d}p{\rm d}\Omega$
  [mb/(GeV/{\it c} sr)] of the production of $\pi^+$'s in $\pi^-$ + C $\rightarrow$ $\pi^+$ + X interactions with $-3.0$~GeV/{\it c} beam momentum; the first error is statistical, the second systematic; $p_{\rm T}$ in GeV/{\it c}, polar angle $\theta$ in degrees.}
 \label{pip.pimc3}
 \begin{center}
 \begin{tabular}{|c||c|c|rcrcr||c|c|rcrcr|}
 \hline
   & \multicolumn{7}{c||}{$20<\theta<30$} & \multicolumn{7}{c|}{$30<\theta<40$} \\
 \hline
 $p_{\rm T}$ & $\langle p_{\rm T} \rangle$ & $\langle \theta \rangle$ & \multicolumn{5}{c||}{${\rm d}^2 \sigma /{\rm d}p{\rm d}\Omega$} &$\langle p_{\rm T} \rangle$ & $\langle \theta \rangle$ & \multicolumn{5}{c|}{${\rm d}^2 \sigma /{\rm d}p{\rm d}\Omega$} \\
 \hline
0.10--0.13 & 0.116 &  24.9 &   37.25 & $\!\!\pm\!\!$ &   1.77 & $\!\!\pm\!\!$ &   2.74 & 0.116 &  34.9 &   32.24 & $\!\!\pm\!\!$ &   1.62 & $\!\!\pm\!\!$ &   2.48 \\
0.13--0.16 & 0.146 &  24.9 &   47.26 & $\!\!\pm\!\!$ &   1.87 & $\!\!\pm\!\!$ &   2.93 & 0.145 &  35.0 &   35.74 & $\!\!\pm\!\!$ &   1.59 & $\!\!\pm\!\!$ &   2.24 \\
0.16--0.20 & 0.180 &  24.9 &   50.61 & $\!\!\pm\!\!$ &   1.59 & $\!\!\pm\!\!$ &   2.68 & 0.180 &  34.8 &   39.56 & $\!\!\pm\!\!$ &   1.41 & $\!\!\pm\!\!$ &   2.12 \\
0.20--0.24 & 0.220 &  24.9 &   54.96 & $\!\!\pm\!\!$ &   1.68 & $\!\!\pm\!\!$ &   2.64 & 0.219 &  35.0 &   43.19 & $\!\!\pm\!\!$ &   1.44 & $\!\!\pm\!\!$ &   2.07 \\
0.24--0.30 & 0.269 &  25.0 &   50.73 & $\!\!\pm\!\!$ &   1.30 & $\!\!\pm\!\!$ &   2.14 & 0.269 &  34.8 &   43.70 & $\!\!\pm\!\!$ &   1.17 & $\!\!\pm\!\!$ &   1.83 \\
0.30--0.36 & 0.328 &  24.9 &   42.10 & $\!\!\pm\!\!$ &   1.16 & $\!\!\pm\!\!$ &   1.69 & 0.328 &  34.7 &   37.11 & $\!\!\pm\!\!$ &   1.09 & $\!\!\pm\!\!$ &   1.46 \\
\hline
 \end{tabular}
 \end{center}
 \end{scriptsize}
 \end{table}
  
 \begin{table}[h]
 \begin{scriptsize}
 \caption{Double-differential inclusive cross-section ${\rm d}^2 \sigma /{\rm d}p{\rm d}\Omega$
  [mb/(GeV/{\it c} sr)] of the production of $\pi^-$'s in $\pi^-$ + C $\rightarrow$ $\pi^-$ + X interactions with $-3.0$~GeV/{\it c} beam momentum; the first error is statistical, the second systematic; $p_{\rm T}$ in GeV/{\it c}, polar angle $\theta$ in degrees.}
 \label{pim.pimc3}
 \begin{center}
 \begin{tabular}{|c||c|c|rcrcr||c|c|rcrcr|}
 \hline
   & \multicolumn{7}{c||}{$20<\theta<30$} & \multicolumn{7}{c|}{$30<\theta<40$} \\
 \hline
 $p_{\rm T}$ & $\langle p_{\rm T} \rangle$ & $\langle \theta \rangle$ & \multicolumn{5}{c||}{${\rm d}^2 \sigma /{\rm d}p{\rm d}\Omega$} &$\langle p_{\rm T} \rangle$ & $\langle \theta \rangle$ & \multicolumn{5}{c|}{${\rm d}^2 \sigma /{\rm d}p{\rm d}\Omega$} \\
 \hline
0.10--0.13 & 0.116 &  25.0 &   62.41 & $\!\!\pm\!\!$ &   2.39 & $\!\!\pm\!\!$ &   4.71 & 0.115 &  35.1 &   50.80 & $\!\!\pm\!\!$ &   2.13 & $\!\!\pm\!\!$ &   3.88 \\
0.13--0.16 & 0.146 &  24.8 &   71.53 & $\!\!\pm\!\!$ &   2.33 & $\!\!\pm\!\!$ &   4.36 & 0.146 &  35.0 &   60.90 & $\!\!\pm\!\!$ &   2.10 & $\!\!\pm\!\!$ &   3.77 \\
0.16--0.20 & 0.181 &  24.8 &   81.94 & $\!\!\pm\!\!$ &   2.09 & $\!\!\pm\!\!$ &   4.23 & 0.180 &  34.8 &   69.99 & $\!\!\pm\!\!$ &   1.91 & $\!\!\pm\!\!$ &   3.70 \\
0.20--0.24 & 0.221 &  24.9 &   85.15 & $\!\!\pm\!\!$ &   2.10 & $\!\!\pm\!\!$ &   3.90 & 0.220 &  34.9 &   69.40 & $\!\!\pm\!\!$ &   1.84 & $\!\!\pm\!\!$ &   3.23 \\
0.24--0.30 & 0.270 &  24.9 &   78.31 & $\!\!\pm\!\!$ &   1.63 & $\!\!\pm\!\!$ &   3.12 & 0.270 &  34.8 &   65.80 & $\!\!\pm\!\!$ &   1.46 & $\!\!\pm\!\!$ &   2.65 \\
0.30--0.36 & 0.330 &  25.0 &   70.07 & $\!\!\pm\!\!$ &   1.54 & $\!\!\pm\!\!$ &   2.60 & 0.331 &  34.8 &   56.43 & $\!\!\pm\!\!$ &   1.35 & $\!\!\pm\!\!$ &   2.10 \\
\hline
 \end{tabular}
 \end{center}
 \end{scriptsize}
 \end{table}
\clearpage

%%%%%%%%%%%%%%%%%%%%%%%%%%%%%%%%%%%%%%
% Here start the 5 GeV/c tables
%%%%%%%%%%%%%%%%%%%%%%%%%%%%%%%%%%%%%%

  
 \begin{table}[h]
 \begin{scriptsize}
 \caption{Double-differential inclusive cross-section ${\rm d}^2 \sigma /{\rm d}p{\rm d}\Omega$
  [mb/(GeV/{\it c} sr)] of the production of protons in p + C $\rightarrow$ p + X interactions with $+5.0$~GeV/{\it c} beam momentum; the first error is statistical, the second systematic; $p_{\rm T}$ in GeV/{\it c}, polar angle $\theta$ in degrees.}
 \label{pro.proc5}
 \begin{center}
 \begin{tabular}{|c||c|c|rcrcr||c|c|rcrcr|}
 \hline
   & \multicolumn{7}{c||}{$20<\theta<30$} & \multicolumn{7}{c|}{$30<\theta<40$} \\
 \hline
 $p_{\rm T}$ & $\langle p_{\rm T} \rangle$ & $\langle \theta \rangle$ & \multicolumn{5}{c||}{${\rm d}^2 \sigma /{\rm d}p{\rm d}\Omega$} &$\langle p_{\rm T} \rangle$ & $\langle \theta \rangle$ & \multicolumn{5}{c|}{${\rm d}^2 \sigma /{\rm d}p{\rm d}\Omega$} \\
 \hline
0.20--0.24 & 0.220 &  24.9 &   76.36 & $\!\!\pm\!\!$ &   2.02 & $\!\!\pm\!\!$ &   4.26 &  &  &  \multicolumn{5}{c|}{ } \\
0.24--0.30 & 0.269 &  24.9 &   73.93 & $\!\!\pm\!\!$ &   1.63 & $\!\!\pm\!\!$ &   3.89 & 0.271 &  35.0 &   72.39 & $\!\!\pm\!\!$ &   1.54 & $\!\!\pm\!\!$ &   3.51 \\
0.30--0.36 & 0.329 &  25.0 &   67.44 & $\!\!\pm\!\!$ &   1.56 & $\!\!\pm\!\!$ &   3.43 & 0.329 &  34.9 &   68.98 & $\!\!\pm\!\!$ &   1.52 & $\!\!\pm\!\!$ &   3.08 \\
0.36--0.42 & 0.390 &  25.1 &   60.92 & $\!\!\pm\!\!$ &   1.49 & $\!\!\pm\!\!$ &   3.08 & 0.390 &  35.0 &   57.19 & $\!\!\pm\!\!$ &   1.45 & $\!\!\pm\!\!$ &   2.87 \\
0.42--0.50 & 0.459 &  25.0 &   55.57 & $\!\!\pm\!\!$ &   1.21 & $\!\!\pm\!\!$ &   2.66 & 0.459 &  34.9 &   46.77 & $\!\!\pm\!\!$ &   1.14 & $\!\!\pm\!\!$ &   2.36 \\
0.50--0.60 & 0.548 &  24.9 &   47.52 & $\!\!\pm\!\!$ &   0.99 & $\!\!\pm\!\!$ &   2.46 & 0.547 &  34.9 &   41.11 & $\!\!\pm\!\!$ &   0.96 & $\!\!\pm\!\!$ &   2.17 \\
\hline
 \end{tabular}
 \end{center}
 \end{scriptsize}
 \end{table}
  
 \begin{table}[h]
 \begin{scriptsize}
 \caption{Double-differential inclusive cross-section ${\rm d}^2 \sigma /{\rm d}p{\rm d}\Omega$
  [mb/(GeV/{\it c} sr)] of the production of $\pi^+$'s in p + C $\rightarrow$ $\pi^+$ + X interactions with $+5.0$~GeV/{\it c} beam momentum; the first error is statistical, the second systematic; $p_{\rm T}$ in GeV/{\it c}, polar angle $\theta$ in degrees.}
 \label{pip.proc5}
 \begin{center}
 \begin{tabular}{|c||c|c|rcrcr||c|c|rcrcr|}
 \hline
   & \multicolumn{7}{c||}{$20<\theta<30$} & \multicolumn{7}{c|}{$30<\theta<40$} \\
 \hline
 $p_{\rm T}$ & $\langle p_{\rm T} \rangle$ & $\langle \theta \rangle$ & \multicolumn{5}{c||}{${\rm d}^2 \sigma /{\rm d}p{\rm d}\Omega$} &$\langle p_{\rm T} \rangle$ & $\langle \theta \rangle$ & \multicolumn{5}{c|}{${\rm d}^2 \sigma /{\rm d}p{\rm d}\Omega$} \\
 \hline
0.10--0.13 & 0.116 &  25.0 &   55.24 & $\!\!\pm\!\!$ &   2.29 & $\!\!\pm\!\!$ &   4.01 & 0.115 &  34.9 &   50.29 & $\!\!\pm\!\!$ &   2.22 & $\!\!\pm\!\!$ &   3.85 \\
0.13--0.16 & 0.145 &  24.8 &   68.05 & $\!\!\pm\!\!$ &   2.32 & $\!\!\pm\!\!$ &   4.09 & 0.146 &  34.8 &   53.06 & $\!\!\pm\!\!$ &   2.06 & $\!\!\pm\!\!$ &   3.22 \\
0.16--0.20 & 0.180 &  24.7 &   77.37 & $\!\!\pm\!\!$ &   2.09 & $\!\!\pm\!\!$ &   4.07 & 0.181 &  34.8 &   56.84 & $\!\!\pm\!\!$ &   1.79 & $\!\!\pm\!\!$ &   2.98 \\
0.20--0.24 & 0.220 &  24.9 &   77.31 & $\!\!\pm\!\!$ &   2.03 & $\!\!\pm\!\!$ &   3.66 & 0.220 &  34.7 &   56.43 & $\!\!\pm\!\!$ &   1.72 & $\!\!\pm\!\!$ &   2.66 \\
0.24--0.30 & 0.269 &  24.8 &   65.47 & $\!\!\pm\!\!$ &   1.51 & $\!\!\pm\!\!$ &   2.76 & 0.270 &  34.9 &   50.67 & $\!\!\pm\!\!$ &   1.33 & $\!\!\pm\!\!$ &   2.12 \\
0.30--0.36 & 0.329 &  25.0 &   56.14 & $\!\!\pm\!\!$ &   1.39 & $\!\!\pm\!\!$ &   2.26 & 0.329 &  34.8 &   40.18 & $\!\!\pm\!\!$ &   1.17 & $\!\!\pm\!\!$ &   1.60 \\
\hline
 \end{tabular}
 \end{center}
 \end{scriptsize}
 \end{table}
  
 \begin{table}[h]
 \begin{scriptsize}
 \caption{Double-differential inclusive cross-section ${\rm d}^2 \sigma /{\rm d}p{\rm d}\Omega$
  [mb/(GeV/{\it c} sr)] of the production of $\pi^-$'s in p + C $\rightarrow$ $\pi^-$ + X interactions with $+5.0$~GeV/{\it c} beam momentum; the first error is statistical, the second systematic; $p_{\rm T}$ in GeV/{\it c}, polar angle $\theta$ in degrees.}
 \label{pim.proc5}
 \begin{center}
 \begin{tabular}{|c||c|c|rcrcr||c|c|rcrcr|}
 \hline
   & \multicolumn{7}{c||}{$20<\theta<30$} & \multicolumn{7}{c|}{$30<\theta<40$} \\
 \hline
 $p_{\rm T}$ & $\langle p_{\rm T} \rangle$ & $\langle \theta \rangle$ & \multicolumn{5}{c||}{${\rm d}^2 \sigma /{\rm d}p{\rm d}\Omega$} &$\langle p_{\rm T} \rangle$ & $\langle \theta \rangle$ & \multicolumn{5}{c|}{${\rm d}^2 \sigma /{\rm d}p{\rm d}\Omega$} \\
 \hline
0.10--0.13 & 0.115 &  25.1 &   44.37 & $\!\!\pm\!\!$ &   1.98 & $\!\!\pm\!\!$ &   3.36 & 0.115 &  34.9 &   36.47 & $\!\!\pm\!\!$ &   1.80 & $\!\!\pm\!\!$ &   2.89 \\
0.13--0.16 & 0.145 &  24.7 &   47.23 & $\!\!\pm\!\!$ &   1.87 & $\!\!\pm\!\!$ &   2.98 & 0.145 &  34.9 &   45.64 & $\!\!\pm\!\!$ &   1.90 & $\!\!\pm\!\!$ &   3.01 \\
0.16--0.20 & 0.179 &  24.6 &   49.82 & $\!\!\pm\!\!$ &   1.59 & $\!\!\pm\!\!$ &   2.69 & 0.179 &  34.8 &   40.47 & $\!\!\pm\!\!$ &   1.45 & $\!\!\pm\!\!$ &   2.24 \\
0.20--0.24 & 0.220 &  24.7 &   48.96 & $\!\!\pm\!\!$ &   1.57 & $\!\!\pm\!\!$ &   2.37 & 0.220 &  34.8 &   39.47 & $\!\!\pm\!\!$ &   1.42 & $\!\!\pm\!\!$ &   1.94 \\
0.24--0.30 & 0.268 &  24.8 &   40.08 & $\!\!\pm\!\!$ &   1.16 & $\!\!\pm\!\!$ &   1.70 & 0.269 &  34.7 &   30.82 & $\!\!\pm\!\!$ &   0.99 & $\!\!\pm\!\!$ &   1.33 \\
0.30--0.36 & 0.329 &  25.1 &   30.67 & $\!\!\pm\!\!$ &   1.01 & $\!\!\pm\!\!$ &   1.25 & 0.329 &  34.7 &   26.68 & $\!\!\pm\!\!$ &   0.95 & $\!\!\pm\!\!$ &   1.10 \\
\hline
 \end{tabular}
 \end{center}
 \end{scriptsize}
 \end{table}
  
 \begin{table}[h]
 \begin{scriptsize}
 \caption{Double-differential inclusive cross-section ${\rm d}^2 \sigma /{\rm d}p{\rm d}\Omega$
  [mb/(GeV/{\it c} sr)] of the production of protons in $\pi^+$ + C $\rightarrow$ p + X interactions with $+5.0$~GeV/{\it c} beam momentum; the first error is statistical, the second systematic; $p_{\rm T}$ in GeV/{\it c}, polar angle $\theta$ in degrees.}
 \label{pro.pipc5}
 \begin{center}
 \begin{tabular}{|c||c|c|rcrcr||c|c|rcrcr|}
 \hline
   & \multicolumn{7}{c||}{$20<\theta<30$} & \multicolumn{7}{c|}{$30<\theta<40$} \\
 \hline
 $p_{\rm T}$ & $\langle p_{\rm T} \rangle$ & $\langle \theta \rangle$ & \multicolumn{5}{c||}{${\rm d}^2 \sigma /{\rm d}p{\rm d}\Omega$} &$\langle p_{\rm T} \rangle$ & $\langle \theta \rangle$ & \multicolumn{5}{c|}{${\rm d}^2 \sigma /{\rm d}p{\rm d}\Omega$} \\
 \hline
0.20--0.24 & 0.220 &  25.0 &   61.58 & $\!\!\pm\!\!$ &   1.61 & $\!\!\pm\!\!$ &   3.45 &  &  &  \multicolumn{5}{c|}{ } \\
0.24--0.30 & 0.270 &  25.0 &   58.04 & $\!\!\pm\!\!$ &   1.29 & $\!\!\pm\!\!$ &   3.09 & 0.270 &  35.0 &   56.52 & $\!\!\pm\!\!$ &   1.21 & $\!\!\pm\!\!$ &   2.76 \\
0.30--0.36 & 0.329 &  25.0 &   50.08 & $\!\!\pm\!\!$ &   1.19 & $\!\!\pm\!\!$ &   2.59 & 0.329 &  35.1 &   54.56 & $\!\!\pm\!\!$ &   1.21 & $\!\!\pm\!\!$ &   2.46 \\
0.36--0.42 & 0.389 &  25.1 &   47.24 & $\!\!\pm\!\!$ &   1.17 & $\!\!\pm\!\!$ &   2.42 & 0.389 &  35.0 &   44.23 & $\!\!\pm\!\!$ &   1.14 & $\!\!\pm\!\!$ &   2.27 \\
0.42--0.50 & 0.459 &  25.0 &   40.81 & $\!\!\pm\!\!$ &   0.90 & $\!\!\pm\!\!$ &   1.96 & 0.459 &  35.0 &   37.34 & $\!\!\pm\!\!$ &   0.90 & $\!\!\pm\!\!$ &   1.95 \\
0.50--0.60 & 0.547 &  24.9 &   33.56 & $\!\!\pm\!\!$ &   0.72 & $\!\!\pm\!\!$ &   1.75 & 0.548 &  35.0 &   30.26 & $\!\!\pm\!\!$ &   0.73 & $\!\!\pm\!\!$ &   1.68 \\
\hline
 \end{tabular}
 \end{center}
 \end{scriptsize}
 \end{table}
  
 \begin{table}[h]
 \begin{scriptsize}
 \caption{Double-differential inclusive cross-section ${\rm d}^2 \sigma /{\rm d}p{\rm d}\Omega$
  [mb/(GeV/{\it c} sr)] of the production of $\pi^+$'s in $\pi^+$ + C $\rightarrow$ $\pi^+$ + X interactions with $+5.0$~GeV/{\it c} beam momentum; the first error is statistical, the second systematic; $p_{\rm T}$ in GeV/{\it c}, polar angle $\theta$ in degrees.}
 \label{pip.pipc5}
 \begin{center}
 \begin{tabular}{|c||c|c|rcrcr||c|c|rcrcr|}
 \hline
   & \multicolumn{7}{c||}{$20<\theta<30$} & \multicolumn{7}{c|}{$30<\theta<40$} \\
 \hline
 $p_{\rm T}$ & $\langle p_{\rm T} \rangle$ & $\langle \theta \rangle$ & \multicolumn{5}{c||}{${\rm d}^2 \sigma /{\rm d}p{\rm d}\Omega$} &$\langle p_{\rm T} \rangle$ & $\langle \theta \rangle$ & \multicolumn{5}{c|}{${\rm d}^2 \sigma /{\rm d}p{\rm d}\Omega$} \\
 \hline
0.10--0.13 & 0.116 &  25.0 &   59.10 & $\!\!\pm\!\!$ &   2.09 & $\!\!\pm\!\!$ &   4.21 & 0.115 &  34.9 &   54.08 & $\!\!\pm\!\!$ &   2.08 & $\!\!\pm\!\!$ &   4.22 \\
0.13--0.16 & 0.145 &  24.8 &   71.32 & $\!\!\pm\!\!$ &   2.10 & $\!\!\pm\!\!$ &   4.21 & 0.146 &  34.8 &   61.43 & $\!\!\pm\!\!$ &   1.98 & $\!\!\pm\!\!$ &   3.69 \\
0.16--0.20 & 0.180 &  24.8 &   87.35 & $\!\!\pm\!\!$ &   1.97 & $\!\!\pm\!\!$ &   4.53 & 0.180 &  34.8 &   68.15 & $\!\!\pm\!\!$ &   1.74 & $\!\!\pm\!\!$ &   3.53 \\
0.20--0.24 & 0.220 &  24.8 &   98.57 & $\!\!\pm\!\!$ &   2.03 & $\!\!\pm\!\!$ &   4.56 & 0.220 &  34.7 &   71.36 & $\!\!\pm\!\!$ &   1.72 & $\!\!\pm\!\!$ &   3.30 \\
0.24--0.30 & 0.270 &  24.8 &   94.42 & $\!\!\pm\!\!$ &   1.62 & $\!\!\pm\!\!$ &   3.89 & 0.270 &  34.8 &   67.45 & $\!\!\pm\!\!$ &   1.36 & $\!\!\pm\!\!$ &   2.76 \\
0.30--0.36 & 0.329 &  24.7 &   82.75 & $\!\!\pm\!\!$ &   1.50 & $\!\!\pm\!\!$ &   3.22 & 0.329 &  34.8 &   61.40 & $\!\!\pm\!\!$ &   1.29 & $\!\!\pm\!\!$ &   2.33 \\
\hline
 \end{tabular}
 \end{center}
 \end{scriptsize}
 \end{table}
  
 \begin{table}[h]
 \begin{scriptsize}
 \caption{Double-differential inclusive cross-section ${\rm d}^2 \sigma /{\rm d}p{\rm d}\Omega$
  [mb/(GeV/{\it c} sr)] of the production of $\pi^-$'s in $\pi^+$ + C $\rightarrow$ $\pi^-$ + X interactions with $+5.0$~GeV/{\it c} beam momentum; the first error is statistical, the second systematic; $p_{\rm T}$ in GeV/{\it c}, polar angle $\theta$ in degrees.}
 \label{pim.pipc5}
 \begin{center}
 \begin{tabular}{|c||c|c|rcrcr||c|c|rcrcr|}
 \hline
   & \multicolumn{7}{c||}{$20<\theta<30$} & \multicolumn{7}{c|}{$30<\theta<40$} \\
 \hline
 $p_{\rm T}$ & $\langle p_{\rm T} \rangle$ & $\langle \theta \rangle$ & \multicolumn{5}{c||}{${\rm d}^2 \sigma /{\rm d}p{\rm d}\Omega$} &$\langle p_{\rm T} \rangle$ & $\langle \theta \rangle$ & \multicolumn{5}{c|}{${\rm d}^2 \sigma /{\rm d}p{\rm d}\Omega$} \\
 \hline
0.10--0.13 & 0.116 &  24.9 &   48.31 & $\!\!\pm\!\!$ &   1.83 & $\!\!\pm\!\!$ &   3.57 & 0.116 &  34.8 &   41.09 & $\!\!\pm\!\!$ &   1.70 & $\!\!\pm\!\!$ &   3.20 \\
0.13--0.16 & 0.146 &  24.8 &   57.80 & $\!\!\pm\!\!$ &   1.87 & $\!\!\pm\!\!$ &   3.54 & 0.145 &  34.9 &   46.08 & $\!\!\pm\!\!$ &   1.70 & $\!\!\pm\!\!$ &   2.96 \\
0.16--0.20 & 0.181 &  24.7 &   63.95 & $\!\!\pm\!\!$ &   1.61 & $\!\!\pm\!\!$ &   3.37 & 0.180 &  34.9 &   46.94 & $\!\!\pm\!\!$ &   1.40 & $\!\!\pm\!\!$ &   2.53 \\
0.20--0.24 & 0.220 &  24.8 &   63.38 & $\!\!\pm\!\!$ &   1.60 & $\!\!\pm\!\!$ &   2.96 & 0.220 &  34.8 &   49.49 & $\!\!\pm\!\!$ &   1.42 & $\!\!\pm\!\!$ &   2.36 \\
0.24--0.30 & 0.270 &  24.7 &   60.54 & $\!\!\pm\!\!$ &   1.27 & $\!\!\pm\!\!$ &   2.46 & 0.270 &  34.8 &   40.95 & $\!\!\pm\!\!$ &   1.02 & $\!\!\pm\!\!$ &   1.70 \\
0.30--0.36 & 0.329 &  24.7 &   49.28 & $\!\!\pm\!\!$ &   1.14 & $\!\!\pm\!\!$ &   1.87 & 0.329 &  34.8 &   40.10 & $\!\!\pm\!\!$ &   1.03 & $\!\!\pm\!\!$ &   1.55 \\
\hline
 \end{tabular}
 \end{center}
 \end{scriptsize}
 \end{table}
  
 \begin{table}[h]
 \begin{scriptsize}
 \caption{Double-differential inclusive cross-section ${\rm d}^2 \sigma /{\rm d}p{\rm d}\Omega$
  [mb/(GeV/{\it c} sr)] of the production of protons in $\pi^-$ + C $\rightarrow$ p + X interactions with $-5.0$~GeV/{\it c} beam momentum; the first error is statistical, the second systematic; $p_{\rm T}$ in GeV/{\it c}, polar angle $\theta$ in degrees.}
 \label{pro.pimc5}
 \begin{center}
 \begin{tabular}{|c||c|c|rcrcr||c|c|rcrcr|}
 \hline
   & \multicolumn{7}{c||}{$20<\theta<30$} & \multicolumn{7}{c|}{$30<\theta<40$} \\
 \hline
 $p_{\rm T}$ & $\langle p_{\rm T} \rangle$ & $\langle \theta \rangle$ & \multicolumn{5}{c||}{${\rm d}^2 \sigma /{\rm d}p{\rm d}\Omega$} &$\langle p_{\rm T} \rangle$ & $\langle \theta \rangle$ & \multicolumn{5}{c|}{${\rm d}^2 \sigma /{\rm d}p{\rm d}\Omega$} \\
 \hline
0.20--0.24 & 0.221 &  25.0 &   46.65 & $\!\!\pm\!\!$ &   1.47 & $\!\!\pm\!\!$ &   2.65 &  &  &  \multicolumn{5}{c|}{ } \\
0.24--0.30 & 0.270 &  25.0 &   47.23 & $\!\!\pm\!\!$ &   1.24 & $\!\!\pm\!\!$ &   2.59 & 0.271 &  35.0 &   47.99 & $\!\!\pm\!\!$ &   1.17 & $\!\!\pm\!\!$ &   2.33 \\
0.30--0.36 & 0.330 &  25.0 &   39.82 & $\!\!\pm\!\!$ &   1.13 & $\!\!\pm\!\!$ &   2.20 & 0.331 &  35.0 &   43.33 & $\!\!\pm\!\!$ &   1.13 & $\!\!\pm\!\!$ &   2.02 \\
0.36--0.42 & 0.391 &  25.1 &   38.33 & $\!\!\pm\!\!$ &   1.12 & $\!\!\pm\!\!$ &   2.03 & 0.391 &  35.1 &   35.93 & $\!\!\pm\!\!$ &   1.08 & $\!\!\pm\!\!$ &   1.87 \\
0.42--0.50 & 0.461 &  25.0 &   32.11 & $\!\!\pm\!\!$ &   0.85 & $\!\!\pm\!\!$ &   1.62 & 0.461 &  34.9 &   29.25 & $\!\!\pm\!\!$ &   0.85 & $\!\!\pm\!\!$ &   1.62 \\
0.50--0.60 & 0.551 &  25.0 &   26.83 & $\!\!\pm\!\!$ &   0.70 & $\!\!\pm\!\!$ &   1.43 & 0.550 &  34.9 &   24.21 & $\!\!\pm\!\!$ &   0.69 & $\!\!\pm\!\!$ &   1.37 \\
\hline
 \end{tabular}
 \end{center}
 \end{scriptsize}
 \end{table}
  
 \begin{table}[h]
 \begin{scriptsize}
 \caption{Double-differential inclusive cross-section ${\rm d}^2 \sigma /{\rm d}p{\rm d}\Omega$
  [mb/(GeV/{\it c} sr)] of the production of $\pi^+$'s in $\pi^-$ + C $\rightarrow$ $\pi^+$ + X interactions with $-5.0$~GeV/{\it c} beam momentum; the first error is statistical, the second systematic; $p_{\rm T}$ in GeV/{\it c}, polar angle $\theta$ in degrees.}
 \label{pip.pimc5}
 \begin{center}
 \begin{tabular}{|c||c|c|rcrcr||c|c|rcrcr|}
 \hline
   & \multicolumn{7}{c||}{$20<\theta<30$} & \multicolumn{7}{c|}{$30<\theta<40$} \\
 \hline
 $p_{\rm T}$ & $\langle p_{\rm T} \rangle$ & $\langle \theta \rangle$ & \multicolumn{5}{c||}{${\rm d}^2 \sigma /{\rm d}p{\rm d}\Omega$} &$\langle p_{\rm T} \rangle$ & $\langle \theta \rangle$ & \multicolumn{5}{c|}{${\rm d}^2 \sigma /{\rm d}p{\rm d}\Omega$} \\
 \hline
0.10--0.13 & 0.115 &  24.9 &   44.58 & $\!\!\pm\!\!$ &   1.87 & $\!\!\pm\!\!$ &   3.29 & 0.116 &  35.0 &   41.43 & $\!\!\pm\!\!$ &   1.85 & $\!\!\pm\!\!$ &   3.26 \\
0.13--0.16 & 0.146 &  24.7 &   55.25 & $\!\!\pm\!\!$ &   1.91 & $\!\!\pm\!\!$ &   3.30 & 0.145 &  34.8 &   43.88 & $\!\!\pm\!\!$ &   1.67 & $\!\!\pm\!\!$ &   2.69 \\
0.16--0.20 & 0.181 &  24.8 &   64.72 & $\!\!\pm\!\!$ &   1.71 & $\!\!\pm\!\!$ &   3.36 & 0.180 &  34.6 &   52.85 & $\!\!\pm\!\!$ &   1.58 & $\!\!\pm\!\!$ &   2.78 \\
0.20--0.24 & 0.220 &  24.7 &   66.54 & $\!\!\pm\!\!$ &   1.71 & $\!\!\pm\!\!$ &   3.11 & 0.221 &  34.7 &   51.21 & $\!\!\pm\!\!$ &   1.48 & $\!\!\pm\!\!$ &   2.40 \\
0.24--0.30 & 0.270 &  24.7 &   68.71 & $\!\!\pm\!\!$ &   1.40 & $\!\!\pm\!\!$ &   2.85 & 0.270 &  34.6 &   47.25 & $\!\!\pm\!\!$ &   1.17 & $\!\!\pm\!\!$ &   1.96 \\
0.30--0.36 & 0.330 &  24.7 &   57.90 & $\!\!\pm\!\!$ &   1.30 & $\!\!\pm\!\!$ &   2.28 & 0.331 &  34.7 &   41.78 & $\!\!\pm\!\!$ &   1.10 & $\!\!\pm\!\!$ &   1.62 \\
\hline
 \end{tabular}
 \end{center}
 \end{scriptsize}
 \end{table}
  
 \begin{table}[h]
 \begin{scriptsize}
 \caption{Double-differential inclusive cross-section ${\rm d}^2 \sigma /{\rm d}p{\rm d}\Omega$
  [mb/(GeV/{\it c} sr)] of the production of $\pi^-$'s in $\pi^-$ + C $\rightarrow$ $\pi^-$ + X interactions with $-5.0$~GeV/{\it c} beam momentum; the first error is statistical, the second systematic; $p_{\rm T}$ in GeV/{\it c}, polar angle $\theta$ in degrees.}
 \label{pim.pimc5}
 \begin{center}
 \begin{tabular}{|c||c|c|rcrcr||c|c|rcrcr|}
 \hline
   & \multicolumn{7}{c||}{$20<\theta<30$} & \multicolumn{7}{c|}{$30<\theta<40$} \\
 \hline
 $p_{\rm T}$ & $\langle p_{\rm T} \rangle$ & $\langle \theta \rangle$ & \multicolumn{5}{c||}{${\rm d}^2 \sigma /{\rm d}p{\rm d}\Omega$} &$\langle p_{\rm T} \rangle$ & $\langle \theta \rangle$ & \multicolumn{5}{c|}{${\rm d}^2 \sigma /{\rm d}p{\rm d}\Omega$} \\
 \hline
0.10--0.13 & 0.116 &  24.7 &   79.37 & $\!\!\pm\!\!$ &   2.53 & $\!\!\pm\!\!$ &   5.82 & 0.115 &  34.8 &   65.20 & $\!\!\pm\!\!$ &   2.32 & $\!\!\pm\!\!$ &   4.97 \\
0.13--0.16 & 0.145 &  24.8 &   92.14 & $\!\!\pm\!\!$ &   2.50 & $\!\!\pm\!\!$ &   5.50 & 0.145 &  34.8 &   74.34 & $\!\!\pm\!\!$ &   2.28 & $\!\!\pm\!\!$ &   4.58 \\
0.16--0.20 & 0.180 &  24.8 &  105.52 & $\!\!\pm\!\!$ &   2.22 & $\!\!\pm\!\!$ &   5.41 & 0.180 &  34.8 &   78.17 & $\!\!\pm\!\!$ &   1.91 & $\!\!\pm\!\!$ &   4.06 \\
0.20--0.24 & 0.219 &  24.9 &  108.37 & $\!\!\pm\!\!$ &   2.22 & $\!\!\pm\!\!$ &   4.90 & 0.219 &  34.6 &   82.80 & $\!\!\pm\!\!$ &   1.93 & $\!\!\pm\!\!$ &   3.79 \\
0.24--0.30 & 0.268 &  24.9 &  101.74 & $\!\!\pm\!\!$ &   1.73 & $\!\!\pm\!\!$ &   4.02 & 0.268 &  34.7 &   73.64 & $\!\!\pm\!\!$ &   1.48 & $\!\!\pm\!\!$ &   2.94 \\
0.30--0.36 & 0.328 &  24.7 &   89.10 & $\!\!\pm\!\!$ &   1.62 & $\!\!\pm\!\!$ &   3.24 & 0.328 &  34.7 &   65.15 & $\!\!\pm\!\!$ &   1.37 & $\!\!\pm\!\!$ &   2.40 \\
\hline
 \end{tabular}
 \end{center}
 \end{scriptsize}
 \end{table}
\clearpage

%%%%%%%%%%%%%%%%%%%%%%%%%%%%%%%%%%%%%%
% Here start the 8 GeV/c tables
%%%%%%%%%%%%%%%%%%%%%%%%%%%%%%%%%%%%%%

  
 \begin{table}[h]
 \begin{scriptsize}
 \caption{Double-differential inclusive cross-section ${\rm d}^2 \sigma /{\rm d}p{\rm d}\Omega$
  [mb/(GeV/{\it c} sr)] of the production of protons in p + C $\rightarrow$ p + X interactions with $+8.0$~GeV/{\it c} beam momentum; the first error is statistical, the second systematic; $p_{\rm T}$ in GeV/{\it c}, polar angle $\theta$ in degrees.}
 \label{pro.proc8}
 \begin{center}
 \begin{tabular}{|c||c|c|rcrcr||c|c|rcrcr|}
 \hline
   & \multicolumn{7}{c||}{$20<\theta<30$} & \multicolumn{7}{c|}{$30<\theta<40$} \\
 \hline
 $p_{\rm T}$ & $\langle p_{\rm T} \rangle$ & $\langle \theta \rangle$ & \multicolumn{5}{c||}{${\rm d}^2 \sigma /{\rm d}p{\rm d}\Omega$} &$\langle p_{\rm T} \rangle$ & $\langle \theta \rangle$ & \multicolumn{5}{c|}{${\rm d}^2 \sigma /{\rm d}p{\rm d}\Omega$} \\
 \hline
0.20--0.24 & 0.220 &  24.8 &   69.70 & $\!\!\pm\!\!$ &   1.83 & $\!\!\pm\!\!$ &   3.80 &  &  &  \multicolumn{5}{c|}{ } \\
0.24--0.30 & 0.270 &  25.0 &   68.40 & $\!\!\pm\!\!$ &   1.49 & $\!\!\pm\!\!$ &   3.47 & 0.270 &  34.9 &   67.75 & $\!\!\pm\!\!$ &   1.45 & $\!\!\pm\!\!$ &   3.17 \\
0.30--0.36 & 0.330 &  25.0 &   63.66 & $\!\!\pm\!\!$ &   1.47 & $\!\!\pm\!\!$ &   3.16 & 0.330 &  34.9 &   61.87 & $\!\!\pm\!\!$ &   1.39 & $\!\!\pm\!\!$ &   2.65 \\
0.36--0.42 & 0.389 &  24.9 &   58.48 & $\!\!\pm\!\!$ &   1.38 & $\!\!\pm\!\!$ &   2.72 & 0.389 &  35.0 &   53.52 & $\!\!\pm\!\!$ &   1.32 & $\!\!\pm\!\!$ &   2.44 \\
0.42--0.50 & 0.458 &  25.0 &   52.22 & $\!\!\pm\!\!$ &   1.12 & $\!\!\pm\!\!$ &   2.28 & 0.459 &  34.9 &   45.93 & $\!\!\pm\!\!$ &   1.08 & $\!\!\pm\!\!$ &   2.07 \\
0.50--0.60 & 0.548 &  24.9 &   45.90 & $\!\!\pm\!\!$ &   0.94 & $\!\!\pm\!\!$ &   1.97 & 0.548 &  34.9 &   39.32 & $\!\!\pm\!\!$ &   0.89 & $\!\!\pm\!\!$ &   1.79 \\
\hline
 \end{tabular}
 \end{center}
 \end{scriptsize}
 \end{table}
  
 \begin{table}[h]
 \begin{scriptsize}
 \caption{Double-differential inclusive cross-section ${\rm d}^2 \sigma /{\rm d}p{\rm d}\Omega$
  [mb/(GeV/{\it c} sr)] of the production of $\pi^+$'s in p + C $\rightarrow$ $\pi^+$ + X interactions with $+8.0$~GeV/{\it c} beam momentum; the first error is statistical, the second systematic; $p_{\rm T}$ in GeV/{\it c}, polar angle $\theta$ in degrees.}
 \label{pip.proc8}
 \begin{center}
 \begin{tabular}{|c||c|c|rcrcr||c|c|rcrcr|}
 \hline
   & \multicolumn{7}{c||}{$20<\theta<30$} & \multicolumn{7}{c|}{$30<\theta<40$} \\
 \hline
 $p_{\rm T}$ & $\langle p_{\rm T} \rangle$ & $\langle \theta \rangle$ & \multicolumn{5}{c||}{${\rm d}^2 \sigma /{\rm d}p{\rm d}\Omega$} &$\langle p_{\rm T} \rangle$ & $\langle \theta \rangle$ & \multicolumn{5}{c|}{${\rm d}^2 \sigma /{\rm d}p{\rm d}\Omega$} \\
 \hline
0.10--0.13 & 0.116 &  24.7 &   67.33 & $\!\!\pm\!\!$ &   2.48 & $\!\!\pm\!\!$ &   4.90 & 0.116 &  34.8 &   47.81 & $\!\!\pm\!\!$ &   1.98 & $\!\!\pm\!\!$ &   3.51 \\
0.13--0.16 & 0.145 &  24.6 &   88.97 & $\!\!\pm\!\!$ &   2.61 & $\!\!\pm\!\!$ &   5.17 & 0.145 &  34.7 &   61.56 & $\!\!\pm\!\!$ &   2.16 & $\!\!\pm\!\!$ &   3.61 \\
0.16--0.20 & 0.180 &  24.5 &  100.88 & $\!\!\pm\!\!$ &   2.31 & $\!\!\pm\!\!$ &   5.09 & 0.180 &  34.8 &   68.63 & $\!\!\pm\!\!$ &   1.92 & $\!\!\pm\!\!$ &   3.47 \\
0.20--0.24 & 0.220 &  24.7 &  104.16 & $\!\!\pm\!\!$ &   2.31 & $\!\!\pm\!\!$ &   4.70 & 0.220 &  34.6 &   68.98 & $\!\!\pm\!\!$ &   1.87 & $\!\!\pm\!\!$ &   3.10 \\
0.24--0.30 & 0.269 &  24.7 &   91.13 & $\!\!\pm\!\!$ &   1.74 & $\!\!\pm\!\!$ &   3.65 & 0.269 &  34.7 &   66.94 & $\!\!\pm\!\!$ &   1.49 & $\!\!\pm\!\!$ &   2.65 \\
0.30--0.36 & 0.329 &  24.8 &   79.00 & $\!\!\pm\!\!$ &   1.62 & $\!\!\pm\!\!$ &   2.92 & 0.330 &  34.7 &   53.00 & $\!\!\pm\!\!$ &   1.32 & $\!\!\pm\!\!$ &   1.94 \\
\hline
 \end{tabular}
 \end{center}
 \end{scriptsize}
 \end{table}
  
 \begin{table}[h]
 \begin{scriptsize}
 \caption{Double-differential inclusive cross-section ${\rm d}^2 \sigma /{\rm d}p{\rm d}\Omega$
  [mb/(GeV/{\it c} sr)] of the production of $\pi^-$'s in p + C $\rightarrow$ $\pi^-$ + X interactions with $+8.0$~GeV/{\it c} beam momentum; the first error is statistical, the second systematic; $p_{\rm T}$ in GeV/{\it c}, polar angle $\theta$ in degrees.}
 \label{pim.proc8}
 \begin{center}
 \begin{tabular}{|c||c|c|rcrcr||c|c|rcrcr|}
 \hline
   & \multicolumn{7}{c||}{$20<\theta<30$} & \multicolumn{7}{c|}{$30<\theta<40$} \\
 \hline
 $p_{\rm T}$ & $\langle p_{\rm T} \rangle$ & $\langle \theta \rangle$ & \multicolumn{5}{c||}{${\rm d}^2 \sigma /{\rm d}p{\rm d}\Omega$} &$\langle p_{\rm T} \rangle$ & $\langle \theta \rangle$ & \multicolumn{5}{c|}{${\rm d}^2 \sigma /{\rm d}p{\rm d}\Omega$} \\
 \hline
0.10--0.13 & 0.116 &  24.8 &   63.34 & $\!\!\pm\!\!$ &   2.33 & $\!\!\pm\!\!$ &   4.58 & 0.115 &  34.7 &   40.44 & $\!\!\pm\!\!$ &   1.75 & $\!\!\pm\!\!$ &   3.06 \\
0.13--0.16 & 0.146 &  24.8 &   70.50 & $\!\!\pm\!\!$ &   2.25 & $\!\!\pm\!\!$ &   4.16 & 0.145 &  34.7 &   50.61 & $\!\!\pm\!\!$ &   1.86 & $\!\!\pm\!\!$ &   3.06 \\
0.16--0.20 & 0.180 &  24.8 &   79.80 & $\!\!\pm\!\!$ &   2.01 & $\!\!\pm\!\!$ &   4.07 & 0.180 &  34.6 &   56.88 & $\!\!\pm\!\!$ &   1.69 & $\!\!\pm\!\!$ &   2.95 \\
0.20--0.24 & 0.220 &  24.7 &   75.05 & $\!\!\pm\!\!$ &   1.92 & $\!\!\pm\!\!$ &   3.37 & 0.220 &  34.7 &   56.93 & $\!\!\pm\!\!$ &   1.66 & $\!\!\pm\!\!$ &   2.59 \\
0.24--0.30 & 0.269 &  24.8 &   65.88 & $\!\!\pm\!\!$ &   1.47 & $\!\!\pm\!\!$ &   2.58 & 0.268 &  34.6 &   53.74 & $\!\!\pm\!\!$ &   1.31 & $\!\!\pm\!\!$ &   2.12 \\
0.30--0.36 & 0.328 &  24.9 &   56.88 & $\!\!\pm\!\!$ &   1.37 & $\!\!\pm\!\!$ &   2.06 & 0.329 &  34.8 &   41.28 & $\!\!\pm\!\!$ &   1.15 & $\!\!\pm\!\!$ &   1.50 \\
\hline
 \end{tabular}
 \end{center}
 \end{scriptsize}
 \end{table}
  
 \begin{table}[h]
 \begin{scriptsize}
 \caption{Double-differential inclusive cross-section ${\rm d}^2 \sigma /{\rm d}p{\rm d}\Omega$
  [mb/(GeV/{\it c} sr)] of the production of protons in $\pi^+$ + C $\rightarrow$ p + X interactions with $+8.0$~GeV/{\it c} beam momentum; the first error is statistical, the second systematic; $p_{\rm T}$ in GeV/{\it c}, polar angle $\theta$ in degrees.}
 \label{pro.pipc8}
 \begin{center}
 \begin{tabular}{|c||c|c|rcrcr||c|c|rcrcr|}
 \hline
   & \multicolumn{7}{c||}{$20<\theta<30$} & \multicolumn{7}{c|}{$30<\theta<40$} \\
 \hline
 $p_{\rm T}$ & $\langle p_{\rm T} \rangle$ & $\langle \theta \rangle$ & \multicolumn{5}{c||}{${\rm d}^2 \sigma /{\rm d}p{\rm d}\Omega$} &$\langle p_{\rm T} \rangle$ & $\langle \theta \rangle$ & \multicolumn{5}{c|}{${\rm d}^2 \sigma /{\rm d}p{\rm d}\Omega$} \\
 \hline
0.20--0.24 & 0.220 &  25.0 &   59.14 & $\!\!\pm\!\!$ &   2.39 & $\!\!\pm\!\!$ &   3.25 &  &  &  \multicolumn{5}{c|}{ } \\
0.24--0.30 & 0.270 &  25.0 &   56.96 & $\!\!\pm\!\!$ &   1.91 & $\!\!\pm\!\!$ &   2.92 & 0.270 &  34.8 &   50.77 & $\!\!\pm\!\!$ &   1.75 & $\!\!\pm\!\!$ &   2.40 \\
0.30--0.36 & 0.330 &  24.9 &   48.65 & $\!\!\pm\!\!$ &   1.79 & $\!\!\pm\!\!$ &   2.44 & 0.330 &  35.0 &   45.74 & $\!\!\pm\!\!$ &   1.68 & $\!\!\pm\!\!$ &   1.98 \\
0.36--0.42 & 0.390 &  25.1 &   41.66 & $\!\!\pm\!\!$ &   1.64 & $\!\!\pm\!\!$ &   1.97 & 0.389 &  35.0 &   38.91 & $\!\!\pm\!\!$ &   1.58 & $\!\!\pm\!\!$ &   1.80 \\
0.42--0.50 & 0.459 &  24.9 &   36.67 & $\!\!\pm\!\!$ &   1.29 & $\!\!\pm\!\!$ &   1.63 & 0.458 &  34.9 &   36.26 & $\!\!\pm\!\!$ &   1.35 & $\!\!\pm\!\!$ &   1.67 \\
0.50--0.60 & 0.548 &  24.9 &   32.99 & $\!\!\pm\!\!$ &   1.10 & $\!\!\pm\!\!$ &   1.44 & 0.548 &  34.9 &   26.81 & $\!\!\pm\!\!$ &   1.02 & $\!\!\pm\!\!$ &   1.24 \\
\hline
 \end{tabular}
 \end{center}
 \end{scriptsize}
 \end{table}
  
 \begin{table}[h]
 \begin{scriptsize}
 \caption{Double-differential inclusive cross-section ${\rm d}^2 \sigma /{\rm d}p{\rm d}\Omega$
  [mb/(GeV/{\it c} sr)] of the production of $\pi^+$'s in $\pi^+$ + C $\rightarrow$ $\pi^+$ + X interactions with $+8.0$~GeV/{\it c} beam momentum; the first error is statistical, the second systematic; $p_{\rm T}$ in GeV/{\it c}, polar angle $\theta$ in degrees.}
 \label{pip.pipc8}
 \begin{center}
 \begin{tabular}{|c||c|c|rcrcr||c|c|rcrcr|}
 \hline
   & \multicolumn{7}{c||}{$20<\theta<30$} & \multicolumn{7}{c|}{$30<\theta<40$} \\
 \hline
 $p_{\rm T}$ & $\langle p_{\rm T} \rangle$ & $\langle \theta \rangle$ & \multicolumn{5}{c||}{${\rm d}^2 \sigma /{\rm d}p{\rm d}\Omega$} &$\langle p_{\rm T} \rangle$ & $\langle \theta \rangle$ & \multicolumn{5}{c|}{${\rm d}^2 \sigma /{\rm d}p{\rm d}\Omega$} \\
 \hline
0.10--0.13 & 0.116 &  24.5 &   73.87 & $\!\!\pm\!\!$ &   3.64 & $\!\!\pm\!\!$ &   5.40 & 0.115 &  35.1 &   43.74 & $\!\!\pm\!\!$ &   2.67 & $\!\!\pm\!\!$ &   3.20 \\
0.13--0.16 & 0.146 &  24.8 &   86.93 & $\!\!\pm\!\!$ &   3.61 & $\!\!\pm\!\!$ &   5.09 & 0.146 &  34.7 &   56.44 & $\!\!\pm\!\!$ &   2.88 & $\!\!\pm\!\!$ &   3.31 \\
0.16--0.20 & 0.180 &  24.6 &  103.34 & $\!\!\pm\!\!$ &   3.28 & $\!\!\pm\!\!$ &   5.25 & 0.181 &  35.1 &   69.00 & $\!\!\pm\!\!$ &   2.70 & $\!\!\pm\!\!$ &   3.51 \\
0.20--0.24 & 0.220 &  24.8 &  113.64 & $\!\!\pm\!\!$ &   3.38 & $\!\!\pm\!\!$ &   5.15 & 0.220 &  34.7 &   74.95 & $\!\!\pm\!\!$ &   2.74 & $\!\!\pm\!\!$ &   3.37 \\
0.24--0.30 & 0.270 &  24.6 &  109.58 & $\!\!\pm\!\!$ &   2.68 & $\!\!\pm\!\!$ &   4.43 & 0.269 &  34.7 &   72.77 & $\!\!\pm\!\!$ &   2.17 & $\!\!\pm\!\!$ &   2.90 \\
0.30--0.36 & 0.330 &  24.8 &   96.55 & $\!\!\pm\!\!$ &   2.51 & $\!\!\pm\!\!$ &   3.59 & 0.330 &  34.5 &   61.52 & $\!\!\pm\!\!$ &   1.99 & $\!\!\pm\!\!$ &   2.26 \\
\hline
 \end{tabular}
 \end{center}
 \end{scriptsize}
 \end{table}
  
 \begin{table}[h]
 \begin{scriptsize}
 \caption{Double-differential inclusive cross-section ${\rm d}^2 \sigma /{\rm d}p{\rm d}\Omega$
  [mb/(GeV/{\it c} sr)] of the production of $\pi^-$'s in $\pi^+$ + C $\rightarrow$ $\pi^-$ + X interactions with $+8.0$~GeV/{\it c} beam momentum; the first error is statistical, the second systematic; $p_{\rm T}$ in GeV/{\it c}, polar angle $\theta$ in degrees.}
 \label{pim.pipc8}
 \begin{center}
 \begin{tabular}{|c||c|c|rcrcr||c|c|rcrcr|}
 \hline
   & \multicolumn{7}{c||}{$20<\theta<30$} & \multicolumn{7}{c|}{$30<\theta<40$} \\
 \hline
 $p_{\rm T}$ & $\langle p_{\rm T} \rangle$ & $\langle \theta \rangle$ & \multicolumn{5}{c||}{${\rm d}^2 \sigma /{\rm d}p{\rm d}\Omega$} &$\langle p_{\rm T} \rangle$ & $\langle \theta \rangle$ & \multicolumn{5}{c|}{${\rm d}^2 \sigma /{\rm d}p{\rm d}\Omega$} \\
 \hline
0.10--0.13 & 0.115 &  24.5 &   58.25 & $\!\!\pm\!\!$ &   3.13 & $\!\!\pm\!\!$ &   4.25 & 0.116 &  34.9 &   36.92 & $\!\!\pm\!\!$ &   2.35 & $\!\!\pm\!\!$ &   2.79 \\
0.13--0.16 & 0.145 &  24.9 &   65.66 & $\!\!\pm\!\!$ &   3.09 & $\!\!\pm\!\!$ &   3.90 & 0.145 &  34.5 &   49.88 & $\!\!\pm\!\!$ &   2.60 & $\!\!\pm\!\!$ &   3.02 \\
0.16--0.20 & 0.180 &  24.8 &   78.25 & $\!\!\pm\!\!$ &   2.82 & $\!\!\pm\!\!$ &   4.00 & 0.180 &  34.7 &   55.01 & $\!\!\pm\!\!$ &   2.33 & $\!\!\pm\!\!$ &   2.86 \\
0.20--0.24 & 0.220 &  24.6 &   75.26 & $\!\!\pm\!\!$ &   2.71 & $\!\!\pm\!\!$ &   3.39 & 0.220 &  34.6 &   51.20 & $\!\!\pm\!\!$ &   2.21 & $\!\!\pm\!\!$ &   2.34 \\
0.24--0.30 & 0.268 &  24.7 &   71.48 & $\!\!\pm\!\!$ &   2.15 & $\!\!\pm\!\!$ &   2.82 & 0.269 &  34.7 &   51.75 & $\!\!\pm\!\!$ &   1.82 & $\!\!\pm\!\!$ &   2.06 \\
0.30--0.36 & 0.329 &  25.0 &   61.31 & $\!\!\pm\!\!$ &   2.00 & $\!\!\pm\!\!$ &   2.24 & 0.329 &  34.6 &   40.08 & $\!\!\pm\!\!$ &   1.59 & $\!\!\pm\!\!$ &   1.46 \\
\hline
 \end{tabular}
 \end{center}
 \end{scriptsize}
 \end{table}
  
 \begin{table}[h]
 \begin{scriptsize}
 \caption{Double-differential inclusive cross-section ${\rm d}^2 \sigma /{\rm d}p{\rm d}\Omega$
  [mb/(GeV/{\it c} sr)] of the production of protons in $\pi^-$ + C $\rightarrow$ p + X interactions with $-8.0$~GeV/{\it c} beam momentum; the first error is statistical, the second systematic; $p_{\rm T}$ in GeV/{\it c}, polar angle $\theta$ in degrees.}
 \label{pro.pimc8}
 \begin{center}
 \begin{tabular}{|c||c|c|rcrcr||c|c|rcrcr|}
 \hline
   & \multicolumn{7}{c||}{$20<\theta<30$} & \multicolumn{7}{c|}{$30<\theta<40$} \\
 \hline
 $p_{\rm T}$ & $\langle p_{\rm T} \rangle$ & $\langle \theta \rangle$ & \multicolumn{5}{c||}{${\rm d}^2 \sigma /{\rm d}p{\rm d}\Omega$} &$\langle p_{\rm T} \rangle$ & $\langle \theta \rangle$ & \multicolumn{5}{c|}{${\rm d}^2 \sigma /{\rm d}p{\rm d}\Omega$} \\
 \hline
0.20--0.24 & 0.220 &  25.1 &   44.49 & $\!\!\pm\!\!$ &   1.35 & $\!\!\pm\!\!$ &   2.52 &  &  &  \multicolumn{5}{c|}{ } \\
0.24--0.30 & 0.268 &  24.9 &   40.03 & $\!\!\pm\!\!$ &   1.03 & $\!\!\pm\!\!$ &   2.11 & 0.270 &  34.9 &   42.54 & $\!\!\pm\!\!$ &   1.03 & $\!\!\pm\!\!$ &   2.09 \\
0.30--0.36 & 0.328 &  25.1 &   36.04 & $\!\!\pm\!\!$ &   0.99 & $\!\!\pm\!\!$ &   1.86 & 0.328 &  34.9 &   37.49 & $\!\!\pm\!\!$ &   0.98 & $\!\!\pm\!\!$ &   1.70 \\
0.36--0.42 & 0.388 &  25.0 &   33.13 & $\!\!\pm\!\!$ &   0.94 & $\!\!\pm\!\!$ &   1.64 & 0.389 &  35.0 &   32.05 & $\!\!\pm\!\!$ &   0.92 & $\!\!\pm\!\!$ &   1.55 \\
0.42--0.50 & 0.456 &  24.9 &   29.65 & $\!\!\pm\!\!$ &   0.76 & $\!\!\pm\!\!$ &   1.40 & 0.457 &  35.0 &   25.59 & $\!\!\pm\!\!$ &   0.73 & $\!\!\pm\!\!$ &   1.26 \\
0.50--0.60 & 0.545 &  25.0 &   24.02 & $\!\!\pm\!\!$ &   0.60 & $\!\!\pm\!\!$ &   1.08 & 0.544 &  34.9 &   21.74 & $\!\!\pm\!\!$ &   0.59 & $\!\!\pm\!\!$ &   1.09 \\
\hline
 \end{tabular}
 \end{center}
 \end{scriptsize}
 \end{table}
\input{table.pip.pimc8.tex}
\input{table.pim.pimc8.tex}
\clearpage

%%%%%%%%%%%%%%%%%%%%%%%%%%%%%%%%%%%%%%
% Here start the 12 GeV/c tables
%%%%%%%%%%%%%%%%%%%%%%%%%%%%%%%%%%%%%%

\input{table.pro.proc12.tex}
\input{table.pip.proc12.tex}
\input{table.pim.proc12.tex}
\input{table.pro.pipc12.tex}
\input{table.pip.pipc12.tex}
\input{table.pim.pipc12.tex}
\input{table.pro.pimc12.tex}
\input{table.pip.pimc12.tex}
\input{table.pim.pimc12.tex}
\clearpage

%%%%%%%%%%%%%%%%%%%%%%%%%%%%%%%%%%%%%%
% Here start the 15 GeV/c tables
%%%%%%%%%%%%%%%%%%%%%%%%%%%%%%%%%%%%%%

\input{table.pro.proc15.tex}
\input{table.pip.proc15.tex}
\input{table.pim.proc15.tex}
\input{table.pro.pipc15.tex}
\input{table.pip.pipc15.tex}
\input{table.pim.pipc15.tex}
\input{table.pro.pimc15.tex}
\input{table.pip.pimc15.tex}
\input{table.pim.pimc15.tex}
\clearpage


\begin{thebibliography}{99}

\bibitem{neutrinofactory} M.~Apollonio {\it et al.}, 
J. Instrum. {\bf 4} (2009) P07001

\bibitem{Beryllium1} A.~Bolshakova {\it et al.}, Eur. Phys. J. {\bf C62} (2009) 293 
(CERN-PH-EP-2008-022, arXiv:0901.3648) 

\bibitem{Beryllium2} A.~Bolshakova {\it et al.}, Eur. Phys. J. {\bf C62} (2009) 697 
(CERN-PH-EP-2008-025, arXiv:0903.2145) 

\bibitem{Tantalum} A.~Bolshakova {\it et al.}, Eur. Phys. J. {\bf C63} (2009) 549
(CERN-PH-EP-2009-009, arXiv:0906.0471) 

\bibitem{Copper} A.~Bolshakova {\it et al.}, Eur. Phys. J. {\bf C64} (2009) 181
(CERN-PH-EP-2009-012, arXiv:0906.3653) 

\bibitem{Lead} A.~Bolshakova {\it et al.}, Eur. Phys. J. {\bf C66} (2010) 57
(CERN-PH-EP-2009-025, arXiv:0912.0378v1)

\bibitem{TPCpub} V.~Ammosov {\it et al.},
Nucl. Instrum. Methods Phys. Res. {\bf A588} (2008) 294

\bibitem{RPCpub} V.~Ammosov {\it et al.},
Nucl. Instrum. Methods Phys. Res. {\bf A578} (2007) 119 

\bibitem{Geant4} S.~Agostinelli {\it et al.}, 
Nucl. Instrum. Methods Phys. Res. {\bf A506} (2003) 250; 
J.~Allison {\it et al.}, IEEE Trans. Nucl. Sci. {\bf 53} (2006) 270

\bibitem{GEANTpub} A.~Bolshakova {\it et al.},   
Eur. Phys. J. {\bf C56} (2008) 323

\bibitem{ASCIItables} A.~Bolshakova {\it et al.}, Tables of cross-sections of large-angle hadron production in proton-- and pion--nucleus interactions VI: 
carbon nuclei and beam momenta from $\pm$3 GeV/{\it c}  to $\pm$15 GeV/{\it c},
CERN--HARP--CDP--2010--001

\bibitem{WebsitePDG2009} http://pdg.lbl.gov/2009/AtomicNuclearProperties

\bibitem{OffLAprotonpaper} M.G.~Catanesi {\it et al.}, Phys. Rev. {\bf C77} (2008) 055207
(arXiv:0805.2871)

\bibitem{OffLApionpaper} M.~Apollonio {\it et al.}, Phys. Rev. {\bf C80} (2009) 065207
(arXiv:0907.1428)

\bibitem{JINSTpub} V.~Ammosov {\it et al.}, 
J. Instrum. {\bf 3} (2008) P01002

\bibitem{EPJCpub} V.~Ammosov {\it et al.},
Eur. Phys. J. {\bf C54} (2008) 169

\bibitem{WhiteBook1} V.~Ammosov {\it et al.}, 
CERN--HARP--CDP--2006--003

\bibitem{WhiteBook2} V.~Ammosov {\it et al.}, 
CERN--HARP--CDP--2006--007

\bibitem{WhiteBook3} V.~Ammosov {\it et al.}, 
CERN--HARP--CDP--2007--001

\bibitem{Meshkovski1}  A.G.~Meshkovski, Ya.Ya.~Shalamov, V.A.~Shebanov,
J. Exp. Theor. Phys. {\bf 33} (1957) 602 (in Russian)

\bibitem{Meshkovski2}  A.G.~Meshkovski {\it et al.},
J. Exp. Theor. Phys. {\bf 31} (1956) 987 (in Russian)

\bibitem{Oganesian}  K.O.~Oganesian,
J. Exp. Theor. Phys. {\bf 54} (1968) 1273 (in Russian)

\end{thebibliography}
\end{document}